\documentclass[a4paper,11pt]{article}
\pdfoutput=1 

\usepackage{jheppub} 

\usepackage[T1]{fontenc} 
\usepackage{amssymb,amsmath,bm,natbib}
\usepackage{color}
\usepackage{slashed}
\usepackage{graphics}
\usepackage{graphicx}
\usepackage[utf8]{inputenc}
\usepackage[caption=false]{subfig}
\usepackage{hyperref}
\usepackage{url}
\usepackage{dsfont}
\usepackage{float}
\usepackage{cancel}
\usepackage{units}
\usepackage{blindtext}
\usepackage[utf8]{inputenc}
\usepackage{upgreek}
\usepackage{booktabs}
\usepackage[dvipsnames,table,xcdraw]{xcolor}
\usepackage{enumerate}

\usepackage[normalem]{ulem}

\renewcommand{\eqref}[1]{\mbox{Eq.~(\ref{#1})}}

\title{Hamiltonian formulation of an effective modified gravity with nondynamical background fields}

\author[a]{Carlos M. Reyes}
\author[b]{and Marco Schreck}

\affiliation[a]{Centro de Ciencias Exactas, Universidad del B\'{i}o-B\'{i}o, Avda. Andr\'es Bello 720, 3800708, Chill\'{a}n, Chile}
\affiliation[b]{Departamento de F\'{i}sica, Universidade Federal do Maranh\~{a}o, Campus Universit\'{a}rio do Bacanga, S\~{a}o Lu\'{i}s (MA), 65085-580, Brazil}

\emailAdd{creyes@ubiobio.cl}
\emailAdd{marco.schreck@ufma.br}

\abstract{The current paper is dedicated to developing a $(3+1)$ decomposition for the minimal gravitational Standard-Model Extension.
Our setting is explicit diffeomorphism violation and we focus on the background fields known in the literature as $u$ and $s^{\mu\nu}$. The Hamiltonian
formalism is developed for these contributions, which amounts to deriving modified Hamiltonian and momentum constraints. We then study the connection
between these modified constraints and the modified Einstein equations. Implications are drawn on the form of the background fields to guarantee
the internal consistency of the corresponding modified-gravity theories. In the course of our analysis, we obtain a set of consistency requirements
for $u$ and certain sectors of $s^{\mu\nu}$. We argue that the constraint structure remains untouched when these conditions are satisfied. Our
results shed light on explicit violations of diffeomorphism invariance and local Lorentz invariance in gravity. They may turn out to be valuable for
developing a better understanding of effective modified-gravity theories.}


\begin{document}
\maketitle
\flushbottom
\section{Introduction}
\label{sec:introduction}
Lorentz invariance is one of the foundations of the current scientific paradigm that has shaped our understanding of
nature at both small and large length scales. The Standard Model (SM) rests on global Lorentz symmetry and provides a
description of elementary particles in terms of quantum fields defined in Minkowski spacetime. This fundamental symmetry implies
that measurements
performed in two identical experiments moving uniformly with respect to each other provide results based on the same laws
of nature connected by a Lorentz transformation. Thus, the form of the laws of nature does not depend on the state of
uniform motion. An analogous property exists for measurements made with an apparatus and an identical rotated one.
The latter viewpoint is called particle Lorentz invariance in the literature.\footnote{The concept ``invariance under active Lorentz
transformations'' is frequently used in Minkowski spacetime, but such a notion does not correspond to particle Lorentz
invariance in the presence of background fields. A background field would transform like a four-tensor under active
Lorentz transformations, but remains fixed under particle Lorentz transformations, since it is beyond control through
experimentalists.}

General Relativity (GR) is a theory based on a dynamical spacetime in which the physical laws are invariant with respect
to diffeomorphisms. It exhibits local Lorentz symmetry in the tangent space $T_p$ at a point $p$ of a spacetime manifold
$\mathcal M$~\cite{Misner:1973,Carroll:1997ar}, such that the results of two measurements of the same quantity are connected
with each other by a local Lorentz transformation. This applies, in particular, when the first is performed in a freely
falling inertial reference frame and the second in a boosted frame with respect to the first.
The same holds true for two freely falling inertial frames whose axes enclose a fixed angle. The vierbein (tetrad) formalism
allows us to transform from a general spacetime frame
described by the metric $g_{\mu\nu}$ of the (curved) manifold $\mathcal{M}$ to a freely falling frame at a particular
spacetime point where the metric corresponds to that of Minkowski spacetime.
Such a transformation makes local Lorentz invariance explicit.

GR also exhibits invariance under general coordinate transformations. The latter relate the same objects in the
manifold, such as points and curves, expressed in different coordinates with each other. In contrast, diffeomorphisms
establish relations between different objects without changing the coordinates.
They can be interpreted as spacetime-dependent translations and are the active counterparts of general coordinate transformations
(see, e.g., p.~133 of~\cite{Carroll:1997ar}). In particular, diffeomorphisms
are represented by smooth maps from a differential manifold onto itself, $\mathcal{M} \to \mathcal{M}$, where the corresponding
inverse maps are also required to be smooth.\footnote{Manifolds that are related by a diffeomorphism can be considered as
geometrically equivalent. A diffeomorphism may change how a manifold is embedded into an
ambient space, but its intrinsic geometry remains untouched.}

Because of diffeomorphism invariance, only two of the ten metric components correspond to physical, propagating degrees of freedom.
This property is made transparent in the $(3+1)$ decomposition of spacetime developed by Arnowitt, Deser,
and Misner (denoted as the ADM decomposition in the remainder of the paper)~\cite{Arnowitt:1962hi,Arnowitt:2008}. In the
ADM formulation, four-dimensional spacetime is foliated with spacelike hypersurfaces
(of constant time) that evolve with respect to time. The foliation itself is governed by a lapse function $N$ and a shift
vector with components $N^i$ and permits constructing a Hamiltonian associated with GR. The conjugate
momenta for the spatial metric components $g_{ij}\equiv q_{ij}$ are nonzero, whereas those associated with $N$ and $N^i$
are zero showing that there are four primary first-class constraints in GR~\cite{Hanson:1976}. The latter exist due to the
covariant structure of this theory. Therefore, $N$ and $N^i$ are unphysical and represent gauge degrees of
freedom. In addition, the Hamiltonian and momentum constraints arise as a set of four secondary first-class
constraints. They generate spacetime diffeomorphisms and spatial diffeomorphisms, respectively.

A violation of (local) Lorentz invariance is the most
prominent signal for physics at the Planck scale, which was shown to arise in particular string field theories~\cite{Kostelecky:1988zi,Kostelecky:1989jp,Kostelecky:1989jw,Kostelecky:1991ak,Kostelecky:1994rn} as well as
in loop quantum gravity~\cite{Gambini:1998it,Bojowald:2004bb}. Furthermore, this effect occurs in other settings such
as noncommutative spacetime structures~\cite{AmelinoCamelia:1999pm,Carroll:2001ws}, spacetime foams~\cite{Klinkhamer:2003ec,Bernadotte:2006ya,Hossenfelder:2014hha},
nontrivial spacetime topologies~\cite{Klinkhamer:1998fa,Klinkhamer:1999zh,Klinkhamer:2002mj,Ghosh:2017iat}, and
Ho\v{r}ava-Lifshitz gravity~\cite{Horava:2009uw}. The strong suppression of Lorentz violation effects at low energy scales
has led to the conception of ultra-sensitive tests for their possible detection \cite{Kostelecky:2008bfz}.

The Standard-Model Extension (SME)~\cite{Colladay:1996iz,Colladay:1998fq,Kostelecky:2000mm} is a comprehensive field
theory framework to parameterize deviations from Lorentz invariance, and it allows us to compare the results of different
experiments with each other. The incorporation of Lorentz violation in the SME is through background fields arising
as vacuum expectation values of tensor-valued fields. The latter imply preferred spacetime directions
and involve controlling coefficients that describe the strength of Lorentz violation. Suitable
contractions of background fields with SM field operators result in expressions invariant under
coordinate transformations (observer Lorentz transformations). The minimal SME includes field operators of mass
dimensions 3 and 4, whereas the nonminimal SME contains field operators
of mass dimensions~$\geq 5$~\cite{Kostelecky:2009zp,Kostelecky:2011gq,Kostelecky:2013rta}. Since the controlling
coefficients in a nongravitational context are usually\footnote{One of the few studies on effects related to
spacetime-dependent coefficients in Minkowski spacetime is provided by~\cite{Lane:2016osk}.} assumed to be independent of the
spacetime coordinates, the SME exhibits translation invariance. As a consequence, energy and momentum are conserved quantities.

In the presence of gravity, the situation is more subtle. The notions of global Lorentz violation and translation
noninvariance in Minkowski spacetime are replaced by two fundamentally distinct concepts: local Lorentz violation and diffeomorphism
violation. To incorporate these concepts into the SME language, a generic background field in a curved spacetime manifold $\mathcal{M}$
can have both contributions defined in a spacetime frame and contributions given in a local inertial reference frame.
A gravitational version of the SME has been put forward in a series of papers~\cite{Kostelecky:2003fs,Bailey:2006fd,Bailey:2014bta,Kostelecky:2016kfm,
Kostelecky:2016uex,Kostelecky:2017zob,Mewes:2019dhj,Kostelecky:2020hbb} to study these aspects.
The spacetime manifold itself is described by the metric tensor $g_{\mu\nu}$ and formal transitions
between a spacetime frame and a local inertial frame are provided by vierbeins
$e_{\mu}^{\phantom{\mu}a}$. In the pure-gravity sector, background fields in spacetime
frames are suitably contracted with objects built from the Riemann
curvature tensor, covariant derivatives, the Levi-Civita tensor, and the
spin connection $\omega_{\mu}^{\phantom{\mu}ab}$ that endows spacetime
with a spin structure. These terms are constructed in a way such that general
coordinate invariance is maintained. The presence of a background field in
a local inertial frame implies that the form of the laws of nature is different in frames that
are boosted or rotated with respect to the original one.

The most recent article~\cite{Kostelecky:2020hbb} rests on a better understanding of Lorentz
violation in gravity acquired since the base of the gravitational SME was laid in~\cite{Kostelecky:2003fs}. It widely
extends the findings in~\cite{Kostelecky:2003fs} and introduces additional concepts such
as global local Lorentz transformations and manifold Lorentz transformations that are
combinations of diffeomorphisms and local Lorentz transformations. The latter can
be considered analogous to global Lorentz transformations in Minkowski spacetime.
An almost flat spacetime setting, which is sufficient for various studies in practice
such as propagating gravitational waves~\cite{Kostelecky:2016kfm,Mewes:2019dhj} or
modified dispersion relations in linearized gravity \cite{Nascimento:2021rlg}, is also introduced.
Therefore,~\cite{Kostelecky:2020hbb} provides many additional possibilities of how to
construct terms leading to observable pure-gravity effects that are not in accordance with GR.
By employing a powerful notation, an infinite number of such
terms covering the mass dimensions $\leq 8$ is stated. Furthermore, contributions involving
individual matter fields as well as those endowed with the entire SM gauge symmetry are compiled, as well.

It is paramount to emphasize that in the setting of nondynamical background fields in gravity, the physics of a
background field depends on whether it is described by a contravariant, covariant
or mixed observer tensor, i.e., the position of spacetime indices plays a crucial role. The
reason is that the spacetime metric changing the index position is a dynamical object
proper, i.e., it must be taken into account in variations of the action. The authors of~\cite{Kostelecky:2020hbb}
emphasize this issue again (although not for the first time). Thus, explicitly Lorentz-violating contributions
formulated in terms of a covariant background field or a contravariant one (in a spacetime frame) must be considered
as distinct models. To avoid conflicts, we will state background fields
with upper indices only as done in~\cite{Kostelecky:2003fs}. Such a distinction is unnecessary for
local Lorentz indices, since the Minkowski metric, which is used to lower and raise these indices, is
a nondynamical object by definition.

In a gravitational field, the notion of a constant background field loses its meaning. While a covariantly
constant background field cannot even be defined in most curved manifolds that are of interest in
gravity~\cite{Kostelecky:2003fs}, even such a field would depend on the spacetime
coordinates. Therefore, apart from local Lorentz violation, the gravitational SME may exhibit
diffeomorphism violation, in general~\cite{Kostelecky:2003fs,Kostelecky:2020hbb}. For the effective modified-gravity
theory\footnote{Reviews on modified-gravity theories are provided by \cite{Petrov:2020,Petrov:2020wgy}. The latter
references also include material on the gravitational SME, but the focus is on models of spontaneous diffeomorphism
and Lorentz violation in this context.} that we will be considering, any variation in the number of degrees
of freedom and any symmetry departure, including a violation of diffeomorphism invariance, are expected to
show up in the constraint structure as well as the Poisson algebra between the canonical variables and the constraints.
This assertion is particularly true when counting the number of degrees of freedom, which depends crucially
on the number of first- and second-class constraints.

In the current paper, we intend to apply the ADM decomposition to two sectors of the minimal
gravitational SME to understand the implications of explicit diffeomorphism
violation on the gravitational degrees of freedom. The ADM decomposition is more than suitable
for such an analysis, since it renders the constraint structure transparent.
Within the effective framework, we consider the following three points as crucial:

\begin{enumerate}

\item[i)] Obtain the standard results of GR in the limit of vanishing controlling coefficients.
\item[ii)]  Maintain the same number of physical, propagating degrees of freedom (2) in the modified-gravity theory.
\item[iii)] Implement the diffeomorphism group $\mathrm{Diff}(\mathcal{M})$ in the sense of
GR as mappings from a manifold onto itself, $\mathcal{M}\to \mathcal{M}$, given by the particle
transformation $f: x^{\mu} \to f^{\mu}(x)$.

\end{enumerate}	

An investigation of the minimal gravitational SME by means of the ADM formulation~\cite{Nilsson:2019eeo,ONeal-Ault:2020ebv} has been
published recently, i.e., our study has some overlap with the latter article. However, we will focus on other aspects --- including the
points mentioned above --- and present the results in a different manner. Note also that our background
fields carry upper spacetime indices compared to those used in the latter papers. Thus, in light of the comments
made previously on the position of spacetime indices in theories with explicit diffeomorphism violation, our model is physically
nonequivalent to the one studied in~\cite{Nilsson:2019eeo,ONeal-Ault:2020ebv}. Furthermore, applying the
ADM formalism does not require working in a weak-field regime, as it was done in~\cite{linham-noncomm} to find the Hamiltonian
using modified Poisson brackets and deformed constraints. Thus, scenarios of strong gravitational fields
in the presence of Lorentz violation could be studied occurring, e.g., during the creation of gravitational
waves~\cite{Xu:2021dcw,Xu:2020zxs}.

The paper is organized as follows. In Sec.~\ref{eq:ADM-decomposition} we explain the concepts and mathematical relationships in
the ADM formalism with an emphasis on those that are of direct relevance for us. Section~\ref{sec:sme-gravity-sector}
provides a summary of the minimal gravitational SME. It is followed by Sec.~\ref{sec:hamiltonian-formulation} that
constitutes the foundation of the article for the subsequent calculations. Here we review how to derive the Hamiltonian of GR and carry out
analogous computations for both the $s^{\mu\nu}$ and the $u$ term of the gravitational SME. These studies imply
modified Hamiltonian and momentum constraints in the presence of the aforementioned background fields.
In Sec.~\ref{eq:field-equations-constraints} we intend to understand how the constraints and the modified Einstein
equations are related with each other. This analysis will enable us to derive requirements for the internal
consistency of a modified-gravity theory resting on explicit diffeomorphism violation. Section~\ref{sec:constraint-structure} is dedicated to a
brief investigation of the constraint structure as well as the Hamilton equations. By evaluating suitable Poisson
brackets between canonical variables and constraints we will demonstrate that the modified constraints still
generate both spacetime diffeomorphisms and spatial diffeomorphisms. Last but not least, our findings are concluded
on in Sec.~\ref{sec:conclusions}.

The main body of the text is dedicated to presenting and interpreting the central results as well as to providing conceptual discussions.
Detailed derivations
and computations are relegated to the appendix. The latter can be skipped by readers who are primarily interested in
the results and their implications, but it may be valuable to researchers who want to base their investigations on the
findings of this article. Appendix~\ref{sec:mathematical-appendix} gives an account on
the most important geometrical formulas that are indispensable to carry out the ADM decomposition of the SME.
Calculational details on constructing the Hamiltonians of both GR and the gravitational SME are presented in
App.~\ref{sec:hamiltonian-formulation-computations}. Appendix~\ref{sec:modified-ADM} states some remarks on a modified
ADM decomposition that plays a role for a subset of the $s^{\mu\nu}$ coefficients.
In the course of the investigations, it has turned out that
suitable boundary terms must be included in the action and the derivations of those are shown in
App.~\ref{sec:boundary-terms}. Appendix~\ref{eq:modified-field-equations-derivation} presents the most crucial
steps in deriving the modified Einstein equations from the action, as they are valuable to understand
the necessity of the boundary terms mentioned before. Subsequently, in
App.~\ref{sec:projection-einstein-equations-modified} we show in a very detailed manner how to relate
the Hamiltonian and momentum constraints to the modified Einstein equations. Useful formulas on the
ADM decomposition of covariant derivatives of background fields are derived here that are probably not to be found
anywhere in the existing literature. Furthermore, in App.~\ref{sec:functional-derivatives-ADM-action-computations} we give
calculational details on how to compute functional derivatives of the ADM-decomposed action. These results
provide further support for our arguments. Last but not least, in App.~\ref{sec:counting-scheme} we introduce a helpful
counting scheme based on
the canonical variables of the ADM decomposition. The latter allowed us to check any relation involving ADM
variables for (dimensional) consistency and turned out to be a useful tool for finding calculational errors.

\section{The ADM decomposition}
\label{eq:ADM-decomposition}
The ADM formulation of General Relativity~\cite{Arnowitt:1962hi,Arnowitt:2008,Misner:1973} furnishes
a decomposition of spacetime into space and time leading to a description of the gravitational phase
space by means of a Hamiltonian.
The Hamiltonian formulation starts
with selecting a special foliation of the generic spacetime manifold $\mathcal M$ that is to be covered
by a chart of coordinates $X^{\mu}$. We start by choosing a temporal coordinate that we call $t$
where $X^0$ does not necessarily correspond to $t$. The spatial coordinates that are employed
for the decomposition will be denoted as $x^i$. By considering a flow of time
\begin{equation}
\label{eq:flow-of-time}
t^{\mu}\equiv \frac{\partial X^{\mu}}{\partial t}\,,
\end{equation}
the four-dimensional manifold $\mathcal{M}$ decomposes into spacelike hypersurfaces $\Sigma_t$ at
fixed times~$t$.

We pick two spacelike hypersurfaces $\Sigma_t$ and $\Sigma_{t+\mathrm{d}t}$ with constant
$t$ and $t+\mathrm{d}t$, respectively. The lapse of proper time between the lower and upper
hypersurface is $N\mathrm{d}t$, which is why the scalar function $N=N(t,x,y,z)$ is called the
lapse in the literature.
We define the unit timelike vector $n^{\mu}$ that indicates the path of proper time $\tau$. In
general, as long as $t$ does not correspond to proper time, there is a misalignment
between $t^{\mu}$ and $n^{\mu}$:
\begin{equation}
\label{eq:shift-four-vector}
N^{\mu}\equiv t^{\mu}-Nn^{\mu}\,.
\end{equation}
The latter vector is called the shift and it depends on both time and the spatial coordinates.

\begin{figure}
\centering
\includegraphics[scale=0.45]{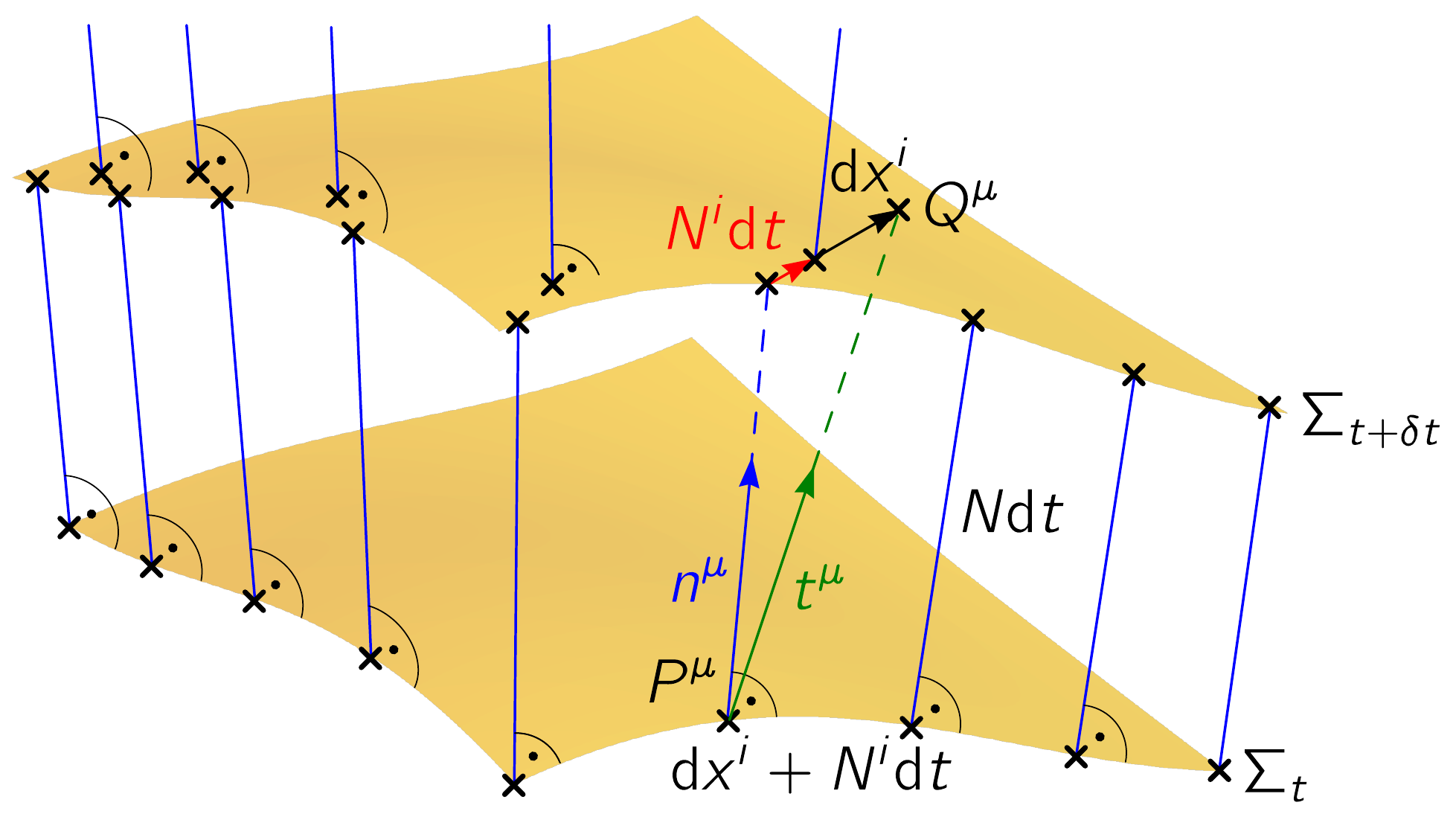}
\caption{Two hypersurfaces $\Sigma_t$ and $\Sigma_{t+\mathrm{d}t}$ that are
linked via connectors illustrated by blue lines. A connector links the point $P^{\mu}=(t,x^i)$
on the lower sheet to the point $Q^{\mu}=(t+\mathrm{d}t,x^i-N^i\mathrm{d}t)$
on the upper one. The axis of a connector points along the direction $n^{\mu}$. The
four-vector $t^{\mu}$ indicates the direction between the point $(t,x^i)$ on the lower
hypersurface and $(t+\mathrm{d}t,x^i+\mathrm{d}x^i)$ on the upper one. The points
where the connectors are wielded are represented by crosses (see also \cite{Misner:1973}).}
\label{fig:adm-construction}
\end{figure}

To grasp a better understanding of the foliation, we provide an illustrative interpretation
of the lapse $N$ and the shift $N^{\mu}$, cf.~Fig.~\ref{fig:adm-construction}.
Let the two hypersurfaces $\Sigma_t$, $\Sigma_{t+\mathrm{d}t}$ be modeled by metal
sheets and let they be separated by connectors wielded at both sheets. The connectors
have a well-defined length $N\mathrm{d}t$ that the lapse is characteristic for. The latter
is not necessarily constant, but depends on which hypersurfaces are connected to each other
as well as where the connectors are placed. We consider a particular connector wielded to the
lower sheet at a point with spatial coordinates $x^i$.

To construct a stable and rigid structure, the connectors must be orthogonal to the lower
sheet at the point $x^i$. As the upper sheet differs from the lower one, this connector
is not necessarily orthogonal to the upper sheet, as well. This also means that the
connector linking $\Sigma_{t+\mathrm{d}t}$ and $\Sigma_{t+2\mathrm{d}t}$ cannot
be placed at the same spatial coordinates $x^i$ on the upper sheet, as it would not
be orthogonal to this sheet at $x^i$. For the connector to be orthogonal, it must be
placed at a point $x^i-N^i\mathrm{d}t$ on $\Sigma_{t+\mathrm{d}t}$. This requirement
introduces a vector $\mathbf{N}$ with components $N^i=N^i(t,x,y,z)$ that is tangent to the sheet
and corresponds to the spatial components of $N^{\mu}$ introduced in \eqref{eq:shift-four-vector}.
We deduce that $N^0=0$, i.e., $N^{\mu}$ is purely spacelike. Also, the shift vector is zero
when the connector is automatically perpendicular to the second sheet such that the next
connector can be placed directly above the previous one. This scenario does not occur in
general, though.

With this construction in mind, we consider a point $(t,x^i)$ on $\Sigma_t$
and move to another point $(t+\mathrm{d}t,x^i+\mathrm{d}x^i)$ on $\Sigma_{t+\mathrm{d}t}$
by following the flow of time; see \eqref{eq:flow-of-time}. Let the geometry
of the hypersurface $\Sigma_t$ be described by the three-metric $q_{ij}=q_{ij}(t,x^i)$.
The latter corresponds to the spatial components of the metric $g_{\mu\nu}=g_{\mu\nu}(t,x^i)$
describing the geometry of the ambient spacetime. To compute the infinitesimal
path length interval squared between the points $(t,x^i)$ and
$(t+\mathrm{d}t,x^i+\mathrm{d}x^i)$, we need the following ingredients.
The infinitesimal distance perpendicular to the lower hypersurface is $N\mathrm{d}t$,
as we have already argued above. Furthermore, the infinitesimal vector pointing from
$x^i$ to $x^i+\mathrm{d}x^i$ in the lower hypersurface is $\mathrm{d}x^i+N^i\mathrm{d}t$.
The path length interval squared then reads
\begin{align}
\label{eq:path-length-interval-ADM-decomposed}
\mathrm{d}s^2&=q_{ij}(\mathrm{d}x^i+N^i\mathrm{d}t)(\mathrm{d}x^j+N^j\mathrm{d}t)-N^2\mathrm{d}t^2 \notag \\
&=q_{ij}\mathrm{d}x^i\mathrm{d}x^j+2N_i\mathrm{d}t\mathrm{d}x^i+(N_aN^a-N^2)\mathrm{d}t^2\,.
\end{align}
The spatial metric $q_{ij}$ is employed to pull indices up and down of the shift vector, as the
latter lives completely in the hypersurface. Identifying the path length interval of
Eq.~(\ref{eq:path-length-interval-ADM-decomposed}) with
\begin{equation}
\mathrm{d}s^2=g_{\mu\nu}\mathrm{d}x^{\mu}\mathrm{d}x^{\nu}\,,
\end{equation}
we identify the decomposed spacetime metric as
\begin{equation}
\label{eq:spacetime-metric}
g_{\mu\nu}=\begin{pmatrix}
N_aN^a-N^2 & N_j \\
N_i & q_{ij} \\
\end{pmatrix}\,.
\end{equation}
Its inverse can be cast into the form
\begin{equation}
g^{\mu\nu}=\begin{pmatrix}
-1/N^2 & N^j/N^2 \\
N^i/N^2 & q^{ij}-N^iN^j/N^2 \\
\end{pmatrix}\,.
\end{equation}
The four-vector $n^{\mu}$ points along the difference between the two
points $P^{\mu}$ and $Q^{\mu}$ lying on top of each other:
\begin{equation}
Q^{\mu}-P^{\mu}\equiv Nn^{\mu}\mathrm{d}t=\begin{pmatrix}
1 \\
-N^i \\
\end{pmatrix}\mathrm{d}t\,,
\end{equation}
which is why
\begin{equation}
\label{eq:nmu-upper}
n^{\mu}=\begin{pmatrix}
1/N \\
-N^i/N \\
\end{pmatrix}\,.
\end{equation}
Lowering the index with the spacetime metric implies that $n_{\mu}$ is purely timelike:
\begin{equation}
\label{eq:nmu-lower}
n_{\mu}=\begin{pmatrix}
-N \\
0^i \\
\end{pmatrix}\,.
\end{equation}
We define
\begin{equation}
\label{eq:induced-metric}
q^{\mu\nu}\equiv g^{\mu\nu}+n^{\mu}n^{\nu}=\begin{pmatrix}
0 & 0^j \\
0^i & q^{ij} \\
\end{pmatrix}\,,
\end{equation}
which can be interpreted as the induced (inverse) metric on a spatial
hypersurface $\Sigma_t$ generalized to $\mathcal{M}$.
The lower $(3\times 3)$ block can contain nonzero entries only.
Pulling down the second index with the metric of \eqref{eq:spacetime-metric} implies
\begin{equation}
\label{eq:projector}
q^{\mu}_{\phantom{\mu}\nu}=\begin{pmatrix}
0 & 0^j \\
N^i & \delta^i_{\phantom{i}j} \\
\end{pmatrix}\,.
\end{equation}
The latter tensor satisfies the important property
\begin{equation}
q^{\mu}_{\phantom{\mu}\nu}q^{\nu}_{\phantom{\nu}\varrho}=q^{\mu}_{\phantom{\mu}\varrho}\,,
\end{equation}
i.e., it is a projector. Furthermore, it obeys
\begin{equation}
q^{\mu}_{\phantom{\mu}\nu}n^{\nu}=0\,,
\end{equation}
which is why it can be employed to project vectors and tensors
defined on $\mathcal{M}$ into the hypersurface $\Sigma_t$. Note that the projector as stated
in \eqref{eq:projector} only involves nondynamical objects by definition.

In this context we would like to comment on a set of coordinates that can be very valuable when dealing with
particular problems: Gaussian normal coordinates (also known as synchronous coordinates). The latter
are characterized by the choices $N=1$ and $N^i=0$, i.e., the lapse function is a coordinate-independent scalar
and the shift vector is discarded. Then, the time coordinate corresponds to proper time for an observer remaining
at fixed spatial coordinates. Furthermore, the unit vector pointing along time is perpendicular to the unit
vectors pointing along each spatial dimension (see, e.g., p.~717 of \cite{Misner:1973}). In Gaussian normal
coordinates it holds that $n^{\mu}=t^{\mu}$ (see \eqref{eq:shift-four-vector}). We will be referring to these
coordinates at some points in the paper.

\section{The SME gravity sector}
\label{sec:sme-gravity-sector}
The action of the minimal gravitational SME is a modification of the Einstein-Hilbert (EH) action that is invariant
with respect to general coordinate transformations~\cite{Kostelecky:2003fs,Kostelecky:2020hbb}. It is
written as
\begin{equation}
S_g=\int _{\mathcal M}\mathrm{d}^4x\,(\mathcal{L}^{(0)}+\mathcal{ L'}  )  \,,
\label{eq:einstein-hilbert-lagrange-density}
\end{equation}
with
\begin{subequations}
\begin{align}
\mathcal{L}^{(0)}&=\frac{\sqrt{-g}}{2\kappa}{}^{(4)}R\,,
\label{eq:lagrange-density-gravitational-sme}
\displaybreak[0]\\[2ex]
\label{eq:lagrange-density-gravitational-sme2}
\mathcal{L'}_{\mathrm{}}&=\frac{\sqrt{-g}}{2\kappa}(k_R)^{\mu\nu\varrho\sigma}{}^{(4)}R_{\mu\nu\varrho\sigma}\,.
\end{align}
\end{subequations}
Here, $\mathcal L^{(0)}$ is the EH Lagrangian without cosmological constant and $\mathcal{L}'$
is a minimal SME term containing a background field $(k_R)^{\mu\nu\varrho\sigma}$ that transforms as a four-tensor under
general coordinate transformations.
Besides, $\kappa=8\pi G_N$ with Newton's constant $G_N$, $g\equiv \det(g_{\mu\nu})$ and ${}^{(4)}R_{\mu\nu\varrho\sigma}$
is the Riemann curvature tensor defined in the four-dimensional spacetime manifold~$\mathcal M$. Its single contraction
${}^{(4)}R_{\mu\nu}\equiv {}^{(4)}R^{\varrho}_{\phantom{\varrho}\mu\varrho\nu}$ is the
Ricci tensor and its double contraction
${}^{(4)}R\equiv {}^{(4)}R^{\mu}_{\phantom{\mu}\mu}$ corresponds to the Ricci scalar.

We work in a scenario of explicit diffeomorphism violation in gravity, i.e., $(k_R)^{\mu\nu\varrho\sigma}$
is a nondynamical tensor-valued background field defined in a spacetime frame of the curved manifold $\mathcal M$.
For simplicity and as we do not consider spontaneous diffeomorphism violation in this article, no confusion should arise by
omitting the bar proposed to be put on top of such coefficients~\cite{Kostelecky:2020hbb}. However, to clarify the physics
in local frames, we can benefit from considering a background vierbein denoted by
$\langle e\rangle_{\mu}^{\phantom{\mu}a}$ that arises, in principle, from
solving the Einstein field equations in the vacuum~\cite{Kostelecky:2020hbb}.

In contrast to a scenario in Minkowski spacetime, a globally constant tensor in the
sense of a vanishing covariant derivative at each point in $\mathcal{M}$ does not exist for a
general manifold. Therefore, we must assume that the background field is
coordinate-dependent: $(k_R)^{\mu\nu\varrho\sigma}=
(k_R)^{\mu\nu\varrho\sigma}(x)$. According
to the third line of Tab.~II in~\cite{Kostelecky:2020hbb}, a term of the form of
Eq.~(\ref{eq:lagrange-density-gravitational-sme2}) does not imply local Lorentz
violation at the level of the Lagrange density. However, suitable combinations
of the coefficients with background vierbeins may give rise to preferred
directions in local frames at each point $x^{\mu}$ of the manifold implying
local Lorentz violation. We will come back to this point later.

The no-go result of the SME gravity
sector~\cite{Kostelecky:2003fs,Kostelecky:2020hbb} implies
that explicit diffeomorphism and local Lorentz violation clash with
specific properties of (pseudo-)Riemannian geometry such as the second Bianchi identity of the Riemann
curvature tensor. This finding requires that certain Noether identities linked to
the invariance under general coordinate transformations must be satisfied for a
consistent setting~\cite{Bluhm:2014oua,Bluhm:2016dzm}. A descriptive interpretation
of the no-go constraint is that a nondynamical background field cannot absorb or emit
momentum, which is why it cannot account for the momentum transfer of a test particle
moving along a geodesic in $\mathcal{M}$ \cite{Bluhm:2014oua}. This issue is avoided neatly by considering
spontaneous Lorentz violation, i.e., a dynamical background field that satisfies its
own field equations. In such scenarios both massless and massive propagating
modes of the background field can be excited. The massless modes correspond
to propagating fluctuations of preferred directions, whereas the massive modes
are interpreted as fluctuations of the strength of Lorentz violation or, in other
words, the size of the controlling coefficients.

Bjorken initially proposed the idea of spontaneous Lorentz violation back in 1963~\cite{Bjorken:1963vg}
to explain the photon as a massless Goldstone boson. In the aftermath, further
physicists took this idea over to gravity to interpret the graviton as a Goldstone
 boson linked to a spontaneous breakdown of local Lorentz and diffeomorphism
  invariance~\cite{Phillips:1966zzc,Ohanian:1970qe}. More recent works in the
  context of electrodynamics are~\cite{Alfaro:2009iv,Escobar:2015gia} and for
  gravity we refer to~\cite{Kraus:2002sa,Berezhiani:2008ue,Kostelecky:2009zr,Carroll:2009mr}
  where the mechanism suggested in~\cite{Kraus:2002sa} is even considered as a solution
  of the cosmological-constant problem. In the context of gravity, it was
  demonstrated that an alternative gravity theory
  called cardinal gravity~\cite{Kostelecky:2009zr,Carroll:2009mr} can be constructed by means of
  a bootstrap method from a linearized theory with a two-tensor field that undergoes
  spontaneous Lorentz violation. At energies much lower than the Planck scale, this
  theory corresponds to GR, but it significantly differs from GR near the Planck energy.

Spontaneous Lorentz violation was investigated in great detail in toy theories known
as bumblebee models in Minkowski
spacetime~\cite{Bluhm:2006im,Bluhm:2007bd,Bluhm:2008yt,Hernaski:2014jsa,Bonder:2015jra}
 as well as in the presence of gravity~\cite{Bluhm:2004ep}. The focus in the latter
 works lies on a profound understanding of the Goldstone and Higgs-like modes.
 A particular model of a purely timelike vector field and its implications for matter
 particles was analyzed in \cite{ArkaniHamed:2004ar}. References~\cite{Altschul:2009ae,Hernaski:2016dyk,Assuncao:2019azw}
 give an account of a scenario of an antisymmetric two-tensor acquiring
 a vacuum expectation value. Papers have also been written on black-hole solutions in
 the presence of bumblebee-type Lorentz violation \cite{Casana:2017jkc,Ding:2019mal,Maluf:2020kgf,Gullu:2020qzu,Carvalho:2021jlp}.
 Note that spontaneous Lorentz violation was
 demonstrated to occur in open-string field theory, too~\cite{Kostelecky:1988zi,
 Kostelecky:1989jp,Kostelecky:1989jw,Kostelecky:1991ak,Kostelecky:1994rn}. The latter
  finding was a motivation for constructing a comprehensive low-energy effective field-theory
  framework for Lorentz violation that we now know as the SME.

An alternative to spontaneous Lorentz and diffeomorphism violation could be explicit
symmetry violation in a more general geometry that does not rely on the quadratic restriction of
path length functionals in Riemannian geometry~\cite{Riemann:2004,Darrigol:2015}.
A promising framework is Finsler geometry~\cite{Finsler:1918,Antonelli:1993,Bao:2000},
as it was shown that classical-particle analogs based on the
SME~\cite{Kostelecky:2010hs} move along geodesics in certain Finsler
spaces~\cite{Kostelecky:2011qz,Kostelecky:2012ac}. This discovery stimulated
a vast series of articles on classical-particle descriptions in Lorentz-violating
background fields as well as their connections to Finsler geometry~\cite{Colladay:2012rv,Russell:2015gwa,Schreck:2014ama,Schreck:2014hga,Colladay:2015wra,Schreck:2015seb,Foster:2015yta,Schreck:2015dsa,Reis:2017ayl,Edwards:2018lsn,Silva:2019qzl,Schreck:2019mmr,Silva:2020tqr,Reis:2021ban}.

However, recent findings suggest that the no-go result is not as restrictive as it was believed to
be for almost 15 years. There may be scenarios of explicit Lorentz violation and/or diffeomorphism
violation in gravity where the St\"{u}ckelberg trick can be used to create massless propagating modes by
introducing a set of additional scalar fields~\cite{Bluhm:2019ato}. This construction
allows for a consistent description of explicit symmetry violations that does not contradict the
Bianchi identity $\nabla_{\mu}G^{\mu\nu}=0$ (where $G^{\mu\nu}$ is the Einstein tensor). Also, it is possible to satisfy the
consistency conditions perturbatively in certain cases~\cite{Bonder:2020fpn}. We
take these findings as a justification for considering explicit diffeomorphism violation in a
Riemannian setting. In this work, we will also provide new insights and precise results in such
a context.

The minimal-gravity modification of Eq.~(\ref{eq:lagrange-density-gravitational-sme2}) is usually
rewritten as follows \cite{Kostelecky:2003fs}:
\begin{subequations}
\label{eq:minimal-gravity-sme-reformulated}
\begin{equation}
\mathcal{L'}=\frac{\sqrt{-g}}{2\kappa}(\mathcal{L'}^{(u)}
+\mathcal{L'}^{(s)}+\mathcal{L'}^{(t)} )\,,
\end{equation}
with
\begin{align}
\label{eq:minimal-gravity-contribution-u}
\mathcal{L'}^{(u)} &=-u{}^{(4)}R\,, \displaybreak[0]\\[2ex]
\label{eq:minimal-gravity-contribution-s}
\mathcal{L'}^{(s)}&=s^{\mu\nu}{}^{(4)}R_{\mu\nu}\,, \displaybreak[0]\\[2ex]
\mathcal{L'}^{(t)}&=t^{\mu\nu\varrho\sigma}{}^{(4)}R_{\mu\nu\varrho\sigma}\,.
\end{align}
\end{subequations}
The decomposition of $\mathcal{L'}$ shown above has turned out to
be valuable. First, $u=u(x)$ is a Lorentz scalar dependent on spacetime
position, which is why it implies diffeomorphism violation. Derivatives of
$u(x)$ for the coordinates can give rise to a preferred direction in a local
frame \cite{Kostelecky:2002ca,Kostelecky:2020hbb}. Therefore, this
contribution is able to induce local Lorentz violation without this being
obvious at a first place. In the setting of spontaneous Lorentz violation with only
weak gravitational fields present, $u$ can be removed by a field redefinition
(see Sec.~III in \cite{Bailey:2006fd}). We will see that the $u$ term is more subtle
in the presence of explicit Lorentz violation. Second, $s^{\mu\nu}=s^{\mu\nu}(x)$ is
a two-tensor-valued background field that can be taken as symmetric due to the
symmetry of the Ricci tensor.

The four-tensor-valued background field $t^{\mu\nu\varrho\sigma}=t^{\mu\nu\varrho\sigma}(x)$
has the symmetries of the Riemann tensor. It does not seem to
play a role in physical observables in the post-Newtonian limit and this interesting
observation was coined the ``t puzzle'' \cite{Bonder:2015maa}. The reason
for that peculiar property seems to be the approximative scheme employed in most
phenomenological studies of Lorentz and diffeomorphism violation in gravity, because it relies
on an asymptotically flat spacetime. If this assumption is not made
such as in cosmology, $t^{\mu\nu\varrho\sigma}$ can give rise to significant effects
providing tight constraints on these coefficients \cite{Bonder:2020fpn,Bonder:2017dpb}.

Note that the following form of the Lagrange density expressed in terms of irreducible
pieces of the Riemann curvature tensor is sometimes employed \cite{Bailey:2006fd}:
\begin{equation}
\label{eq:minimal-gravity-sme-alternative-form}
\mathcal{L}''=\frac{\sqrt{-g}}{2\kappa}(-u{}^{(4)}R+s^{\mu\nu}{}^{(4)}R_{\mu\nu}^T+t^{\mu\nu\varrho\sigma}{}^{(4)}C_{\mu\nu\varrho\sigma})\,,
\end{equation}
with the trace-free Ricci tensor ${}^{(4)}R_{\mu\nu}^T$ and the four-dimensional Weyl
tensor ${}^{(4)}C_{\mu\nu\varrho\sigma}$. The latter inherits all the symmetries from
the Riemann tensor, but it does not involve nonvanishing traces, anymore. Hence, by
using \eqref{eq:minimal-gravity-sme-alternative-form}, the trace of $s^{\mu\nu}$ as well
as the single and double traces of $t^{\mu\nu\varrho\sigma}$
have been extracted, which leaves 9 independent components of $s^{\mu\nu}$ and 10 of
$t^{\mu\nu\varrho\sigma}$ \cite{Kostelecky:2003fs}. However, throughout the paper we will be employing the form
of \eqref{eq:minimal-gravity-sme-reformulated}. As will become clear later, special care
has to be taken in our setting when performing field redefinitions to move such
traces from one term to another. In the context of spontaneous diffeomorphism violation,
\eqref{eq:minimal-gravity-sme-alternative-form} can be interpreted as following directly
from \eqref{eq:lagrange-density-gravitational-sme2} by extracting the single and double
traces of the Riemann curvature tensor. When diffeomorphism invariance is violated
explicitly, though, the background fields $u$, $s^{\mu\nu}$, and $t^{\mu\nu\varrho\sigma}$ in
\eqref{eq:minimal-gravity-sme-reformulated} should be taken as definitions independent
of $(k_R)^{\mu\nu\varrho\sigma}$ occurring in \eqref{eq:lagrange-density-gravitational-sme2}.
Furthermore, $u$, $s^{\mu\nu}$, and $t^{\mu\nu\varrho\sigma}$ of \eqref{eq:minimal-gravity-sme-reformulated}
are assumed to be independent of each other, i.e., we will leave traces where they are
and avoid transferring them between different contributions.

Although the background fields $u$, $s^{\mu\nu}$, and $t^{\mu\nu\varrho\sigma}$
do not imply local Lorentz violation at the level of the Lagrange density, quantities like
 $s^{ab}\equiv s^{\mu\nu}\langle e\rangle_{\mu}^{\phantom{\mu}a}\langle e
 \rangle_{\nu}^{\phantom{\nu}b}$ and $t^{abcd}\equiv t^{\mu\nu\varrho\sigma}
 \langle e\rangle_{\mu}^{\phantom{\mu}a}\langle e\rangle_{\nu}^{\phantom{\nu}b}
 \langle e\rangle_{\varrho}^{\phantom{\varrho}c}\langle e\rangle_{\sigma}^{\phantom
 {\sigma}d}$ (that must be interpreted as mere definitions) can give rise to preferred
 orientations in a local frame where an experiment is performed \cite{Kostelecky:2020hbb}.
Considering the modification of the dispersion relation of gravitational waves in the
regime of weak gravitational fields, $t^{\mu\nu\varrho\sigma}$ leads to birefringence
 in contrast to $s^{\mu\nu}$ that does not do so \cite{Kostelecky:2016kfm}. Therefore,
 there is an analogy between $s^{\mu\nu}$ ($t^{\mu\nu\varrho\sigma}$) and the
 nonbirefringent (birefringent) part of the {\em CPT}-even background field $(k_F)^{\mu\nu\varrho\sigma}$
 of the SME photon sector. Indeed, appropriate field redefinitions allow for transforming
 between photon sector and gravity sector coefficients such that matter-gravity
 experiments are only sensitive to combinations of such coefficients~\cite{Bluhm:2019ato}.
\section{Hamiltonian formulation of a gravity theory}
\label{sec:hamiltonian-formulation}
The Hamiltonian formulation of GR is the foundation for many formulations
of quantum gravity \cite{Thiemann:2007zz} and the Hamiltonian itself is a
powerful tool to define the total mass, momentum, and angular momentum
of a gravitational system. The ADM formulation provides a set of suitable
canonical variables and a means to obtain the GR Hamiltonian. It permits
a better understanding of the physics on hypersurfaces characteristic for a
particular spacetime, e.g., the event horizon of a black hole, which is a null
hypersurface in the corresponding four-dimensional spacetime \cite{Poisson:2002,Poisson:2004}.

The Hamiltonian formulation also uncovers that GR is characterized by
constraints~\cite{Hanson:1976}. Theories subject to constraints play a
pivotal role in physics \cite{Henneaux:1992}. A constraint is a relation between the canonical variables
 that reduces the number of variables that can be considered as physical.
 Hence, in a constrained theory not every canonical variable does necessarily
 describe a physical degree of freedom, but there are some variables that
 correspond to mere gauge degrees of freedom. Constraints appear, e.g., in
 classical (non)relativistic mechanics, electrodynamics, and GR \cite{Hanson:1976}.
 In the following sections, we will first of all obtain modified constraints where their
 structure will only by revealed in Sec.~\ref{sec:constraint-structure} towards the end of the main text.

Note that the constraints of GR are given in terms of the canonical variables. However,
a Hamiltonian in the context of the ADM formulation is first obtained as a function of the
extrinsic curvature describing the embedding of spacelike hypersurfaces $\Sigma_t$ in
the spacetime manifold $\mathcal{M}$. The latter must be eliminated in favor of the canonical
momentum density to obtain the relationships that are usually interpreted as constraints in
phase space. Nevertheless, we will sometimes also employ the terminology ``constraint'' for
the original relationships depending on the extrinsic curvature.

\subsection{General Relativity}
We start by briefly reviewing how to derive the Hamiltonian associated with GR in the context of
the ADM formalism. Consider the EH action without cosmological constant:
\begin{equation}
\label{eq:einstein-hilbert-action}
S^{(0)}=\int_{\mathcal M}\mathrm{d}^4x\frac{\sqrt{-g}}{2\kappa}\,{}^{(4)}R\,.
\end{equation}
With the help of the expression in \eqref{eq:decomposition-ricci-scalar-covariant-derivative}, the EH action can be written as~\cite{Thiemann:2007zz}
\begin{equation}
\label{eq:einstein-hilbert-action-ADM-decomposed}
S^{(0)}=\int_{\mathcal M}\mathrm{d}^4x
\frac{N\sqrt{q}}{2\kappa}\,\left[R-K^2+K_{ij}K^{ij}+2\nabla_{\mu}(n^{\mu}K-a^{\mu})\right]\,,
\end{equation}
where $q\equiv\det(q_{ij})$, $R$ is the Ricci scalar associated with $\Sigma_t$, and $K_{ij}$ the exterior-curvature tensor defined by
\begin{equation}
\label{eq:extrinsic-curvature}
K_{ij}=\frac{1}{2N}(\dot{q}_{ij}-D_iN_j-D_jN_i)\,.
\end{equation}
Furthermore, $K\equiv K^i_{\phantom{i}i}$ corresponds to the trace of the latter and $a_{\mu}$
is the acceleration that is linked to the derivative of the lapse function (see Eq.~(\ref{eq:acceleration})).
In what follows, we will discard the total covariant derivative (last term) in \eqref{eq:einstein-hilbert-action-ADM-decomposed},
which leads to a boundary term when the integral over $\Sigma_t$ is computed.
Although this procedure seems to be of minor importance, it turns out to
be an essential point and we will come back to it in Sec.~\ref{sec:GHY-boundary-term-generalized}.

There are 10 canonical variables in GR: the lapse function $N$, the three shift
vector components $N^i$ and the six spatial-metric
components $q_{ij}$. To obtain the Hamilton density associated
with the EH Lagrange density, we need the time
derivatives $\dot{q}_{ij}$ that follow from \eqref{eq:extrinsic-curvature}
and are given by
\begin{equation}
\label{eq:canonical-momenta-q-dot}
\dot{q}_{ij}=2NK_{ij}+D_iN_j+D_jN_i\,.
\end{equation}
The canonical momentum densities associated with $\dot{q}_{ij}$ are
\begin{equation}
(\pi_0)^{ij}=\frac{\partial \mathcal{L}^{ (0)}}{\partial \dot{q}_{ij}}=\frac{\sqrt{q}}{2\kappa}(K^{ij}-q^{ij}K)\,,
\end{equation}
where calculational details can be found in App.~\ref{sec:hamiltonian-einstein-hilbert}.

Considering the Legendre transformation and using \eqref{eq:canonical-momenta-q-dot}, we obtain a preliminary form of the Hamilton density
\begin{equation}
\label{eq:einstein-hilbert-hamilton-density}
{\mathcal{H}}^{(0)}=-\frac{\sqrt{q}}{2\kappa}\left[N(R+K^2-K_{ij}K^{ij})+2(KD_iN^i-K^{ij}D_iN_j)\right]\,.
\end{equation}
 An integration over
the spatial hypersurface $\Sigma_t$ provides the Hamiltonian. Performing suitable partial integrations
and expressing the extrinsic curvature in terms of the canonical momentum leads to
\begin{subequations}
\label{eq:hamiltonian-einstein-hilbert}
\begin{equation}
H^{(0)}=\int_{\Sigma_t}\mathrm{d}^3x\,\mathcal{H}^{(0)}=
-\int_{\Sigma_t} \mathrm{d}^3x\,\frac{\sqrt{q}}{2\kappa}(N \mathcal C_0+N^i \mathcal C_i)\,,
\end{equation}
with
\begin{align}
\label{eq:parameter-c0-GR}
\mathcal C_0&=R+K^2-K_{ij}K^{ij}\,, \\[2ex]
\label{eq:parameter-ci-GR}
\mathcal C_i&=2(D_jK_i^{\phantom{i}j}-D_iK)  \,.
\end{align}
\end{subequations}
To carry out the canonical analysis of the EH Hamilton density in Eq.~(\ref{eq:hamiltonian-einstein-hilbert}),
we must express the extrinsic curvature in terms of the phase space variables, in particular, the canonical
momentum $\pi^{ij}$. By doing so, we obtain the Hamilton density
\begin{subequations}
\begin{equation}
\label{eq:hamilton-density-EH-constraints}
\mathcal{H}^{(0)}=N{C}_0+N^i{C}_i\,,
\end{equation}
with the Hamiltonian and momentum constraint
\begin{align}
\label{eq:hamiltonian-constraint}
{C}_0&=\frac{2\kappa}{\sqrt{q}}\left[(\pi_0)^{ij}(\pi_0)_{ij}-\frac{(\pi_0)^2}{2}\right]-\frac{\sqrt{q}}{2\kappa}R\,, \\[2ex]
\label{eq:momentum-constraint}
{C}_i&=-2D_j(\pi_0)^j_{\phantom{j}i}\,,
\end{align}
\end{subequations}
where $\pi_0$ is understood as the trace of the canonical momentum: $\pi_0\equiv (\pi_0)^i_{\phantom{i}i}$.
Note that the prefactor in \eqref{eq:hamiltonian-einstein-hilbert} has
been absorbed into ${C}_0$ and ${C}_i$. The part
proportional to the lapse function involves the spatial part of the Ricci
scalar as well as the extrinsic curvature. The part linear in the shift vector
does not depend on the internal geometry of the spatial hypersurface, but
only on the way how it is embedded into the four-dimensional manifold.
\subsection{Minimal gravitational SME}
\label{sec:hamiltonians-smunu}
We will focus on the $s^{\mu\nu}$ sector defined by~\eqref{eq:minimal-gravity-contribution-s} of the
minimal gravitational SME. The background field $s^{\mu\nu}$ can be decomposed into three parts
\begin{equation}
\label{eq:decomposition-s}
s^{\alpha\beta}=q^{\alpha}_{\phantom{\alpha}\mu}q^{\beta}_{\phantom{\beta}\nu}
s^{\mu\nu}-(q^{\alpha}_{\phantom{\alpha}\nu}n^{\beta}+q^{\beta}_{\phantom{\beta}\nu}
n^{\alpha})s^{\nu \mathbf{n}}+n^{\alpha}n^{\beta}s^{\mathbf{nn}}\,,
\end{equation}
where we introduce the notation $s^{ij}\equiv q^i_{\phantom{i}\mu}q^j_{\phantom{j}\nu}s^{\mu\nu}$ for the part
projected entirely into $\Sigma_t$. In addition, we define a mixed (vectorial) part via $s^{i\mathbf{n}}\equiv
q^i_{\phantom{i}\mu}n_{\nu}s^{\mu\nu}$ as well as a scalar part $s^{\mathbf{n}
\mathbf{n}}\equiv n_{\mu}n_{\nu}s^{\mu\nu}$ projected completely along the
direction perpendicular to $\Sigma_t$. Equation~(\ref{eq:decomposition-s}) has to be considered as an identity.
Note that we will interpret $s^{i\mathbf{n}}$ and $s^{\mathbf{nn}}$ as new degrees of freedom
independent of $s^{\mu\nu}$ and the lapse function $N$. As long as diffeomorphism invariance is
violated explicitly by the
nondynamical field $s^{\mu\nu}$, we can think of $s^{i\mathbf{n}}$ and $s^{\mathbf{nn}}$ as
being {\em defined} in the manner above. When diffeomorphism invariance is violated spontaneously,
it could be questioned whether such definitions make sense, as they would mix dynamical and
nondynamical objects.

In the second contribution to the Lagrange density $\mathcal{L'}$
of \eqref{eq:minimal-gravity-sme-reformulated} the background field $s^{\mu\nu}$
is contracted with the four-dimensional Ricci tensor. As a first step, Eq.~(\ref{eq:induced-metric})
enables us to express the Lagrange density in terms of quantities defined on $\Sigma_t$:
\begin{align}
\label{eq:contraction-s-ricci}
s^{\alpha\beta}{}^{(4)}R_{\alpha\beta}&=g^{\alpha\gamma}g^{\beta\delta}s_{\gamma\delta}{}^{(4)}
R_{\alpha\beta}=(q^{\alpha\gamma}-n^{\alpha}n^{\gamma})(q^{\beta\delta}-n^{\beta}n^{\delta})
s_{\gamma\delta}{}^{(4)}R_{\alpha\beta} \notag \\
&=s^{\gamma\delta}q^{\alpha}_{\phantom{\alpha}\gamma}q^{\beta}_{\phantom{\beta}\delta}{}^{(4)}
R_{\alpha\beta}-2n_{\delta}s^{\gamma\delta}q^{\alpha}_{\phantom{\alpha}\gamma}n^{\beta}{}^{(4)}
R_{\alpha\beta}+n^{\gamma}n^{\delta}s_{\gamma\delta}{}^{(4)}R_{\alpha\beta}n^{\alpha}n^{\beta}\,.
\end{align}
In the second step, we will benefit from the following relations that give possible contractions of the
Ricci tensor with the projector $q^{\mu}_{\phantom{\mu}\nu}$ and the four-vector $n^{\mu}$:
\begin{subequations}
\begin{align}
\label{eq:projection-equation-1}
q^{\nu}_{\phantom{\nu}\beta}q^{\sigma}_{\phantom{\sigma}\delta}{}^{(4)}R_{\nu\sigma}&
=\frac{1}{N}\mathcal{L}_mK_{\beta\delta}-\frac{1}{N}D_{\beta}D_{\delta}N+R_{\beta\delta}
+KK_{\beta\delta}-2K^{\mu}_{\phantom{\mu}\beta}K_{\mu\delta}\,, \\[2ex]
\label{eq:projection-equation-2}
q^{\mu}_{\phantom{\mu}\beta}n^{\nu}{}^{(4)}R_{\mu\nu}&=D_{\mu}K^{\mu}_{\phantom{\mu}\beta}
-D_{\beta}K\,, \\[2ex]
\label{eq:projection-equation-3}
n^{\nu}n^{\sigma}{}^{(4)}R_{\nu\sigma}&=-\frac{1}{N}\mathcal{L}_mK+\frac{1}{N}D_{\beta}
D^{\beta}N-K^{\mu\nu}K_{\mu\nu}\,,
\end{align}
\end{subequations}
where $\mathcal{L}_m$ is the Lie derivative (see~Eqs.~(5.31), (5.32)
of~\cite{Carroll:1997ar}) along the four-vector $m^{\mu}\equiv Nn^{\mu}$ and $D_{\mu}$
denotes the three-dimensional covariant derivative. We refer to App.~\ref{sec:mathematical-appendix}
as well as~\cite{Gourgoulhon:2007ue,Gourgoulhon:2012,Thiemann:2007zz} for details on their derivation.
Equation~(\ref{eq:projection-equation-1}) describes how the four-dimensional Ricci tensor
 is projected completely into a spacelike hypersurface $\Sigma_t$. The result involves the
 three-dimensional Ricci tensor, products of the extrinsic-curvature tensor as well as suitable
 derivatives of the extrinsic curvature and the lapse function. Equation~(\ref{eq:projection-equation-2})
  is the contracted Codazzi-Mainardi relation. It describes a partial projection of the Ricci tensor
  into $\Sigma_t$ and involves three-dimensional covariant derivatives of the extrinsic curvature
  only. Last but not least, \eqref{eq:projection-equation-3} states the complete projection of the
  Ricci tensor along the direction orthogonal to $\Sigma_t$. This particular projection contains
  contributions similar to those in~\eqref{eq:projection-equation-1}, but it is devoid of the
  three-dimensional Ricci tensor.

Performing the decomposition of Eq.~(\ref{eq:minimal-gravity-contribution-s}) implies
\begin{subequations}
\begin{equation}
\label{eq:lagrange-density-LV}
\mathcal{L}'=\sum_{\alpha=1}^4 \mathcal{L}'^{(\alpha)}  \,,
\end{equation}
where
\begin{align}
\label{eq:lagrange-density-LV-1}
\mathcal{L}'^{(1)}   &=\frac{N\sqrt{q}}{2\kappa}s^{ij}\left(\frac{1}{N}\mathcal{L}_m
K_{ij}-\frac{1}{N}D_iD_jN+R_{ij}+KK_{ij}-2K_i^{\phantom{i}l}K_{lj}\right)   \,, \displaybreak[0]\\[2ex]
\label{eq:lagrange-density-LV-2}
\mathcal{L}'^{(2)} &=\frac{N\sqrt{q}}{2\kappa}\left[2s^{i\mathbf{n}}(D_iK-
D_lK^l_{\phantom{l}i})\right]\,, \displaybreak[0]\\[2ex]
\label{eq:lagrange-density-LV-3}
\mathcal{L}'^{(3)} &=\frac{N\sqrt{q}}{2\kappa}s^{\mathbf{n}\mathbf{n}}\left(-\frac{1}
 {N}\mathcal{L}_mK+\frac{1}{N}D_iD^iN-K^{ij}K_{ij}\right)\,, \displaybreak[0]\\[2ex]
\label{eq:lagrange-density-LV-4}
\mathcal{L}'^{(4)} &=-\frac{N\sqrt{q}}{2\kappa}u\left(\frac{2}{N}\mathcal{L}_mK-\frac{2}{N}D_iD^iN+R+K^2+K_{ij}K^{ij}\right)\,.
\end{align}
\end{subequations}
Now the associated canonical momentum density follows from
\begin{subequations}
\begin{align}
(\pi')^{ij}&\equiv\frac{\partial\mathcal{L}'}{\partial\dot{q}_{ij}}=\sum_{\alpha=1}^4 \pi^{(\alpha)ij} \,.
\end{align}
\end{subequations}
Starting from these relations, we will obtain the Hamilton density
\begin{equation}
\mathcal{H}'=(\pi')^{ij}\dot{q}_{ij}-\mathcal{L}'\,,
\end{equation}
where calculational details will be relegated to App.~\ref{sec:hamiltonian-SME}
unless they are worthwhile to be mentioned in the main text.

\subsubsection{Decoupling the sectors}
\label{sec:decoupling-sectors}
Based on \eqref{eq:decomposition-s}, $s^{\mu\nu}$ can be decomposed into three sectors.
In what follows, we intend to analyze these sectors independently of each other. This can
be accomplished by looking at particular observer frames where only one of the three sets $s^{ij}$,
$s^{i\mathbf{n}}$, and $s^{\mathbf{nn}}$ provides nonzero coefficients. For example, if one is interested
in $s^{\mathbf{nn}}$ only, we would consider an observer frame where all components of $s^{\mu\nu}$ vanish
except of $s^{00}$. We will precisely follow this strategy, as the three sectors have distinct
characteristics that can be exploited in the computations. Hence, we will just focus on a single sector and
turn off the remaining ones to simplify the analysis without loosing generality.

\subsubsection{Purely spacelike sector}
Consider the purely spacelike modification given by \eqref{eq:lagrange-density-LV-1} (to be added to the
EH Lagrangian later) with the remaining modifications turned off. In general, Lie derivatives
with respect to $m^{\mu}$ involve time derivatives, which is why the term $\mathcal{L}_mK_{ij}$
implies time derivatives of the extrinsic curvature according to
\begin{equation}
\mathcal{L}_mK_{ij}=\dot{K}_{ij}-\mathcal{L}_NK_{ij}\,.
\end{equation}
Such time derivatives give rise to additional second-order time derivatives of the induced metric. This finding
would eventually force us to consider the Ostrogradski formalism of higher-derivative theories~\cite{Ostro:1850,Borneas:1959,Riahi:1972,Woodard:2015,Bollini:1986am}.
To avoid this problem, we follow the method introduced in~\cite{Deruelle:2009zk} and employ an identity
that allows for shifting time derivatives from the extrinsic curvature to the Lorentz-violating background
\begin{equation}
\label{eq:relation-deruelle-1}
\frac{1}{N}s^{ij} \mathcal{L}_mK_{ij}=\nabla_{\mu}
(n^{\mu}K_{ij}s^{ij})-KK_{ij}s^{ij}-\frac{1}{N}K_{ij}\mathcal{L}_ms^{ij} \,.
\end{equation}
Here, the Lie derivative of the purely spacelike background field with respect to $m^{\mu}$ reads
\begin{subequations}
\begin{align}
\label{eq:lie-derivative-s}
\mathcal{L}_ms^{ij}&=\dot{s}^{ij}-\mathcal{L}_Ns^{ij}\,, \\[2ex]
\label{eq:lie-derivative-purely-spacelike-part}
\mathcal{L}_Ns^{ij}&=N^kD_ks^{ij}-(D_kN^i)s^{kj}-(D_kN^j)s^{ik}\,.
\end{align}
\end{subequations}
Let us now rewrite~\eqref{eq:lagrange-density-LV-1}:
\begin{align}
\label{eq:f1-reformulated}
\mathcal{L'}^{(1)}&=\frac{N\sqrt{q}}{2\kappa}\left[\nabla_{\mu}(n^{\mu}K_{ij}s^{ij})-\frac{1}{N}K_{ij} \mathcal{L}_ms^{ij}\right. \notag \\
&\phantom{{}={}}\left.{}\hspace{1.1cm}-s^{ij}\left(\frac{1}{N}D_iD_jN-R_{ij}+2K_i^{\phantom{i}l}K_{lj}\right)\right]\,.
\end{align}
Integrating over the first contribution above within the action leads to another boundary term
(cf.~Eq.~(\ref{eq:einstein-hilbert-action-ADM-decomposed})) that we discard.
This step will turn out to be crucial to understand the results (see Sec.~\ref{sec:GHY-boundary-term-generalized}).
So we omit the covariant-derivative term and consider instead
\begin{equation}
\label{eq:f1-mod-total-derivative}
{\mathcal{L}}^{(1)}=\frac{\sqrt{q}}{2\kappa}\left[-K_{ij} \mathcal{L}_ms^{ij}-s^{ij}D_iD_jN+N s^{ij}(R_{ij}-2K_i^{\phantom{i}l}K_{lj})\right]\,.
\end{equation}
After some calculation (for details we refer to App.~\ref{sec:hamiltonian-SME-purely-spacelike-details}), the Hamiltonian associated with the Lagrange density is given by
\begin{subequations}
\label{eq:hamiltonian-part1}
\begin{equation}
H^{(1)}=-\int_{\Sigma_t} \mathrm{d}^3x\,\frac{\sqrt{q}}{2\kappa}(N\mathcal{C}_0^{(1)}+N^i\mathcal{C}_i^{(1)})\,,
\end{equation}
with
\begin{align}
\label{eq:hamiltonian-part1-constraint1}
\mathcal{C}_0^{(1)}&=s^{ij}(R_{ij}+2K^l_{\phantom{l}i}K_{lj})-D_iD_js^{ij}\,, \\[2ex]
\label{eq:hamiltonian-part1-constraint2}
\mathcal{C}_i^{(1)}&=-q_{ij}D_k\left[\frac{1}{N}\mathcal{L}_ms^{kj}+
2(s^{kl}K^j_{\phantom{j}l}+s^{jl}K^k_{\phantom{k}l})\right]\,.
\end{align}
\end{subequations}
We see that $H^{(1)}$ has a structure similar to that of the EH
Hamiltonian in \eqref{eq:hamiltonian-einstein-hilbert}. Time derivatives of the
lapse function and the shift vector do not occur. Note the presence of the problematic
term $\mathcal{L}_ms^{ij}$ that occurs in the modification of $\mathcal{C}_i$ of
\eqref{eq:parameter-ci-GR} and that is
proportional to the inverse of the lapse function. Therefore, it does
not fit into the usual structure of the Hamiltonian. Understanding the
physical implications of this contribution will turn out to be of paramount importance
for the internal consistency of the theory and we will return to this problem later.

The next step is to consider the total Hamiltonian $H=H^{(0)}+H^{(1)}$ with $H^{(0)}$
of \eqref{eq:hamiltonian-einstein-hilbert} and $H^{(1)}$ stated in \eqref{eq:hamiltonian-part1}.
Our goal now is to eliminate the extrinsic-curvature tensor in $H$ in favor of the canonical
momentum $\pi^{ij}$ associated with $H$. To do so, we first need an expression for
$\pi^{ij}$ that is obtained in Eq.~(\ref{eq:canonicel-momentum-sector1}). As $K_{ij}$ is
symmetric, the canonical momentum employed in the Legendre transformation can be symmetrized,
whereupon we cast the latter result into the form
\begin{subequations}
\begin{align}
\pi^{ij}&=\frac{\sqrt{q}}{2\kappa}\left[K^{ij}-q^{ij}K-\frac{1}{2N}\mathcal{L}_ms^{ij}-
(s^{il}K^j_{\phantom{j}l}+s^{jl}K^i_{\phantom{i}l})\right] \notag \\
&=\frac{\sqrt{q}}{2\kappa}\left(G^{ijab}K_{ab}-\frac{1}{2N}\mathcal{L}_ms^{ij}\right)\,,
\end{align}
with the four-tensor (symmetrized in the first and second pair of indices):
\begin{equation}
\label{eq:four-component-G-object}
G^{ijab}=\frac{1}{2}(q^{ia}q^{jb}+q^{ib}q^{ja})-q^{ij}q^{ab}-
\frac{1}{2}(s^{ia}q^{jb}+s^{ja}q^{ib}+s^{ib}q^{ja}+s^{jb}q^{ia})\,.
\end{equation}
\end{subequations}
Inverting the canonical momentum for the extrinsic curvature and its trace gives
\begin{subequations}
\begin{align}
\label{eq:extrinsic-curvature-canonical-momentum-sector1}
K_{ab}&=G_{abij}\left(\frac{2\kappa}{\sqrt{q}}\pi^{ij}+\frac{1}{2N}\mathcal{L}_ms^{ij}\right)\,, \\[2ex]
K&=q^{ab}G_{abij}\left(\frac{2\kappa}{\sqrt{q}}\pi^{ij}+\frac{1}{2N}\mathcal{L}_ms^{ij}\right)\,,
\end{align}
\end{subequations}
with the inverse tensor $G_{abij}$ satisfying
\begin{equation}
\label{eq:tensor-time-inverse}
G_{cdij}G^{ijab}=\frac{1}{2}(\delta_c^{\phantom{c}a}
\delta_d^{\phantom{d}b}+\delta_d^{\phantom{d}a}\delta_c^{\phantom{c}b})\,.
\end{equation}
Note that the fourth-rank tensor in \eqref{eq:four-component-G-object} for $s^{ij}=0$
is proportional to an object known as the Wheeler-DeWitt metric in the literature
(cf.,~for example, Eq.~(7.45) in~\cite{Hanson:1976}). Its inverse in the
diffeomorphism-invariant setting is readily found to be
\begin{equation}
\label{eq:wheeler-de-witt-zero-lv}
G_{cdij}^{(0)}=\frac{1}{2}(q_{ci}q_{dj}+q_{di}q_{cj}-q_{cd}q_{ij})\,,
\end{equation}
reproducing the results given in
Eqs.~(\ref{eq:trace-canonical-momentum-general-relativity}),
(\ref{eq:extrinsic-curvature-canonical-momentum-GR}).
For a generic $s^{ij}$ it seems challenging the find an inverse in closed
form. There are two possibilities of proceeding. First, a special case
 for $s^{ij}$ could be considered, e.g., a decomposition into suitably
 chosen four-vectors. The exact inverse $G_{cdij}$ can be computed
 for such cases. Second, we were able to determine the inverse
 $G_{cdij}$ for a generic $s^{ij}$ at first order in the controlling coefficients.
 To be as general as possible, we choose the second approach. It is
 reasonable to propose a suitable \textit{Ansatz} for $G_{cdij}$ that involves
  all fourth-rank tensors constructed from $q^{ij}$ and $s^{ij}$ at first order
  in $s^{ij}$. Contracting the latter with \eqref{eq:four-component-G-object}
  and requiring \eqref{eq:tensor-time-inverse} at first order in Lorentz violation
  implies a linear system of equations for the parameters. Solving this system
  provides the parameters of the \textit{Ansatz}. The computation is performed
   best with computer algebra and the result reads
\begin{equation}
G_{cdij}^{(1)}=\frac{1}{2}[q_{ci}q_{dj}+q_{di}q_{cj}-
(1-s^l_{\phantom{l}l})q_{cd}q_{ij}]+s_{cj}q_{di}+s_{di}q_{cj}-(s_{cd}q_{ij}+s_{ij}q_{cd})\,,
\end{equation}
where the latter reproduces \eqref{eq:wheeler-de-witt-zero-lv} for vanishing Lorentz
violation. We will mow restrict our consideration to background fields that satisfy the
requirement $\mathcal{L}_ms^{ij}=0$ such that the constraint structure is standard.
This particular choice is to be discussed at the end of Sec.~\ref{sec:hamiltonians-smunu}
and will find substantial additional motivation in Sec.~\ref{eq:field-equations-constraints}.

By considering the Hamilton density $\mathcal{H}^{(1)}$ associated with $H^{(1)}$
and expressing $K_{ij}$ in terms of $\pi^{ij}$ in \eqref{eq:hamiltonian-part1-constraint1},
we deduce the total Hamilton density
\begin{equation}
\mathcal{H}^{(0)}+\mathcal{H}^{(1)}=NC_0^{(1)}+N^iC^{(1)}_{i}\,,
\end{equation}
with the modified Hamiltonian constraint
\begin{subequations}
\label{eq:hamiltonian-constraint-C01}
\begin{align}
{C}^{(1)}_0&={C}_0+\delta {C}^{(1)}_0\,, \displaybreak[0]\\[2ex]
\delta {C}_0^{(1)}&=\frac{\kappa}{\sqrt{q}}\left[4s^{ij}(\pi_{ij}\pi-
\pi_i^{\phantom{i}k}\pi_{jk})-s^i_{\phantom{i}i}\pi^2\right]
+\frac{\sqrt{q}}{2\kappa}(s^{ij}R_{ij}-D_jD_is^{ij})\,,
\end{align}
\end{subequations}
where $C_0$ is given by Eq.~(\ref{eq:hamiltonian-constraint}). The latter result is valid at
first order in the controlling coefficients and under the requirement that $\mathcal{L}_ms^{ij}=0$.
Performing the analogous replacement in \eqref{eq:hamiltonian-part1-constraint2},
we make an interesting discovery. The standard part implies
\begin{align}
\label{eq:momentum-constraint-sector1-part1}
-\frac{\sqrt{q}}{\kappa}(D_kK^{ki}-D^iK)&=-2\left[D_k\pi^{ki}+s^{kl}
D_k\pi^i_{\phantom{i}l}+s^{ik}(D_l\pi_k^{\phantom{k}l}-D_k\pi)
+\pi^{kl}D_ks^i_{\phantom{i}l}\right. \notag \\
&\phantom{{}={}}\hspace{0.6cm}\left.{}+\pi^{ik}D_ls_k^{\phantom{k}l}-\pi D_ks^{ik}\right]+\dots\,,
\end{align}
where all contributions beyond linear order in the controlling coefficients
 have been dropped. In the modification given by
 \eqref{eq:hamiltonian-part1-constraint2}, it is sufficient to
 employ the standard \eqref{eq:extrinsic-curvature-canonical-momentum-GR}:
\begin{align}
\label{eq:momentum-constraint-sector1-part2}
\frac{\sqrt{q}}{\kappa}D_k(s^{kl}K^i_{\phantom{i}l}+s^{il}K^k_{\phantom{k}l})&=
2D_k\left[s^{kl}\left(\pi^i_{\phantom{i}l}-\frac{\pi}{2}\delta^i_{\phantom{i}l}\right)+
s^{il}\left(\pi^k_{\phantom{k}l}-\frac{\pi}{2}\delta^k_{\phantom{k}l}\right)\right] \notag \\
&=2D_k\left[s^{kl}\pi^i_{\phantom{i}l}+s^{il}\pi^k_{\phantom{k}l}-s^{ik}\pi\right]\,.
\end{align}
Thus, all diffeomorphism-violating contributions of Eqs.~(\ref{eq:momentum-constraint-sector1-part1}),
 (\ref{eq:momentum-constraint-sector1-part2}) cancel each other.
 Then, the momentum constraint remains unmodified at first order in Lorentz violation:
\begin{equation}
\label{eq:momentum-constraint-Ci1}
C^{(1)}_i={C}_i=2D_j\pi^j_{\phantom{j}i}\,,
\end{equation}
with $C_i$ stated in Eq.~(\ref{eq:momentum-constraint}).
\subsubsection{Mixed sector}
\label{sec:hamiltonian-mixed-sector}
In contrast to the Lagrange density of the purely spacelike sector, $\mathcal{L}^{(2)}$
given by Eq.~(\ref{eq:lagrange-density-LV-2}) involves
the covariant derivative defined on the spacelike hypersurface. The latter does not give
rise to time derivatives of the extrinsic curvature. However, to be capable of shifting
the spatial covariant derivatives to the vector-valued background field $s^{i\mathbf{n}}$,
we add a suitable boundary term to the Lagrange density. Note that the latter is not
integrated over the spacetime boundary (such as those to be considered in Sec.~\ref{sec:GHY-boundary-term-generalized}),
but over the boundary of the spacelike
hypersurface $\Sigma_t$. Hence, the corresponding integral is two-dimensional and
runs over the coordinates $z$ employed to parameterize the boundary $\partial\Sigma_t$.
Then, the Lagrangian reads
\begin{align}
\label{eq:boundary-term-mixed-sector}
L^{(2)}&=\frac{1}{\kappa}\bigg[\int_{\Sigma_t}\mathrm{d}^3x \sqrt{q}\,Ns^{i\mathbf{n}}(D_iK-D_l
K^l_{\phantom{l}i})-\oint_{\partial\Sigma_t}\mathrm{d}^2z \sqrt{q}
\,r_iN(Ks^{i\mathbf{n}}-K^i_{\phantom{i}j}s^{j\mathbf{n}})\bigg] \notag \\
&=\frac{1}{\kappa}\int_{\Sigma_t}\mathrm{d}^3x \sqrt{q}\,N\left\{s^{i\mathbf{n}}(D_iK-D_l
K^l_{\phantom{l}i})-\frac{1}{N}D_i[N(Ks^{i\mathbf{n}}-K^i_{\phantom{i}j}s^{j\mathbf{n}})]\right\}\,,
\end{align}
with a properly normalized three-vector $\mathbf{r}$ that is orthogonal to $\partial\Sigma_t$. Then,
the associated Lagrange density has the form
\begin{align}
\label{eq:eq:lagrange-density-LV-2-with-boundary-term}
\mathcal{L}^{(2)}&=\frac{\sqrt{q}}{\kappa}\left\{Ns^{i\mathbf{n}}(D_iK-
D_lK^l_{\phantom{l}i})-D_i[N(Ks^{i\mathbf{n}}-K^i_{\phantom{i}j}s^{j\mathbf{n}})]\right\} \notag \\
&=\frac{\sqrt{q}}{\kappa}N\left[K^i_{\phantom{i}j}D_is^{j\mathbf{n}}-K
D_is^{i\mathbf{n}}+a_i(K^i_{\phantom{i}j}s^{j\mathbf{n}}-Ks^{i\mathbf{n}})\right]\,.
\end{align}
Based on the latter result, we can directly obtain the canonical momentum
and the Hamilton density via a Legendre transformation. Details of the computation
are relegated to App.~\ref{sec:derivation-hamiltonian-mixed-sector}. The Hamiltonian then has the form
\begin{subequations}
\label{eq:hamiltonian-part2}
\begin{equation}
H^{(2)}=-\int_{\Sigma_t} \mathrm{d}^3x\,\frac{\sqrt{q}}{2\kappa}(N\mathcal{C}_0^{(2)}+N^a\mathcal{C}_a^{(2)})\,,
\end{equation}
with
\begin{align}
\mathcal{C}_0^{(2)}&=0\,, \\[2ex]
\mathcal{C}_a^{(2)}&=q_{ab}[D_r(a^r+D^r)s^{b\mathbf{n}}+D_r(a^b+D^b)s^{r\mathbf{n}}-2D^b(a_i+D_i)s^{i\mathbf{n}}]\,.
\end{align}
\end{subequations}
Interestingly, $\mathcal{C}_0$ in \eqref{eq:parameter-c0-GR} is not affected
by the coefficients $s^{i\mathbf{n}}$ and the modification of $\mathcal{C}_i$ of
\eqref{eq:parameter-ci-GR} is independent of the exterior curvature. The form of the Lagrange
density $\mathcal{L}^{(2)}$ suggests that it must be interpreted as a constraint that does
not affect the dynamics. A possible explanation is given by the intriguing
finding that $\mathcal{L}^{(2)}$ can be generated at first order in the controlling
coefficients via a modified ADM decomposition (see App.~\ref{sec:modified-ADM}).
The latter is characterized by an effective shift vector that includes the controlling
coefficients $s^{i\mathbf{n}}$. Thus, we conclude that the mixed coefficients
$s^{i\mathbf{n}}$ are mere gauge degrees of freedom.

Despite of this result, expressing the previous constraints in terms of the canonical momentum
density $\pi^{ij}$ may still provide further insight. Considering
$\mathcal{H}^{(0)}+\mathcal{H}^{(2)}$ with the Hamilton density
$\mathcal{H}^{(2)}$ of the mixed sector, the total canonical momentum of
\eqref{eq:canonical-momentum-mixed-sector} can be inverted for the
exterior curvature when we write
\begin{align}
\pi^{ij}&=\frac{\sqrt{q}}{2\kappa}(K^{ij}-q^{ij}K+\tilde \pi^{ij})\,, \\[2ex]
\label{eq:shift-canonical-momentum-mixed-sector}
\tilde  \pi^{ij}&=\frac{1}{2}\left[(a^r+D^r)s^{s\mathbf{n}}+(a^s+D^s)s^{r\mathbf{n}}\right]-q^{rs}(a_i+D_i)s^{i\mathbf{n}}\,.
\end{align}
Note that $\tilde  \pi^{ij}$ does not depend on the exterior curvature.
By employing the inverse Wheeler-DeWitt metric of \eqref{eq:wheeler-de-witt-zero-lv}, we obtain
\begin{equation}
K_{ab}=G_{abij}^{(0)}\left(\frac{2\kappa}{\sqrt{q}}\pi^{ij}-\tilde  \pi^{ij}\right)
=\frac{2\kappa}{\sqrt{q}}\left(\pi_{ab}-\frac{\pi}{2}q_{ab}\right)-\tilde  \pi_{ab}+\frac{\tilde  \pi}{2}q_{ab}\,.
\end{equation}
The total Hamilton density then has the form
\begin{subequations}
\begin{equation}
\mathcal{H}^{(0)}+\mathcal{H}^{(2)}=N{C}_0^{(2)}+N^a{C}_a^{(2)}\,,
\end{equation}
with the Hamiltonian constraint
\begin{align}
\label{eq:hamiltonian-constraint-mixed-sector}
{C}_0^{(2)}&={C}_0+\delta {C}_0^{(2)}\,, \\[2ex]
\delta {C}_0^{(2)}&=\tilde  \pi\pi-2 \tilde  \pi_{ij}\pi^{ij}+\frac{\sqrt{q}}{2\kappa}\left(\tilde  \pi^{ij} \tilde  \pi_{ij}-\frac{\tilde  \pi^2}{2}\right)\,,
\end{align}
and the momentum constraint
\begin{equation}
\label{eq:momentum-constraint-mixed-sector}
{C}_i^{(2)}=-2D_jP_i^{\phantom{i}j}\,,\quad P^{ij}\equiv \pi^{ij}-\frac{\sqrt{q}}{2\kappa} \tilde  \pi^{ij}\,.
\end{equation}
\end{subequations}
In principle, the latter can be interpreted as a redefined momentum
constraint with the redefined momentum density $P^{ij}$. A short computation then also reveals that
\begin{equation}
{C}_0^{(2)}=\frac{2\kappa}{\sqrt{q}}\left(P^{ij}P_{ij}-\frac{P^2}{2}\right)-\frac{\sqrt{q}}{2\kappa}R\,.
\end{equation}
These results are another indication for $s^{i\mathbf{n}}$ not
conveying any physical information. We were able to reproduce the
Hamiltonian and momentum constraint of EH theory stated in
Eqs.~(\ref{eq:hamiltonian-constraint}), (\ref{eq:momentum-constraint})
simply by absorbing the controlling coefficients $s^{i\mathbf{n}}$ into the momentum density.
The only caveat is that we are putting nondynamical coefficients $s^{i\mathbf{n}}$ into the
canonical momentum $P^{ij}$, which is a dynamical entity. However, this procedure corresponds
to performing a mere shift of the original canonical momentum $\pi^{ij}$ that does not even
depend on the exterior curvature (see $\tilde{\pi}^{ij}$ in \eqref{eq:shift-canonical-momentum-mixed-sector}).

\subsubsection{Purely timelike sector}
To obtain the Hamiltonian associated with the purely timelike sector based on
Eq.~(\ref{eq:lagrange-density-LV-3}), we employ the relation
\begin{equation}
\label{eq:relation-deruelle-2}
\frac{1}{N}s^{\mathbf{nn}}(\dot{K}-\mathcal{L}_NK)=\nabla_{\mu}(n^{\mu}K
s^{\mathbf{nn}})-K^2s^{\mathbf{nn}}-\frac{1}{N}K(\dot{s}^{\mathbf{nn}}
-\mathcal{L}_Ns^{\mathbf{nn}})\,,
\end{equation}
which is similar to \eqref{eq:relation-deruelle-1}. Inserting the latter into ${\mathcal L'}^{(3)}$ leads to
\begin{equation}
\label{eq:f3-reformulated}
\mathcal{L'}^{(3)}=\frac{N\sqrt{q}}{2\kappa}\left[-\nabla_{\mu}(n^{\mu}Ks^{\mathbf{nn}})
+\frac{1}{N}K\mathcal{L}_ms^{\mathbf{nn}}+s^{\mathbf{nn}}\left(\frac{1}{N}D_iD^iN-K^{ij}K_{ij}+K^2\right)\right]\,.
\end{equation}
Note the similarity with \eqref{eq:f1-reformulated}. Again, we discard the
covariant-derivative contribution that would provide a boundary term in the action
(see Sec.~\ref{sec:GHY-boundary-term-generalized}). So we consider
\begin{subequations}
\begin{align}
\label{eq:f3-mod-total-derivative}
   \mathcal{L}^{(3)} = \frac{\sqrt{q}}{2\kappa}\left[K\mathcal{L}_m
   s^{\mathbf{nn}}+s^{\mathbf{nn}}\left(D_iD^iN-N K^{ij}K_{ij}+N K^2\right)\right]\,,
\end{align}
where the Lie derivative of the scalar background field $s^{\mathbf{nn}}$ can simply be understood as a directional derivative:
\begin{equation}
\label{eq:lie-derivative-scalar}
\frac{1}{N}\mathcal{L}_ms^{\mathbf{nn}}=n^{\mu}\nabla_{\mu}s^{\mathbf{nn}}\,.
\end{equation}
\end{subequations}
The Hamiltonian corresponding to $\mathcal{L}^{(3)}$ is derived
in App.~\ref{sec:derivation-hamiltonian-purely-timelike}. It reads
\begin{subequations}
\label{eq:hamiltonian-part3}
\begin{equation}
H^{(3)}=-\int_{\Sigma_t} \mathrm{d}^3x\,\frac{\sqrt{q}}{2\kappa}(N\mathcal{C}_0^{(3)}+N^a\mathcal{C}_a^{(3)})\,,
\end{equation}
with
\begin{align}
\label{eq:hamiltonian-constraint-timelike-sector}
\mathcal{C}_0^{(3)}&=D_iD^is^{\mathbf{nn}}+s^{\mathbf{nn}}(K^{ij}K_{ij}-K^2)\,, \\[2ex]
\label{eq:momentum-constraint-timelike-sector}
\mathcal{C}^{(3)}_i&=D_i\left(\frac{1}{N}\mathcal{L}_ms^{\mathbf{nn}}+2
s^{\mathbf{nn}}K\right)-2D_j(s^{\mathbf{nn}}K^j_{\phantom{j}i})\,.
\end{align}
\end{subequations}
Note the similarities, but also the differences of the latter modifications to $\mathcal{C}_0,\mathcal{C}_i$ in comparison
to those of Eqs.~(\ref{eq:hamiltonian-part1-constraint1}), (\ref{eq:hamiltonian-part1-constraint2}).
Similarly, the contribution $\mathcal{L}_ms^{\mathbf{nn}}$ deviates from the conventional
constraint structure, as it is proportional to the inverse of the lapse function. This term will
also play a pivotal role for the consistency of the purely timelike sector, as will become evident later.

Let us now introduce the total Hamiltonian $H=H^{(0)}+H^{(3)}$ with the total canonical momentum $\pi^{ij}$
associated. The modified $\pi^{ij}$ is obtained by adding Eqs.~(\ref{eq:canonical-momentum-general-relativity}),
(\ref{eq:canonical-momentum-purely-timelike}). Inverting the latter via the inverse of the Wheeler-deWitt
metric in Eq.~(\ref{eq:wheeler-de-witt-zero-lv}) implies the extrinsic curvature in terms of the canonical momentum:
\begin{subequations}
\label{eq:extrinsic-curvature-timelike-case}
\begin{align}
K_{ab}&=\frac{G^{(0)}_{abij}}{1-s^{\mathbf{nn}}}\left(\frac{2\kappa}{\sqrt{q}}\pi^{ij}-\frac{\Xi}{2}q^{ij}\right)=\frac{1}{1-s^{\mathbf{nn}}}\left[\frac{2\kappa}{\sqrt{q}}\left(\pi_{ab}-\frac{\pi}{2}q_{ab}\right)+\frac{\Xi}{4}q_{ab}\right]\,.
\end{align}
The covariant directional derivative of $s^{\mathbf{nn}}$ with respect to $n^{\mu}$ (cf.~\eqref{eq:lie-derivative-scalar}),
\begin{equation}
\Xi\equiv \frac{1}{N}\mathcal{L}_ms^{\mathbf{nn}}\,,
\end{equation}
\end{subequations}
had to be isolated before computing the inverse. Therefore, the total Hamiltonian in canonical
variables follows from adding Eqs.~(\ref{eq:hamiltonian-einstein-hilbert}), (\ref{eq:hamiltonian-part3})
and expressing the extrinsic curvature in terms of the canonical momentum
 via Eq.~(\ref{eq:extrinsic-curvature-timelike-case}):
\begin{subequations}
\begin{equation}
\mathcal{H}^{(0)}+\mathcal{H}^{(3)}=N{C}_0^{(3)}+N^a{C}_a^{(3)}\,,
\end{equation}
with the modified Hamiltonian constraint
\begin{align}
\label{eq:hamiltonian-constraint-C03}
{C}_0^{(3)}&=\frac{2\kappa}{\sqrt{q}(1-s^{\mathbf{nn}})}\left(\pi^{ij}\pi_{ij}-\frac{\pi^2}{2}\right)-\frac{\sqrt{q}}{2\kappa}(R+D_aD^as^{\mathbf{nn}}) \notag \\
&\phantom{{}={}}+\frac{\Xi}{2(1-s^{\mathbf{nn}})}\left(\pi-\frac{3\sqrt{q}}{8\kappa}\Xi\right)\,.
\end{align}
and the momentum constraint
\begin{equation}
\label{eq:momentum-constraint-Ci3}
{C}_a^{(3)}=C_a=-2D_b\pi_a^{\phantom{a}b}\,.
\end{equation}
\end{subequations}
As before, the Lorentz-violating contributions in the momentum constraint, which follows
from Eqs.~(\ref{eq:parameter-ci-GR}), (\ref{eq:momentum-constraint-timelike-sector}), cancel
when the latter is written as a function of the total canonical momentum.
\subsubsection{Scalar sector}
According to Eq.~(\ref{eq:minimal-gravity-contribution-u}), the minimal gravitational SME also
contains a scalar background field called $u$. In the context of spontaneous diffeomorphism violation, $u$
can be eliminated by redefining the gravitational field, i.e., $1-u$ is merely a scaling factor in this case.
However, in the current section, we will demonstrate that the fate of $u$ in the setting of explicit diffeomorphism violation is much
more subtle. As $u$ comes together with the Ricci scalar, the corresponding Lagrange density of \eqref{eq:minimal-gravity-contribution-u}
can be decomposed by applying Eqs.~(\ref{eq:gauss-relation-scalar}), (\ref{eq:ricci-totally-orthogonal-projection}).
The result is given by \eqref{eq:lagrange-density-LV-4}. Now, the identity
\begin{equation}
\label{eq:relation-deruelle-3}
\frac{1}{N}u\mathcal{L}_mK=\nabla_{\mu}(n^{\mu}Ku)-K^2u-\frac{1}{N}K\mathcal{L}_mu\,,
\end{equation}
which is analogous to Eqs.~(\ref{eq:relation-deruelle-1}) and (\ref{eq:relation-deruelle-2}), allows us to move derivatives from the exterior curvature to the background scalar $u$ modulo a total covariant derivative:
\begin{align}
\label{eq:f4-reformulated}
\frac{2\kappa}{N\sqrt{q}}\mathcal{L}'^{(4)}&=-2\nabla_{\mu}(n^{\mu}Ku)+2K^2u+\frac{2}{N}K\mathcal{L}_mu-\left(-\frac{2}{N}D_iD^iN+R+K^2+K_{ij}K^{ij}\right)u \notag \\
&=-2\nabla_{\mu}(n^{\mu}Ku)+\frac{2}{N}(K\mathcal{L}_mu+uD_iD^iN)-(R-K^2+K_{ij}K^{ij})u\,.
\end{align}
A suitable boundary term added to the action eliminates the total derivative (see Sec.~\ref{sec:GHY-boundary-term-generalized}). Taking this boundary term into account, implies the form of the Lagrange density that we are going to work with:
\begin{equation}
\label{eq:lagrange-density-u}
\mathcal{L}^{(4)}=\frac{N\sqrt{q}}{2\kappa}\left[\frac{2}{N}(K\mathcal{L}_mu+uD_iD^iN)-(R-K^2+K_{ij}K^{ij})u\right]\,.
\end{equation}
A Legendre transformation (see Sec.~\ref{sec:derivation-hamiltonian-scalar}) leads to the Hamiltonian
\begin{subequations}
\label{eq:hamiltonian-part4}
\begin{equation}
H^{(4)}=-\int_{\Sigma_t}\mathrm{d}^3x\,\frac{\sqrt{q}}{2\kappa}(\mathcal{C}_0^{(4)}N+\mathcal{C}_l^{(4)}N^l)\,,
\end{equation}
where
\begin{align}
\label{eq:hamiltonian-constraint-scalar-sector}
\mathcal{C}_0^{(4)}&=-(R+K^2-K_{ij}K^{ij})u+2D_iD^iu\,, \\[2ex]
\label{eq:momentum-constraint-scalar-sector}
\mathcal{C}_l^{(4)}&=2\left[D_l\left(\frac{1}{N}\mathcal{L}_mu\right)-D_k(uK^k_{\phantom{k}l})+D_l(uK)\right]\,.
\end{align}
\end{subequations}
We now consider the theory based on the total Hamiltonian $H=H^{(0)}+H^{(4)}$. The total canonical momentum can be inverted for the exterior curvature via the inverse Wheeler-DeWitt metric of Eq.~(\ref{eq:wheeler-de-witt-zero-lv}):
\begin{subequations}
\begin{equation}
\label{eq:extrinsic-curvature-scalar-sector}
K_{ab}=\frac{G^{(0)}_{abij}}{1-u}\left(\frac{2\kappa}{\sqrt{q}}\pi^{ij}-\Upsilon q^{ij}\right)=\frac{1}{1-u}\left[\frac{2\kappa}{\sqrt{q}}\left(\pi_{ab}-\frac{\pi}{2}q_{ab}\right)+\frac{\Upsilon}{2}q_{ab}\right]\,,
\end{equation}
where we introduced a symbol for the Lie derivative of the background field:
\begin{equation}
\Upsilon\equiv\frac{1}{N}\mathcal{L}_mu\,.
\end{equation}
\end{subequations}
Now, the total Hamilton density reads
\begin{subequations}
\begin{equation}
\mathcal{H}^{(0)}+\mathcal{H}^{(4)}=NC_0^{(4)}+N^iC_i^{(4)}\,,
\end{equation}
with the modified Hamiltonian and momentum constraint:
\begin{align}
\label{eq:hamiltonian-constraint-C04}
C_0^{(4)}&=\frac{2\kappa}{\sqrt{q}(1-u)}\left(\pi_{ij}\pi^{ij}-\frac{\pi^2}{2}\right)-\frac{\sqrt{q}}{2\kappa}\left[(1-u)R+2D_iD^iu\right] \notag \\
&\phantom{{}={}}+\frac{\Upsilon}{1-u}\left(\pi-\frac{3\sqrt{q}}{4\kappa}\Upsilon\right)\,, \\[2ex]
\label{eq:momentum-constraint-Ci4}
C_k^{(4)}&=-2D_i\pi^i_{\phantom{i}k}\,.
\end{align}
\end{subequations}
The momentum constraint is unaffected by diffeomorphism violation such as for the purely spacelike and timelike sectors of $s^{\mu\nu}$; cf.~Eqs.~(\ref{eq:momentum-constraint-Ci1}), (\ref{eq:momentum-constraint-Ci3}). Note the parallels to \eqref{eq:hamiltonian-constraint-C03}, although no curvature term is induced by $s^{\mathbf{nn}}$ in contrast to $u$.

A further interesting conclusion can be drawn from supposing that $u$ arises from a nonvanishing trace of $s^{\mu\nu}$. In the case of spontaneous diffeomorphism violation, this argument is usually developed to disregard the trace of $s^{\mu\nu}$ as an unobservable contribution. We then choose $s^{\mu\nu}=ug^{\mu\nu}$ and use Gaussian normal coordinates where $s^{ij}=uq^{ij}$. Inserting the latter into the Lagrange densities of the purely spacelike and purely timelike sector of Eqs.~(\ref{eq:f1-mod-total-derivative}), (\ref{eq:f3-mod-total-derivative}) results in:
\begin{subequations}
\begin{align}
\frac{2\kappa}{N\sqrt{q}}\mathcal{L}^{(1)}&=-\frac{1}{N}[K_{ij}\mathcal{L}_m(uq^{ij})+uq^{ij}D_iD_jN]+uq^{ij}(R_{ij}-2K_i^{\phantom{i}l}K_{lj}) \notag \\
&=2uK_{ij}K^{ij}-\frac{1}{N}(K\mathcal{L}_mu+uD_iD^iN)+uR-2uK^{ij}K_{ij} \notag \\
&=-\frac{1}{N}(K\mathcal{L}_mu+uD_iD^iN)+uR\,, \displaybreak[0]\\[2ex]
\frac{2\kappa}{N\sqrt{q}}\mathcal{L}^{(3)}&=-\frac{1}{N}(K\mathcal{L}_mu+uD_iD^iN)+uK^{ij}K_{ij}-uK^2\,.
\end{align}
\end{subequations}
The sum of both corresponds to the negative of Eq.~(\ref{eq:lagrange-density-u}), as expected. In contrast, if we insert $s^{\mu\nu}=ug^{\mu\nu}$ in Gaussian normal coordinates into the modifications of $\mathcal{C}_0$ given by Eqs.~(\ref{eq:hamiltonian-part1-constraint1}), (\ref{eq:hamiltonian-constraint-timelike-sector}), we obtain:
\begin{align}
\label{eq:hamiltonian-constraint-u}
\mathcal{C}_0^{(u)}&\equiv \mathcal{C}_0^{(1)}+\mathcal{C}_0^{(3)} \notag \\
&=u(R+2K^{ij}K_{ij})-D_iD^iu-D_iD^iu-u(K^{ij}K_{ij}-K^2) \notag \\
&=u(R+K^2+K^{ij}K_{ij})-2D_iD^iu\,,
\end{align}
which is off from the negative of \eqref{eq:hamiltonian-constraint-scalar-sector} by a term $2uK^{ij}K_{ij}$. The
reason for this mismatch is found in the Lie derivative term of \eqref{eq:f1-mod-total-derivative}. The
canonical momentum provides an additional contribution:
\begin{equation}
\pi^{kl}=\frac{\partial\mathcal{L}^{(1)}}{\partial\dot{q}_{kl}}\supset \frac{N\sqrt{q}}{2\kappa}\left(\frac{1}{N}
uK_{ij}\delta^i_{\phantom{i}k}\delta^j_{\phantom{j}l}\right)=\frac{\sqrt{q}}{2\kappa}uK_{kl}\,,
\end{equation}
where we can use that $\dot{q}^{lk}=-q^{li}\dot{q}_{ij}q^{jk}$. Thus, $\mathcal{C}_0^{(u)}$ of
\eqref{eq:hamiltonian-constraint-u} has to be endowed with a correction term given by
\begin{equation}
\pi^{kl}\dot{q}_{kl}\supset \frac{\sqrt{q}}{2\kappa}uK_{kl}2NK^{kl}=\frac{N\sqrt{q}}{2\kappa}(2uK_{kl}K^{kl})\,.
\end{equation}
Note that the prefactor and a global minus sign was extracted from Eqs.~(\ref{eq:hamiltonian-part1-constraint2}),
(\ref{eq:hamiltonian-constraint-timelike-sector}). So we reproduce \eqref{eq:hamiltonian-constraint-scalar-sector} only under these circumstances.
This finding teaches us a crucial lesson. In the setting of explicit Lorentz violation, a statement like $s^{\mu\nu}=ug^{\mu\nu}$ is
simply meaningless, as the background fields are nondynamical, but the metric is a dynamical object
(see the discussion below \eqref{eq:minimal-gravity-sme-alternative-form}).\footnote{An analogous
situation occurs when including the extended Chern-Simons term $\epsilon^{\mu\nu\alpha}
 \Box A_{\mu}\partial_{\nu}  A_{\alpha}$ into $(1+2)$-dimensional electrodynamics~\cite{testing}.
This effective term can be absorbed into the gauge field via the redefinition $\bar A_{\mu}\equiv A_{\mu}+\epsilon_{\mu \nu \alpha}\partial^{\nu}A^{\alpha}$ in order to
rewrite the Lagrange density in terms of a new field strength tensor as $-\frac{1}{4} \bar F_{\mu \nu} \bar F^{\mu \nu}$ with $\bar F_{\mu \nu}=\partial_{\mu}\bar A_{\nu}-\partial_{\nu}\bar A_{\mu}$. In spite of this form hiding the additional degrees of freedom, the latter still provides a parity-violating theory. The example presented demonstrates that field redefinitions have to be carried out and interpreted with care.}

Hence, it is then also not possible to absorb the $u$ term of \eqref{eq:minimal-gravity-contribution-u} into the gravitational field to eliminate it. The important message is that $u$ becomes a physical object when explicit diffeomorphism violation is considered. One cannot get rid of it by a simple field redefinition.

\subsection{Generalized Gibbons-Hawking-York boundary term}
\label{sec:GHY-boundary-term-generalized}
In what follows, we will comment on the time derivatives of the extrinsic curvature that occur in the ADM-decomposed EH Lagrange density of Eq.~(\ref{eq:einstein-hilbert-lagrange-density}) as well as in the modifications of Eqs.~(\ref{eq:lagrange-density-LV-1}), (\ref{eq:lagrange-density-LV-3}), and (\ref{eq:lagrange-density-LV-4}) via the Lie derivative $\mathcal{L}_m$. These time derivatives imply that the Lagrangians contain second-order time derivatives of the metric, which is puzzling, as the (modified) Einstein equations themselves are of second order in time (cf.~Eqs.~(62), (63) in \cite{Kostelecky:2003fs} and Eqs.~(6), (7) in \cite{Bailey:2006fd}). Although the EH action contains second-order time derivatives of the metric (see the definition of the Ricci scalar), the Einstein equations themselves do not involve time derivatives of the metric higher than 2. To gain a better understanding of this peculiar property, we consult Ref.~\cite{Landau:1980} (see pp.~297) that provides a powerful decomposition of the EH action as follows:
\begin{subequations}
\label{eq:einstein-hilbert-action-extracting-time-derivatives}
\begin{equation}
\label{eq:einstein-hilbert-action-extracting-time-derivatives-1}
\int_{\mathcal{M}}\mathrm{d}^4x\,\frac{\sqrt{-g}}{2\kappa}{}^{(4)}R=\int_{\mathcal{M}}\mathrm{d}^4x\,\frac{\sqrt{-g}}{2\kappa}W+\frac{1}{2\kappa}\int_{\mathcal{M}}\mathrm{d}^4x\,\frac{\partial(\sqrt{-g}w^{\lambda})}{\partial x^{\lambda}}\,,
\end{equation}
with the quantities $W$ and $w^{\mu}$ given by
\begin{align}
W&=g^{\mu\nu}(\Gamma^{\varrho}_{\phantom{\varrho}\mu\sigma}\Gamma^{\sigma}_{\phantom{\sigma}\nu\varrho}-\Gamma^{\sigma}_{\phantom{\sigma}\mu\nu}\Gamma^{\varrho}_{\phantom{\varrho}\sigma\varrho})\,, \displaybreak[0]\\[2ex]
w^{\lambda}&=g^{\alpha\beta}\Gamma^{\lambda}_{\phantom{\lambda}\alpha\beta}-g^{\lambda\alpha}\Gamma^{\nu}_{\phantom{\nu}\alpha\nu}\,,
\end{align}
\end{subequations}
where $\Gamma^{\mu}_{\phantom{\mu}\nu\varrho}$ are the Christoffel symbols of four-dimensional spacetime. By following this procedure, $\sqrt{-g}W$ involves only first-order derivatives of the metric, whereas all second-order derivatives of the metric are put into $\partial_{\lambda}(\sqrt{-g}w^{\lambda})$. This decomposition works, as the Ricci scalar is linear in the second-order time derivatives of the metric. Note that $W$ is not a Lorentz scalar and $w^{\lambda}$ is not a four-vector.

An explicit computation (for example, done with the powerful \textit{Mathematica} package \textit{xTensor} \cite{xTensor:2020}) demonstrates that a variation of the first term on the right-hand side of Eq.~(\ref{eq:einstein-hilbert-action-extracting-time-derivatives-1}) with respect to the metric leads to $\sqrt{-g}G_{\mu\nu}$ with the Einstein tensor $G_{\mu\nu}$. Interestingly, \cite{Landau:1980} claims that a variation of the second term is zero, as it is a term on the boundary $\partial\mathcal{M}$ of the spacetime manifold $\mathcal{M}$. Therefore, it is not expected to contribute to the field equations. However, research done in the 1970s revealed that the situation is more subtle. As the second term on the right-hand side of Eq.~(\ref{eq:einstein-hilbert-action-extracting-time-derivatives-1}) depends on second-order derivatives of the metric, the corresponding surface term still contains first-order time derivatives of the dynamical field $g_{\mu\nu}$. In general, Hamilton's principle requires that variations of dynamical fields vanish on the boundary, which means
$\delta g_{\mu\nu}|_{\partial\mathcal{M}}=0$ for GR. However, the requirement that first-order derivatives of these variations also vanish on the boundary is too strong and should not be implemented, if one does not want to change Hamilton's principle. So $\partial_{\varrho}(\delta g_{\mu\nu})|_{\partial\mathcal{M}}\neq 0$ must be assumed. Then the boundary term cannot simply be set to zero, which is a particular situation in GR, as the EH action already involves second-order derivatives of the dynamical fields.

Since there is a contribution on the boundary, the latter can only be canceled by subtracting a suitable term from the action. It is called the Gibbons-Hawking-York (GHY) term \cite{York:1972sj,Gibbons:1976ue,Blau:2020} and it has the form
\begin{equation}
\label{eq:GHY-boundary-term}
S_{\mathrm{GHY}}=\frac{\varepsilon}{\kappa}\oint_{\partial\mathcal{M}} \mathrm{d}^3y\,\sqrt{q}\,K\,,
\end{equation}
where $\varepsilon=\mp 1$ for a spacelike (timelike) boundary and $y$ are the coordinates on the boundary. As its derivation is not found in a number of GR books, we present the essential arguments and calculational steps in App.~\ref{sec:gibbons-hawking-york}. By doing so, the reader will also be able to understand better how the computation must be adapted to the settings of the $s^{\mu\nu}$ and $u$ terms of the gravitational SME.

The ADM decomposition of the EH Lagrange density based on \eqref{eq:einstein-hilbert-lagrange-density} provides an alternative explanation of the GHY boundary term. The Lagrange density involves a first-order time derivative of the metric within the extrinsic curvature tensor via Eqs.~(\ref{eq:extrinsic-curvature}), (\ref{eq:canonical-momenta-q-dot}). Second-order time derivatives of the metric occur in the covariant-derivative term of \eqref{eq:einstein-hilbert-lagrange-density}. By considering this contribution inside the action with $N\sqrt{q}=\sqrt{-g}$, we have
\begin{equation}
\frac{1}{\kappa}\int_{\mathcal{M}}\mathrm{d}^4x\,\sqrt{-g}\,\nabla_{\mu}(n^{\mu}K-a^{\mu})=\frac{\varepsilon}{\kappa}\oint_{\partial\mathcal{M}} \mathrm{d}^3y\,\sqrt{q}\,K\,,
\end{equation}
where we employ $n\cdot a=0$ and $n^2=\varepsilon$.\footnote{Note that the four-vector $n^{\mu}$ orthogonal to $\Sigma_t$ can be employed as a vector normal to the boundary. The relevant parts of $\partial\mathcal{M}$ in this context are interpreted as hypersurfaces $\Sigma_t$ being timelike or spacelike.} The result corresponds to Eq.~(\ref{eq:GHY-boundary-term}). Therefore, the ADM formalism gives rise to the GHY boundary term automatically. The latter is needed to cancel the first-order derivatives of the metric on the boundary.

Let us now focus on the Lorentz-violating modification of the EH Lagrange density given by $\mathcal{L}'^{(s)}$ in \eqref{eq:minimal-gravity-contribution-s}. We observed that the Lagrange density of the mixed sector of $\mathcal{L}'^{(s)}$ stated in Eq.~(\ref{eq:lagrange-density-LV-2}) is completely devoid of second-order time derivatives of the metric, as it only contains covariant derivatives defined in the spacelike hypersurface. In fact, we introduced a boundary term for this sector in Eq.~(\ref{eq:boundary-term-mixed-sector}). However, the latter lived on the boundary $\partial\Sigma_t$ of a spacelike hypersurface as opposed to the boundary $\partial\mathcal{M}$ of the spacetime manifold. Furthermore, the motivation for introducing this term was completely different and did not have any relation with second-order time derivatives of the metric.

In contrast, the situation is quite different for both the purely timelike and the purely spacelike sector whose Lagrange densities are given by Eq.~(\ref{eq:lagrange-density-LV-1}) and Eq.~(\ref{eq:lagrange-density-LV-3}), respectively. They involve first-order time derivatives of $K_{ij}$ and $K$ via the Lie derivative along $m^{\mu}$.
Therefore, second-order time derivatives of $q_{ij}$ are implied. Formally, a similar decomposition as that of \eqref{eq:einstein-hilbert-action-extracting-time-derivatives} can still be carried out, since $\mathcal{L}^{(s)}$ is linear in the second-order time derivatives of the metric, as well:
\begin{subequations}
\label{eq:modified-action-extracting-time-derivatives}
\begin{equation}
\label{eq:modified-action-extracting-time-derivatives-1}
\int_{\mathcal{M}}\mathrm{d}^4x\,\frac{\sqrt{-g}}{2\kappa}s^{\mu\nu}{}^{(4)}R_{\mu\nu}=\int_{\mathcal{M}}\mathrm{d}^4x\,\frac{\sqrt{-g}}{2\kappa}W^{(s)}+\frac{1}{2\kappa}\int_{\mathcal{M}}\mathrm{d}^4x\,\frac{\partial(\sqrt{-g}w^{(s)\lambda})}{\partial x^{\lambda}}\,,
\end{equation}
with modified quantities $W^{(s)}$ and $w^{(s)\mu}$:
\begin{align}
W^{(s)}&=s^{\mu\nu}\Big\{\Gamma^{\sigma}_{\phantom{\sigma}\mu\nu}\Gamma^{\varrho}_{\phantom{\varrho}\sigma\varrho}-\Gamma^{\varrho}_{\phantom{\varrho}\mu\sigma}\Gamma^{\sigma}_{\phantom{\sigma}\nu\varrho}+\frac{1}{2g}(\Gamma^{\varrho}_{\phantom{\varrho}\mu\varrho}\partial_{\nu}g-\Gamma^{\varrho}_{\phantom{\varrho}\mu\nu}\partial_{\varrho}g)\Big\} \notag \\
&\phantom{{}={}}+\Gamma^{\varrho}_{\phantom{\varrho}\mu\varrho}\partial_{\nu}s^{\mu\nu}-\Gamma^{\varrho}_{\phantom{\varrho}\mu\nu}\partial_{\varrho}s^{\mu\nu}\,, \displaybreak[0]\\[2ex]
w^{(s)\lambda}&=s^{\alpha\beta}\Gamma^{\lambda}_{\phantom{\lambda}\alpha\beta}-s^{\lambda\alpha}\Gamma^{\nu}_{\phantom{\nu}\alpha\nu}\,.
\end{align}
\end{subequations}
The second-order derivatives of the metric are absorbed in $\partial_{\lambda}(\sqrt{-g}w^{(s)\lambda})$, whereas $W^{(s)}$ only involves first-order time derivatives. A study analogous to that done before shows that there are nonvanishing contributions on the boundary originating from a variation of the second term on the right-hand side of Eq.~(\ref{eq:modified-action-extracting-time-derivatives-1}). The outcome is that modified GHY terms must be introduced to compensate these effects. Details of the procedure are presented in App.~\ref{sec:gibbons-hawking-york-modified}. We also take into account $\mathcal{L}'^{(u)}$ in \eqref{eq:minimal-gravity-contribution-u}, which is straightforward, since it has the same structure as the EH Lagrangian. The indispensable boundary terms are then found to be given by
\begin{equation}
\label{eq:gibbons-hawking-york-modified}
S_{\substack{\mathrm{mod} \\ \mathrm{GHY}}}=\frac{\varepsilon}{2\kappa}\oint_{\partial\mathcal{M}} \mathrm{d}^3y\,\sqrt{q}\,\left[s^{ij}K_{ij}-(s^{\mathbf{nn}}+2u)K\right]\,.
\end{equation}
Hence, there is a boundary term for the purely spacelike part of $s^{\mu\nu}$ governed by the controlling coefficients $s^{ij}$, a second one for the purely timelike part parameterized by $s^{\mathbf{nn}}$, and a third one for $u$. The mixed part of $s^{\mu\nu}$ does not have an associated boundary term of this form, though (cf.~Eq.~(\ref{eq:boundary-term-mixed-sector})). Note that no additional global factor of 2 occurs for $s^{ij}$ as well as $s^{\mathbf{nn}}$ in comparison to the GHY term in Eq.~(\ref{eq:GHY-boundary-term}).

At this point we may look at the ADM formalism again. Integrating the total-derivative contributions in Eqs.~(\ref{eq:f1-reformulated}), (\ref{eq:f3-reformulated}), and (\ref{eq:f4-reformulated}) gives rise to exactly the same surface terms that we found above:
\begin{subequations}
\begin{align}
\label{eq:boundary-term-spacelike-sector}
\frac{1}{2\kappa}\int_{\mathcal{M}}\mathrm{d}^4x\,\sqrt{-g}\,\nabla_{\mu}(n^{\mu}K_{ij}s^{ij})&=\frac{\varepsilon}{2\kappa}\oint_{\partial\mathcal{M}}\mathrm{d}^3y\,\sqrt{q}\,K_{ij}s^{ij}\,, \displaybreak[0]\\[2ex]
\label{eq:boundary-term-timelike-sector}
-\frac{1}{2\kappa}\int_{\mathcal{M}}\mathrm{d}^4x\,\sqrt{-g}\,\nabla_{\mu}(n^{\mu}Ks^{\mathbf{nn}})&=-\frac{\varepsilon}{2\kappa}\oint_{\partial\mathcal{M}}\mathrm{d}^3y\,\sqrt{q}\,Ks^{\mathbf{nn}}\,, \displaybreak[0]\\[2ex]
\label{eq:boundary-term-scalar-sector}
-\frac{1}{\kappa}\int_{\mathcal{M}}\mathrm{d}^4x\,\sqrt{-g}\nabla_{\mu}(n^{\mu}Ku)&=-\frac{\varepsilon}{\kappa}\oint_{\partial\mathcal{M}}\mathrm{d}^3y\,\sqrt{q}Ku\,.
\end{align}
\end{subequations}
In the setting of the $u$ and $s^{\mu\nu}$ coefficients, the ADM formalism still correctly provides suitable surface terms that are generalizations of the GHY term. They are necessary to compensate the additional first-order time derivatives of the metric on the boundary that arise due to diffeomorphism violation. Note that modified GHY boundary terms associated with the minimal gravitational SME were also considered in \cite{Bonder:2015maa}.

We are interested in the boundary contributions introduced above to understand how to treat the second-order time derivatives of the metric properly that occur in Eqs.~(\ref{eq:lagrange-density-LV-1}), (\ref{eq:lagrange-density-LV-3}), and (\ref{eq:lagrange-density-LV-4}). Second- and higher-order time derivatives of the dynamical fields may lead to additional propagating degrees of freedom that are unphysical (such as ghosts). In the nonminimal SME such unphysical degrees of freedom are very common. They are often neglected in phenomenological analyses in the low-energy limit of quantum field theories based on the Lee-Wick procedure \cite{Lee:1969fy,Lee:1970iw}. However, ghosts cannot simply be discarded when quantum corrections are taken into account and the internal consistency of such theories must be questioned and investigated (see, e.g., \cite{Reyes:2010pv,Lopez-Sarrion:2013kxa,Reyes:2013nca,Maniatis:2014xja,Schreck:2013kja,Schreck:2014qka,Reyes:2016pus,Casana:2018rhg,Balart:2018bbr,Ferreira:2019lpu,Mariz:2018yvo,Avila:2019xdn} where this list is not claimed to be exhaustive).

In the minimal SME, Lorentz-violating contributions can introduce additional first-order time derivatives of the dynamical fields. For example, additional time derivatives are known to occur in the Dirac fermion sector for certain choices of the dimensionless $c$, $d$, $e$, $f$, and $g$ coefficients \cite{Colladay:1998fq,Kostelecky:2000mm}. These time derivatives spoil the conventional time evolution of spinor solutions of the Dirac equation, but it is well-known that they can be removed by suitable field redefinitions in spinor space~\cite{Colladay:2001wk,Reis:2016hzu,Schreck:2017isa}. In spite of that, the minimal (nongravitational) SME does not exhibit second-order time derivatives of dynamical fields.

In general, second-order time derivatives of the dynamical fields must be treated with the method developed by Ostrogradsky~\cite{Ostro:1850,Borneas:1959,Riahi:1972,Woodard:2015,Bollini:1986am}. This procedure is usually
employed in the context of the nonminimal SME only. Hence, it must be considered as more than surprising that this
approach should be necessary to deal with the second-order time derivatives of the metric in
Eqs.~(\ref{eq:lagrange-density-LV-1}), (\ref{eq:lagrange-density-LV-3}), and (\ref{eq:lagrange-density-LV-4}), which are based on the minimal SME.
The argument made via Eq.~(\ref{eq:modified-action-extracting-time-derivatives}) is a justification for
transferring the first-order time derivatives from $K_{ij}$ to $s^{ij}$ via Eq.~(\ref{eq:relation-deruelle-1}), from $K$ to $s^{\mathbf{nn}}$ with the help of Eq.~(\ref{eq:relation-deruelle-2}), and from $K$ to $u$ by means of \eqref{eq:relation-deruelle-3}. In this process, generalized GHY
boundary terms are introduced to cancel the first-order time derivatives of the metric on the boundary.
Thus, the Ostrogradsky method is not needed.

As a consequence of our procedure, first-order time derivatives of the Lorentz-violating fields $s^{ij}$, $s^{\mathbf{nn}}$, and $u$ arise, which reveals two important properties of these contributions. First, the number of degrees of freedom must be conserved when applying Eqs.~(\ref{eq:relation-deruelle-1}), (\ref{eq:relation-deruelle-2}), and (\ref{eq:relation-deruelle-3}). In principle, the procedure transfers the degrees of freedom that come with the additional time derivatives of the extrinsic curvature to $s^{ij}$, $s^{\mathbf{nn}}$, and $u$. For $s^{ij}$ and $u$ this means that the background fields must somehow absorb these degrees of freedom.

Recall that we work in the setting of explicit diffeomorphism violation, wherewith $s^{ij}$ must be considered as a nondynamical tensor-valued function that is projected into $\Sigma_t$ from an initially chosen $s^{\mu\nu}$. Hence, $s^{ij}$ is not capable of absorbing any dynamical degrees of freedom, which indicates a mismatch. This behavior is how the well-known clash between explicit diffeomorphism violation and Riemannian geometry~\cite{Kostelecky:2003fs} manifests itself within the ADM formalism applied to this particular sector. The argument is similar for $u$. However, as will be discussed in Sec.~\ref{eq:field-equations-constraints}, this mismatch can possibly be resolved when taking into account a certain set of consistency conditions for the background fields.

Moreover, when the time derivatives are transferred from $K$ to $s^{\mathbf{nn}}$ in the purely timelike sector (see Eq.~(\ref{eq:relation-deruelle-2})), they do not only act on $s^{00}$, but they imply time derivatives of the lapse function. An interpretation of this behavior is that some of the gauge degrees of freedom of GR can become dynamical in this sector. The significance of that observation is highly obscure. On the other hand, both the purely spacelike sector and the scalar sector do not involve any time derivatives of the lapse function or the shift vector, i.e., these gauge degrees of freedom remain nondynamical in the presence of $s^{ij}$ and $u$, respectively. This additional problem specific to the purely timelike sector can also be tackled by introducing a suitable requirement for $s^{\mathbf{nn}}$, as will become clear in Sec.~\ref{eq:field-equations-constraints}.

In short, based on the previous discussion as well as the form of the modifications of $\mathcal{C}_i$ in Eqs.~(\ref{eq:hamiltonian-part1-constraint2}), (\ref{eq:momentum-constraint-scalar-sector}) one might be tempted to restrict the purely spacelike and the scalar sector to such background fields satisfying $\mathcal{L}_ms^{ij}=0$ and $\mathcal{L}_mu=0$, respectively. These conditions mean that $s^{ij}$ and $u$, respectively, are generated by the flow defined by the four-vector $m^{\mu}$ and they are necessary requirements for the internal consistency of these sectors. The conclusion is that $s^{ij}$ and $u$ chosen suitably in this manner could, indeed, imply consistent sectors of the minimal gravitational SME that violate diffeomorphism invariance explicitly. Furthermore, a similar requirement $\mathcal{L}_ms^{\mathbf{nn}}=0$ could be employed (see also the modification of $\mathcal{C}_i$ obtained in Eq.~(\ref{eq:momentum-constraint-timelike-sector})). Note that we consider $s^{\mathbf{nn}}$ as a new degree of freedom independent of $s^{\mu\nu}$ and $N$, i.e., we will not think of it as a quantity that involves time derivatives $\dot{s}^{00}$ and $\dot{N}$ separately. The forthcoming section will provide further substance to these (preliminary) claims.

\section{Field equations and constraints}
\label{eq:field-equations-constraints}
It is remarkable that in the ADM formalism there are direct relationships between the Einstein equations and the Hamiltonian constraint as well as the momentum constraints. These relations involve suitable contractions of the Einstein equations with the projector $q^{\mu}_{\phantom{\mu}\nu}$ and the four-vector $n^{\mu}$. We intend to give a brief summary of how this procedure works for GR. After that, we will be trying to tackle the diffeomorphism-violating modifications in an analogous way. Although the method applied to GR is rather unproblematic, its application to scenarios with explicit diffeomorphism violation has turned out to be a formidable task requiring tedious computations. A large part of those will be moved to App.~\ref{sec:projection-einstein-equations-modified}. However, the procedure will eventually imply several rewarding findings.

\subsection{General Relativity}
The structure of GR allows us to relate the field equations with the Hamiltonian and momentum constraints. We will consider the Einstein equations of GR
\begin{subequations}
\begin{equation}
(T_{\mathrm{mat}})^{\alpha\beta}={}^{(4)}G^{\alpha\beta}\,,
\end{equation}
where
\begin{equation}
{}^{(4)}G^{\alpha\beta}={}^{(4)}R^{\alpha\beta}-\frac{1}{2}g^{\alpha\beta}{}^{(4)}R\,,\quad (T_{\mathrm{mat}})^{\alpha\beta}=\frac{2}{\sqrt{-g}}\frac{\delta\mathcal{L}_{\mathrm{mat}}}{\delta g_{\alpha\beta}}\,,
\end{equation}
\end{subequations}
with the Einstein tensor ${}^{(4)}G^{\alpha\beta}$ in four-dimensional spacetime and the Belinfante energy-momentum tensor $(T_{\mathrm{mat}})^{\alpha\beta}$ linked to a matter Lagrange density $\mathcal{L}_{\mathrm{mat}}$. Since we are not taking matter into account, we will set $(T_{\mathrm{mat}})^{\alpha\beta}=0$.

Suitable projections of the Einstein equations imply the Hamiltonian and momentum
constraints (expressed in terms of the exterior curvature). First, a total projection along the direction orthogonal to $\Sigma_t$
leads to \eqref{eq:parameter-c0-GR}:
\begin{align}
2n_{\alpha}n_{\beta}{}^{(4)}G^{\alpha\beta}&=2n_{\alpha}n_{\beta}{}^{(4)}R^{\alpha\beta}-n^2{}^{(4)}R \notag \\
&=2n_{\alpha}n_{\beta}{}^{(4)}R^{\alpha\beta}+(R+K^2-K_{ij}K^{ij}-2n_{\alpha}n_{\beta}{}^{(4)}R^{\alpha\beta})=\mathcal{C}_0\,,
\end{align}
where we employed Eqs.~(\ref{eq:gauss-relation-scalar}),
(\ref{eq:ricci-totally-orthogonal-projection}). Second, a mixed projection parallel to
$\Sigma_t$ and along the direction orthogonal to $\Sigma_t$ implies \eqref{eq:parameter-ci-GR}:
\begin{equation}
2q^i_{\phantom{i}\alpha}n_{\beta}{}^{(4)}G^{\alpha\beta}=2q^i_{\phantom{i}\alpha}n_{\beta}{}^{(4)}
R^{\alpha\beta}-2q^i_{\phantom{i}\alpha}n^{\alpha}{}^{(4)}R=2(D_jK^{ji}-D^iK)=\mathcal{C}^i\,.
\end{equation}
Here we used the contracted Codazzi-Mainardi relation of \eqref{eq:codazzi-mainardi-contracted}.
\subsection{Minimal gravitational SME: $s^{\mu\nu}$ term}
\label{sec:projections-field-equations-s-term}
The modification of the Einstein equations follows from varying the action
\begin{equation}
\label{eq:action-s-term}
S'^{(s)}=\int_{\mathcal{M}}\mathrm{d}^4x\,\mathcal{L}'^{(s)}\,,
\end{equation}
with $\mathcal{L}'^{(s)}$ given by Eq.~(\ref{eq:minimal-gravity-contribution-s}). The result is well-known and is
stated in Eqs.~(6), (7) of~\cite{Bailey:2006fd}. Without a matter source and for a nonzero $s^{\mu\nu}$ only, the
modified Einstein equations amount to
\begin{subequations}
\label{eq:einstein-equations-modified}
\begin{align}
0&={}^{(4)}G^{\alpha\beta}-(T^{Rs})^{\alpha\beta}\,, \displaybreak[0]\\[2ex]
\label{eq:TsLIV}
(T^{Rs})^{\alpha\beta}&=\frac{1}{2}\left[g^{\alpha\beta}s^{\mu\nu}R_{\mu\nu}+\nabla_{\nu}\nabla^{\alpha}s^{\nu\beta}+\nabla_{\nu}\nabla^{\beta}s^{\nu\alpha}-\nabla^2s^{\alpha \beta}-g^{\alpha\beta}\nabla_{\mu}\nabla_{\nu}s^{\mu\nu}\right]\,.
\end{align}
\end{subequations}
Note that the modified Einstein equations stated in Eqs.~(62), (63) of~\cite{Kostelecky:2003fs} have a slightly different form where
the corresponding $(\tilde{T}^{Rs})^{\alpha\beta}$ has been reprinted in \eqref{eq:einstein-equation-modification-lower-indices} for completeness.
The latter field equations are valid for a modification of GR given by the action
\begin{equation}
\label{eq:action-tilde-s-term}
S''^{(s)}=\int_{\mathcal{M}}\mathrm{d}^4x\frac{\sqrt{-g}}{2\kappa}s_{\mu\nu}{}^{(4)}R^{\mu\nu}\,,
\end{equation}
i.e., for a background field with lower indices (although the field equations are expressed in terms of $s^{\mu\nu}$ with both indices raised by the metric). This finding shows that for explicit diffeomorphism violation, the field theories defined by the action $S'^{(s)}$ of \eqref{eq:action-s-term} and $S''^{(s)}$ of \eqref{eq:action-tilde-s-term} are not equivalent. More emphasis is put on this property in the recent work~\cite{Kostelecky:2020hbb} (see also the remarks made in Sec.~\ref{sec:introduction}). Since our setting is based on the action of \eqref{eq:action-s-term}, our analysis will be resting on the modified Einstein equations~(\ref{eq:einstein-equations-modified}). For the purpose of clarification, a short derivation of the modified field equations is provided in App.~\ref{eq:modified-field-equations-derivation}.

We now intend to find out whether there are possible connections between the modified Einstein equations and the constraints derived in Sec.~\ref{sec:hamiltonians-smunu}. To do so, we will have to compute suitable contractions of $(T^{Rs})^{\alpha\beta}$ in \eqref{eq:TsLIV} with $n_{\alpha}$ and $q^{\alpha}_{\phantom{\alpha}\beta}$, respectively. The computations turned out to be challenging and revealed further interesting insights. Details are presented in App.~\ref{sec:projection-einstein-equations-modified}.

There is an additional peculiarity with respect to \eqref{eq:decomposition-s} that we have employed to decompose $s^{\mu\nu}$ into a purely spacelike, a mixed, and a purely timelike sector. The latter decomposition must be considered as an identity and the individual parts depend on the coordinates chosen for the ADM decomposition. The situation is most clear for the purely spacelike part. According to its definition as $s^{ij}\equiv q^i_{\phantom{i}\mu}q^j_{\phantom{j}\nu}s^{\mu\nu}$, the purely spacelike part definitely involves coefficients of $s^{\mu\nu}$ with spacelike indices only --- independently of the exact form of the projectors. However, the same does not hold true for the mixed and purely timelike sectors. The problem is best understood by looking at the following explicit example. At first, the only coefficient of $s^{\mu\nu}$ that contributes to the purely timelike sector is $s^{00}$, as $s^{\mathbf{nn}}=s^{\mu\nu}n_{\mu}n_{\nu}=N^2s^{00}$. However, for a nonzero shift vector the purely timelike sector of the decomposition in Eq.~(\ref{eq:decomposition-s}) then reads
\begin{equation}
(s^{\alpha\beta})|_{\substack{\text{purely} \\ \text{timelike}}}=(n^{\alpha}n^{\beta}s^{\mathbf{nn}})=\begin{pmatrix}
1 & -N^i \\
-N^j & N^iN^j \\
\end{pmatrix}s^{00}\,,
\end{equation}
which obviously involves coefficients other than $s^{00}$. The behavior is similar for the mixed sector. Thus, a nonzero shift vector implies that the purely timelike and mixed sector couple to each other as well as to the purely spacelike one. If one does not want to pick particular observer frames with specific forms of $s^{\mu\nu}$ (see Sec.~\ref{sec:decoupling-sectors}), such couplings between different sectors can be avoided by employing Gaussian normal coordinates \cite{Gourgoulhon:2007ue,Gourgoulhon:2012} where $N=1$ and $N^i=0$. A slight generalization of Gaussian normal coordinates including a nontrivial lapse function $N$ can also be considered. It is indispensable, though, to work with a zero shift vector if one wants to avoid that different sectors couple to each other.

\subsubsection{Purely spacelike sector}
In what follows, we will obtain the suitable projections of the modified Einstein equations. Calculational details are shown in Sec.~\ref{eq:projections-purely-spacelike-details}. A complete projection of \eqref{eq:TsLIV} along the direction orthogonal to $\Sigma_t$ implies
\begin{equation}
2n_{\alpha}n_{\beta}(T^{Rs})^{\alpha\beta}=-(K^l_{\phantom{l}i}K_{jl}s^{ij}+KK_{ij}s^{ij}+s^{ij}R_{ij}-q^{\mu}_{\phantom{\mu}\alpha}q^{\nu}_{\phantom{\nu}\beta}\nabla_{\mu}\nabla_{\nu}s^{\alpha\beta})\,.
\end{equation}
The final term is involved. Evaluating it with care gives rise to
\begin{equation}
q^{\mu}_{\phantom{\mu}\alpha}q^{\nu}_{\phantom{\nu}\beta}\nabla_{\mu}\nabla_{\nu}s^{\alpha\beta}=D_iD_js^{ij}-K^l_{\phantom{l}i}K_{jl}s^{ij}+KK_{ij}s^{ij}-\frac{1}{N}K_{ij}\mathcal{L}_ms^{ij}\,,
\end{equation}
with the Lie derivative of the purely spacelike background given by \eqref{eq:lie-derivative-purely-spacelike-part}. Now we obtain the following intriguing result:
\begin{align}
\label{eq:orthogonal-projection-purely-spacelike}
2n_{\alpha}n_{\beta}(T^{Rs})^{\alpha\beta}&=-\left[s^{ij}(R_{ij}+2K^l_{\phantom{l}i}K_{lj})-D_iD_js^{ij}+\frac{1}{N}K_{ij}\mathcal{L}_ms^{ij}\right] \notag \\
&=-\left(\mathcal{C}_0^{(1)}+\frac{1}{N}K_{ij}\mathcal{L}_ms^{ij}\right)\,,
\end{align}
with $\mathcal{C}_0^{(1)}$ given by \eqref{eq:hamiltonian-part1-constraint1}. Thus, the modification of the Einstein equations completely projected along the direction orthogonal to $\Sigma_t$ almost equals the modification of $\mathcal{C}_0$, but there is an additional contribution given by the Lie derivative of the purely spacelike background tensor with respect to $m^{\mu}$.

Evaluating the mixed projection of \eqref{eq:TsLIV} is even more involved. An intermediate result reads
\begin{align}
\label{eq:mixed-projection-purely-spacelike-intermediate}
2q^k_{\phantom{k}\alpha}n_{\beta}(T^{Rs})^{\alpha\beta}&=-s^{ij}D_iK_j^{\phantom{j}k}-K_{ij}D^ks^{ij}-a_iK^i_{\phantom{i}j}s^{jk}+2K_{ij}D^is^{jk}+KD_is^{ik} \notag \\
&\phantom{{}={}}-(K\gamma^k_{\phantom{k}\lambda}+K^k_{\phantom{k}\lambda})\nabla_{\alpha}s^{\alpha\lambda}+n_{\alpha}q^{\nu}_{\phantom{\nu}\lambda}q^k_{\phantom{k}\sigma}\nabla_{\nu}\nabla^{\alpha}s^{\lambda\sigma}+s^{k\lambda}\nabla_{\alpha}K^{\alpha}_{\phantom{\alpha}\lambda}\,,
\end{align}
and finally, we obtain
\begin{align}
\label{eq:mixed-projection-purely-spacelike}
2q^k_{\phantom{k}\alpha}n_{\beta}(T^{Rs})^{\alpha\beta}&=-\left\{-D_i\left[\frac{1}{N}\mathcal{L}_m s^{ik}+2s^{lk}K^i_{\phantom{i}l}\right]+K_{ij}D^ks^{ij}\right\} \notag \\
&=-\left[q^{kl}\mathcal{C}^{(1)}_l+2D_i(s^{il}K^k_{\phantom{k}l})+K_{ij}D^ks^{ij}\right]\,,
\end{align}
with the modification $\mathcal{C}_l^{(1)}$ of $\mathcal{C}_l$ given by \eqref{eq:hamiltonian-part1-constraint2}.

\subsubsection{Mixed sector}
In Sec.~\ref{sec:hamiltonian-mixed-sector} we have brought up convincing arguments for the mixed sector involving only gauge degrees of freedom. Therefore, we do not consider it worthwhile to compute projections of $(T^{Rs})^{\alpha\beta}$ for this particular sector.

\subsubsection{Purely timelike sector}
To avoid couplings with the other sectors, we will be working with coordinates such that $N^i=0$. For the purely timelike sector there is the peculiarity that an additional contribution must be taken into account for the field equations that comes from varying the action. We have that
\begin{equation}
\delta(s^{\mathbf{nn}}n^{\mu}n^{\nu}R_{\mu\nu})\supset s^{\mathbf{nn}}R_{\mu\nu}\delta(n^{\mu}n^{\nu})=s^{\mathbf{nn}}R_{\mu\nu}\delta(q^{\mu\nu}-g^{\mu\nu})=-s^{\mathbf{nn}}R_{\mu\nu}\delta g^{\mu\nu}\,,
\end{equation}
according to \eqref{eq:induced-metric}. Thus, an extra term emerges within the diffeomorphism-violating modification of \eqref{eq:TsLIV} where the minus sign is extracted:
\begin{equation}
\label{eq:purely-timelike-sector-T-extra-term}
(T^{Rs})^{\alpha\beta}\mapsto (T^{Rs})^{\alpha\beta}+s^{\mathbf{nn}}g^{\alpha\mu}g^{\beta\nu}R_{\mu\nu}=(T^{Rs})^{\alpha\beta}+s^{\mathbf{nn}}R^{\alpha\beta}\,.
\end{equation}
Thus, a complete projection along the direction perpendicular to $\Sigma_t$ implies
\begin{align}
\label{eq:orthogonal-projection-purely-timelike}
2n_{\alpha}n_{\beta}(T^{Rs})^{\alpha\beta}&=-\left[D_iD^is^{\mathbf{nn}}+s^{\mathbf{nn}}(K^{ij}K_{ij}-K^2)+Kn^{\mu}\nabla_{\mu}s^{\mathbf{nn}}\right] \notag \\
&=-\left(\mathcal{C}_0^{(3)}-\frac{1}{N}K\mathcal{L}_ms^{\mathbf{nn}}\right)\,,
\end{align}
with $\mathcal{C}_0^{(3)}$ stated in \eqref{eq:hamiltonian-constraint-timelike-sector}. Carrying out a projection along $\Sigma_t$ with respect to one index and a projection along $n^{\mu}$ for the second index results in
\begin{align}
\label{eq:mixed-projection-purely-timelike}
2q^k_{\phantom{k}\alpha}n_{\beta}(T^{Rs})^{\alpha\beta}&=-\left[D^k(n^{\mu}\nabla_{\mu}s^{\mathbf{nn}}+2s^{\mathbf{nn}}K)-2D_j(s^{\mathbf{nn}}K^{jk})-KD^ks^{\mathbf{nn}}\right] \notag \\
&=-\left(q^{kl}\mathcal{C}^{(3)}_l-KD^ks^{\mathbf{nn}}\right)\,,
\end{align}
where $\mathcal{C}_l^{(3)}$ of \eqref{eq:momentum-constraint-timelike-sector} can be employed here. Details on how to arrive at these results are relegated to App.~\ref{eq:projections-purely-timelike-details}.

\subsection{Minimal gravitational SME: $u$ term}

Without a matter source and only the $u$ term present, the modification of the Einstein equations is obtained from varying
\begin{equation}
\label{eq:action-u-term}
S'^{(u)}=\int_{\mathcal{M}}\mathrm{d}^4x\,\mathcal{L}'^{(u)}\,,
\end{equation}
with $\mathcal{L}'^{(u)}$ given by Eq.~(\ref{eq:minimal-gravity-contribution-u}). Then,
\begin{subequations}
\label{eq:einstein-equations-modified-u}
\begin{align}
0&={}^{(4)}G^{\alpha\beta}-(T^{Ru})^{\alpha\beta}\,, \displaybreak[0]\\[2ex]
\label{eq:TuLIV}
(T^{Ru})^{\alpha\beta}&=-\frac{1}{2}(\nabla^{\alpha}\nabla^{\beta}u+\nabla^{\beta}\nabla^{\alpha}u)+g^{\alpha\beta}\nabla^2u+u{}^{(4)}G^{\alpha\beta}\,.
\end{align}
\end{subequations}
A short derivation of this result is also presented in App.~\ref{eq:modified-field-equations-derivation}. As we did before, we can compute suitable projections of the modified Einstein equations completely orthogonal to $\Sigma_t$ and partially into $\Sigma_t$:
\begin{subequations}
\begin{align}
\label{eq:orthogonal-projection-scalar}
2n_{\alpha}n_{\beta}(T^{Ru})^{\alpha\beta}&=-\left(\mathcal{C}_0^{(4)}-\frac{2}{N}K\mathcal{L}_mu\right)\,, \\[2ex]
\label{eq:mixed-projection-scalar}
2q^k_{\phantom{k}\alpha}n_{\beta}(T^{Ru})^{\alpha\beta}&=-\left(q^{kl}\mathcal{C}_l^{(4)}-2KD^ku\right)\,,
\end{align}
\end{subequations}
with $\mathcal{C}_0^{(4)}$, $\mathcal{C}_l^{(4)}$ given by \eqref{eq:hamiltonian-constraint-scalar-sector} and \eqref{eq:momentum-constraint-scalar-sector}, respectively. Computational details are presented in App.~\ref{eq:projections-scalar-details}.

\subsection{Concluding remarks}

To summarize, projections of the diffeomorphism-violating modifications $(T^{Rs})^{\alpha\beta}$, $(T^{Ru})^{\alpha\beta}$ for the purely spacelike, the purely timelike, and the scalar sector with $q^{\mu}_{\phantom{\mu}\nu}$ and $n^{\mu}$ do not completely provide the Hamiltonian and momentum constraints (expressed in terms of the exterior curvature). In contrast to GR, there are correction terms. To substantiate these outcomes, we will be taking a deeper look at additional properties of the ADM action in the forthcoming section.

\subsection{Functional derivatives of ADM action}
\label{sec:functional-derivatives-ADM-action}

In the current section we intend to compute functional derivatives of the ADM action with respect to the lapse function and the shift vector. The first is expected to be connected to the Hamiltonian constraint, whereas the second is associated with the momentum constraint~\cite{Bertschinger:2002}. Calculational details are shown in App.~\ref{sec:functional-derivatives-ADM-action-computations}. First, for the ADM-decomposed EH action of \eqref{eq:einstein-hilbert-action-ADM-decomposed} we obtain
\begin{subequations}
\begin{align}
\label{eq:functional-derivative-EH-lapse}
\frac{\delta S^{(0)}}{\delta N}&=\frac{\sqrt{q}}{2\kappa}\mathcal{C}_0\,, \\[2ex]
\label{eq:functional-derivative-EH-shift}
\frac{\delta S^{(0)}}{\delta N^k}&=\frac{\sqrt{q}}{2\kappa}\mathcal{C}_k\,,
\end{align}
\end{subequations}
with $\mathcal{C}_0$, $\mathcal{C}_k$ given by \eqref{eq:parameter-c0-GR} and \eqref{eq:parameter-ci-GR}, respectively. To find out whether or not analogous relationships exist in the context of the background fields $s^{\mu\nu}$ and $u$, we consider the modifications of the GR action within the ADM formalism. The latter read
\begin{equation}
S'^{(i)}=\int_{\mathcal{M}}\mathrm{d}^4x\,\mathcal{L}'^{(i)}\,,
\end{equation}
with the Lagrangians given by Eqs.~(\ref{eq:lagrange-density-LV-1}) -- (\ref{eq:lagrange-density-LV-4}) for $i\in \{1,2,3,4\}$. Computing the functional derivatives implies
\begin{subequations}
\begin{align}
\label{eq:functional-derivatives-purely-spacelike}
\frac{\delta S'^{(1)}}{\delta N}&=\frac{\sqrt{q}}{2\kappa}\left(\mathcal{C}_0^{(1)}+\frac{1}{N}K_{ij}\mathcal{L}_ms^{ij}\right)\,, \displaybreak[0]\\[2ex]
\label{eq:functional-derivative-purely-spacelike-shift}
\frac{\delta S'^{(1)}}{\delta N^k}
&=\frac{\sqrt{q}}{2\kappa}\left[\mathcal{C}^{(1)}_k+K_{ij}D_ks^{ij}+2D_i(s^{ij}K_{kj})\right]\,, \displaybreak[0]\\[2ex]
\label{eq:functional-derivative-mixed-lapse}
\frac{\delta S'^{(2)}}{\delta N}&=\frac{\sqrt{q}}{2\kappa}\left\{\mathcal{C}_0^{(2)}-\frac{2}{N}D_i\left[N(s^{j\mathbf{n}}K^i_{\phantom{i}j}-s^{i\mathbf{n}}K)\right]\right\}\,, \displaybreak[0]\\[2ex]
\label{eq:functional-derivative-mixed-shift}
\frac{\delta S'^{(2)}}{\delta N^k}&=\frac{\sqrt{q}}{2\kappa}\mathcal{C}_k^{(2)}\,, \displaybreak[0]\\[2ex]
\label{eq:functional-derivaties-purely-timelike}
\frac{\delta S'^{(3)}}{\delta N}&=\frac{\sqrt{q}}{2\kappa}\left(\mathcal{C}_0^{(3)}-\frac{1}{N}K\mathcal{L}_ms^{\mathbf{nn}}\right)\,, \displaybreak[0]\\[2ex]
\label{eq:functional-derivative-purely-timelike-shift}
\frac{\delta S'^{(3)}}{\delta N^k}&=\frac{\sqrt{q}}{2\kappa}\left(\mathcal{C}_k^{(3)}-KD_ks^{\mathbf{nn}}\right)\,, \displaybreak[0]\\[2ex]
\frac{\delta S'^{(4)}}{\delta N}&=\frac{\sqrt{q}}{2\kappa}\left(\mathcal{C}_0^{(4)}-\frac{2}{N}K\mathcal{L}_mu\right)\,, \displaybreak[0]\\[2ex]
\frac{\delta S'^{(4)}}{\delta N^k}&=\frac{\sqrt{q}}{2\kappa}\left(\mathcal{C}_k^{(4)}-2KD_ku\right)\,.
\end{align}
\end{subequations}
Several observations are in order. First, in contrast to GR, there is no complete match between the functional derivatives and the constraints for the purely spacelike, the purely timelike, and the scalar sector. There are additional contributions that involve products of the extrinsic curvature and Lie derivatives or covariant derivatives of the background fields. Second, there is a match for the Hamiltonian constraint of the mixed sector modulo a boundary term on $\partial\Sigma_t$. Interestingly, the momentum constraint of the mixed sector matches perfectly without any additional contributions. Third, note the similarities between the purely timelike and the scalar sectors that are also evident in the boundary terms of Eqs.~(\ref{eq:boundary-term-timelike-sector}), (\ref{eq:boundary-term-scalar-sector}).

\subsection{Consistency requirements}

In what follows, we will draw some intriguing conclusions from the projections of the modified Einstein equations and
the functional derivatives of the ADM actions. We do so for the purely spacelike and the purely timelike sector of $s^{\mu\nu}$
as well as for $u$. Since the mixed sector of $s^{\mu\nu}$ involves gauge degrees of freedom only, it will not
be taken into consideration. For the purely spacelike sector we start by
comparing Eqs.~(\ref{eq:orthogonal-projection-purely-spacelike}), (\ref{eq:mixed-projection-purely-spacelike}) with the corresponding
functional derivatives of the ADM action:
\begin{subequations}
\begin{align}
\frac{\delta S'^{(1)}}{\delta N}&=-\frac{\sqrt{q}}{\kappa}n_{\alpha}
n_{\beta}(T^{Rs})^{\alpha\beta}\,, \displaybreak[0]\\[2ex]
\frac{\delta S'^{(1)}}{\delta N_k}&=-\frac{\sqrt{q}}{\kappa}
q^k_{\phantom{k}\alpha}n_{\beta}(T^{Rs})^{\alpha\beta}\,.
\end{align}
\end{subequations}
For the purely timelike sector, we
compare Eqs.~(\ref{eq:orthogonal-projection-purely-timelike}), (\ref{eq:mixed-projection-purely-timelike}) with the suitable functional derivatives:
\begin{subequations}
\begin{align}
\frac{\delta S'^{(3)}}{\delta N}&=-\frac{\sqrt{q}}{\kappa}n_{\alpha}n_{\beta}(T^{Rs})^{\alpha\beta}\,, \displaybreak[0]\\[2ex]
\frac{\delta S'^{(3)}}{\delta N_k}&=-\frac{\sqrt{q}}{\kappa}q^k_{\phantom{k}\alpha}n_{\beta}(T^{Rs})^{\alpha\beta}\,.
\end{align}
\end{subequations}
Finally, we take Eqs.~(\ref{eq:orthogonal-projection-scalar}), (\ref{eq:mixed-projection-scalar}) from the scalar sector and compare those to the corresponding functional derivatives:
\begin{subequations}
\begin{align}
\frac{\delta S'^{(4)}}{\delta N}&=-\frac{\sqrt{q}}{\kappa}n_{\alpha}n_{\beta}(T^{Ru})^{\alpha\beta}\,, \displaybreak[0]\\[2ex]
\frac{\delta S'^{(4)}}{\delta N_k}&=-\frac{\sqrt{q}}{\kappa}q^k_{\phantom{k}\alpha}n_{\beta}(T^{Ru})^{\alpha\beta}\,.
\end{align}
\end{subequations}
Thus, we conclude that the functional derivatives of the ADM-decomposed actions with respect to the lapse function are
proportional to the associated modifications of the Einstein equations
projected along the direction orthogonal to $\Sigma_t$. Furthermore,
the functional derivatives for the shift covector are proportional to the mixed
projections of the modifications. In this context, discrepancies do not arise for
the purely spacelike, the purely timelike, and the scalar sector.

Next, we compare the latter findings directly to the Hamiltonian and momentum
constraints. For the purely spacelike sector we establish the correspondences
\begin{subequations}
\begin{align}
\frac{\delta}{\delta N}\left(S'^{(1)}\big|_{\mathcal{L}_ms^{ij}=0}\right)&
=\mathcal{C}_0^{(1)}\,, \displaybreak[0]\\[2ex]
\frac{\delta}{\delta N^k}\left(S'^{(1)}\big|_{\mathcal{L}_ms^{ij}=0}\right)&
=\mathcal{C}_k^{(1)}\big|_{\mathcal{L}_ms^{ij}=0}\,,
\end{align}
\end{subequations}
with $\mathcal{C}_0^{(1)}$ and $\mathcal{C}_k^{(1)}$ given by Eqs.~(\ref{eq:hamiltonian-part1-constraint1}),
(\ref{eq:hamiltonian-part1-constraint2}). For the purely timelike sector it has to hold that
\begin{subequations}
\begin{align}
\frac{\delta}{\delta N}\left(S'^{(3)}\big|_{\mathcal{L}_ms^{\mathbf{nn}} =0}\right)&=\mathcal{C}_0^{(3)}\,, \displaybreak[0]\\[2ex]
\frac{\delta}{\delta N^k}\left(S'^{(3)}\big|_{\mathcal{L}_ms^{\mathbf{nn}}=0}\right)&=\mathcal{C}_k^{(3)}\big|_{\mathcal{L}_ms^{\mathbf{nn}}=0}\,,
\end{align}
\end{subequations}
with $\mathcal{C}_0^{(3)}$ and $\mathcal{C}_k^{(3)}$ stated in Eqs.~(\ref{eq:hamiltonian-constraint-timelike-sector}),
(\ref{eq:momentum-constraint-timelike-sector}). Last but not least, for the scalar sector we deduce
\begin{subequations}
\begin{align}
\frac{\delta}{\delta N}\left(S'^{(4)}\big|_{\mathcal{L}_mu=0}\right)&=\mathcal{C}_0^{(4)}\,, \displaybreak[0]\\[2ex]
\frac{\delta}{\delta N^k}\left(S'^{(4)}\big|_{\mathcal{L}_mu=0}\right)&=\mathcal{C}_k^{(4)}\big|_{\mathcal{L}_mu=0}\,,
\end{align}
\end{subequations}
where $\mathcal{C}_0^{(4)}$ and $\mathcal{C}_k^{(4)}$ must be taken from Eqs.~(\ref{eq:hamiltonian-constraint-scalar-sector}), (\ref{eq:momentum-constraint-scalar-sector}). Hence, we conclude that for these three sectors being internally consistent, the following conditions must be required:
\begin{subequations}
\label{eq:consistency-conditions}
\begin{align}
\mathcal{L}_ms^{ij}=0\,, \\[2ex]
\mathcal{L}_ms^{\mathbf{nn}}=0\,, \\[2ex]
\mathcal{L}_mu=0\,.
\end{align}
\end{subequations}
In what follows, the latter will be denoted as consistency conditions. In the context of the ADM decomposition, a gravity theory endowed with $s^{ij}$-, $s^{\mathbf{nn}}$- or $u$-type background fields, which violate diffeomorphism invariance explicitly, is likely to be internally consistent as long as the Lie derivatives of these backgrounds with respect to the vectorfield $m^{\mu}$ vanish. These consistency requirements are directly connected to diffeomorphisms acting on the underlying spacetime manifold. Diffeomorphisms affect tensorfields in the tangent bundle and are generated by Lie derivatives along arbitrary vectorfields. Thus, Eqs.~(\ref{eq:consistency-conditions}) mean that the corresponding background fields must be invariant under diffeomorphisms generated by $m^{\mu}$ within the ADM formalism such that the modified-gravity theory is consistent.

The recent findings demonstrate how a background field violating diffeomorphism invariance explicitly must be understood within gravity. A background field in the nongravitational SME is {\em defined} to not transform covariantly under particle Lorentz transformations, i.e., it is {\em defined} as fixed under such transformations. Note that diffeomorphisms in Minkowski spacetime are equivalent to translations induced by a constant four-vector $\zeta^{\mu}$. Then, the Lie derivative of a two-tensor-valued background field $k^{\mu\nu}$ simply corresponds to the directional derivative along $\zeta^{\mu}$. An analogous consistency requirement would then amount to
\begin{equation}
\label{eq:condition-translation-invariance}
\mathcal{L}_{\zeta}k^{\mu\nu}=\zeta^{\varrho}\nabla_{\varrho}k^{\mu\nu}=0\,.
\end{equation}
For an arbitrary $\zeta^{\mu}$ this condition is satisfied for backgrounds $k^{\mu\nu}$ that do depend on the spacetime position (at least when expressed in terms of Cartesian coordinates). In the context of the nongravitational SME, constant controlling coefficients are usually employed for two reasons. First, it is simpler to think of a background as being constant. More importantly, coefficients depending on the spacetime position violate translation invariance and, therefore, energy-momentum conservation due to Noether's theorem. This property would imply additional complications that are beyond studies of Lorentz violation.

However, in curved spacetime, a dependence of controlling coefficients on spacetime position must usually be assumed. For example, $\nabla_{\varrho}s^{\mu\nu}=0$ would only hold in spacetimes known as parallelizable \cite{Kostelecky:2003fs}. The latter are very special choices and of lesser interest in the context of gravity. Within the ADM formalism in gravity, \eqref{eq:consistency-conditions} can be interpreted as generalizations of \eqref{eq:condition-translation-invariance} where the latter implies energy-momentum conservation in Minkowski spacetime.

A pictorial interpretation of the problems that arise with explicit diffeomorphism violation in gravity is as follows. The arguments to be made rely on test particles being present in the curved spacetime manifold. Note that we have not introduced a coupling term with matter in the action, so far. Hence, rigorous studies of the interplay between matter and diffeomorphism-violating background fields will be done elsewhere. Nevertheless, we can make some physical arguments to interpret the significance of \eqref{eq:consistency-conditions}.

We can consider a test particle propagating in a curved spacetime. The particle moving between two distinct points follows a geodesic along of which it is in free fall, i.e., its acceleration vanishes. A background field giving rise to an explicit violation of diffeomorphism invariance modifies the geodesic equation, whereupon particle motion is affected.\footnote{Note that classical-particle analogs subject to certain types of Lorentz violation described by the (nongravitational) SME were shown to follow geodesics associated with Finsler geometries \cite{Kostelecky:2011qz}.} Then, the momentum of the particle will change in a way that is not described by GR, but that has to be accounted for by the background field. However, as the latter is nondynamical, it is incapable of absorbing or emitting momentum \cite{Bluhm:2014oua}.

Similar arguments can be developed for light rays propagating through a background field in curved spacetime. As long as the gravitational field is weak enough, the eikonal approximation is a suitable approach (see, e.g.,~\cite{Born:1999,Wu:1988}). Then, the curved spacetime manifold is approximately described by an inhomogeneous optical medium, i.e., its refractive index is position-dependent (and may also depend on polarization). The presence of a diffeomorphism-violating background field leads to additional optical effects such as anisotropy, dispersion, and birefringence. In this context, a background field violating diffeomorphism symmetry explicitly corresponds to a nondynamical optical medium on top of the optical medium ascribed to the curved spacetime manifold $\mathcal{M}$. Whenever a light ray changes its propagation direction, its wave vector changes, whereupon this change must be transferred between the light ray and the diffeomorphism-violating medium. Furthermore, in the presence of an explicitly time-dependent medium, even energy is to be transferred between both entities. However, a nondynamical medium neither accomplishes the first nor the second. Similar arguments were already developed in~\cite{Schreck:2015dsa}.

To solve the aforementioned problems, \eqref{eq:consistency-conditions} seem to be the necessary requirements that render a nondynamical background capable of incorporating energy-momentum transfer between a test particle and the background consistently. There may be a certain notion of energy-momentum that is conserved even for a nondynamical background satisfying \eqref{eq:consistency-conditions}. Whether or not these quantities correspond to the Killing energy and Killing momentum, which are associated with isometries of the underlying spacetime manifold, remains an interesting open question to be studied in the future.
\section{Analysis of constraints and Hamilton field equations}
\label{sec:constraint-structure}
Finally, let us analyze the structure of modified Hamiltonian and momentum constraints that we obtained for Eqs.~(\ref{eq:minimal-gravity-contribution-u}), (\ref{eq:minimal-gravity-contribution-s}). In general, our analysis has been based on a field theory described by a Lagrange density $\mathcal{L}$. A constraint is called primary when it follows directly from the form of $\mathcal{L}$. Such constraints occur for a certain canonical variable, say $\Phi$, when the Lagrange density does not involve the time derivative of the latter. The canonical momentum
\begin{equation}
\Pi_{\Phi}\equiv \frac{\partial\mathcal{L}}{\partial\Phi}\,,
\end{equation}
associated with $\Phi$ is then equal to zero. Therefore, it is not possible to express $\dot{\Phi}$ in terms of $\Pi_{\Phi}$, which does not permit deriving a Hamilton density via the Legendre transform. Then a Hamilton density $\mathcal{H}$ is obtained from $\mathcal{L}$ without taking the constrained variable into consideration. Subsequently the constraint is added to $\mathcal{H}$ via a Lagrange multiplier $\xi$ to define an extended Hamiltonian of the form $\mathcal{H}^{\mathrm{(ext)}}\equiv \mathcal{H}+\xi\Pi_{\Phi}$. In the literature, a primary constraint is written as
\begin{equation}
\Pi_{\Phi}\approx 0\,,
\end{equation}
where $\approx$ means ``weakly equal to zero.'' This notation is used to indicate that $\Pi$ is only taken to be zero when the constraint is satisfied, which is not necessarily assumed in computations right from the start. For example, setting $\Pi_{\Phi}=0$ (``strongly equal to zero'') in $\mathcal{H}^{\mathrm{(ext)}}$ would imply that the term added with the Lagrange multiplier does not contribute at all, which is undesired.

In what follows, the constraint structure of GR shall be reviewed briefly (see, e.g., \cite{Hanson:1976}). We start with a set of 10 canonical variables $X_i=\{N,N^i,q_{ij}\}$. Note that the index of $X_i$ is defined as a lower one although it involves the shift vector components $N^i$ with upper indices. The canonical momenta follow from the generic Lagrange density $\mathcal{L}$ via
\begin{equation}
\Pi^i\equiv \frac{\partial\mathcal{L}}{\partial X_i}\,,
\end{equation}
leading to the set of 10 canonical momenta $\Pi^i=\{\pi_N,\pi_i,\pi^{ij}\}$ given as
\begin{equation}
\pi_N\equiv \frac{\partial\mathcal{L}}{\partial\dot{N}}\,,\quad \pi_i\equiv\frac{\partial\mathcal{L}}{\partial\dot{N}^i}\,,\quad \pi^{ij}\equiv\frac{\partial\mathcal{L}}{\partial\dot{q}_{ij}}\,.
\end{equation}
By employing the canonical variables and momenta, we define the Poisson bracket of two quantities $F=F(x)$, $G=G(x')$ via
\begin{equation}
\label{eq:poisson-bracket}
\{F,G\}\equiv \int_{\Sigma_t}\mathrm{d}^3y\,\left[\frac{\delta F}{\delta X_i(y)}\frac{\delta G}{\delta\Pi^i(y)}-\frac{\delta F}{\delta\Pi^i(y)}\frac{\delta G}{\delta X_i(y)}\right]\,,
\end{equation}
where $\delta/\delta\Phi$ denotes the variational derivative with respect to the variable $\Phi$.
\subsection{General Relativity}
As the EH Lagrange density of \eqref{eq:lagrange-density-gravitational-sme} and contained in \eqref{eq:einstein-hilbert-action-ADM-decomposed} does not involve time
derivatives of both the lapse function and the shift vector, the associated canonical momenta vanish. Thus, according
to the introductory explanations, we have already identified a set of 4 primary constraints in GR:
\begin{equation}
\label{eq:primary-constraints-GR}
\pi_N\approx 0\,,\quad \pi_i\approx 0\,,
\end{equation}
whereupon we define
\begin{equation}
\mathcal{H}^{\mathrm{(ext)}}\equiv\mathcal{H}^{(0)}+\eta\pi_N+\theta^i\pi_i\,,
\end{equation}
with the four Lagrange multipliers $\eta$ and $\theta^i$. The time evolution of constraints is governed
by suitable Poisson brackets with the extended Hamilton density. As the dependence of $\mathcal{H}^{\mathrm{(ext)}}$ on the lapse function and the shift vector is transparent, we quickly arrive at
\begin{subequations}
\label{eq:poisson-brackets-constraints-EH}
\begin{align}
\{\pi_N,H^{\mathrm{(ext)}}\}&=-C_0\,, \displaybreak[0]\\[2ex]
\{\pi_i,H^{\mathrm{(ext)}}\}&=-C_i\,.
\end{align}
\end{subequations}
Each primary constraint $\Pi_{\Phi}\approx 0$ should be preserved with time to not change the constraint
structure. Its time evolution is governed by
\begin{equation}
\frac{\mathrm{d}\Pi_{\Phi}}{\mathrm{d}t}=\{\Pi_{\Phi},\mathcal{H}^{(\mathrm{ext})}\}+\frac{\partial\Pi_{\Phi}}{\partial t}\,.
\end{equation}
The partial time derivative on the right-hand side is only needed when the primary constraint depends on time explicitly.
To preserve this primary constraint in time, we must require that
\begin{equation}
\frac{\mathrm{d}\Pi_{\Phi}}{\mathrm{d}t}\approx 0\,,
\end{equation}
which implies a further constraint that is called a secondary one. Note that we must again talk of this secondary
constraint as being weakly equal to zero. Inserting $\Pi_{\Phi}=0$ directly would trivially result in a vanishing time
derivative. If the secondary constraint is not automatically weakly equal to zero when the primary constraint satisfies
this property, this new constraint must be included into the Hamilton density via another Lagrange multiplier~$\zeta$:
\begin{equation}
\mathcal{H}^{\mathrm{(ext)}}\mapsto \tilde{\mathcal{H}}^{\mathrm{(ext)}}=\mathcal{H}^{(0)}+\xi\Pi_{\Phi}+\zeta\frac{\mathrm{d}\Pi_{\Phi}}{\mathrm{d}t}\,.
\end{equation}
The procedure continues in this manner and may provide even further constraints. Hence, in the context of GR, for
the primary constraints of \eqref{eq:primary-constraints-GR} to be conserved, we must impose the following secondary constraints:
\begin{subequations}
\label{eq:secondary-constraints-GR}
\begin{align}
\{\pi_N,H^{\mathrm{(ext)}}\}&=-C_0\approx 0\,, \displaybreak[0]\\[2ex]
\{\pi_i,H^{\mathrm{(ext)}}\}&=-C_i\approx 0\,,
\end{align}
\end{subequations}
with $C_0$, $C_i$ of \eqref{eq:hamiltonian-constraint} and \eqref{eq:momentum-constraint}, respectively. The latter finding
now clearly demonstrates why $C_0$, $C_i$ are considered as constraints in GR.

The last step and perhaps the most essential one to derive the number of degrees of freedom in a constrained theory is to identify
the first- and second-class constraints \cite{Hanson:1976,Henneaux:1992}. Imagine that we have a set of $n$ constraints $\{\phi_a\}$ with $a=1,2 \dots n$. A
constraint $Q$ is called first-class, if it has weakly vanishing Poisson brackets with each member of the set $\{\phi_a\}$, i.e.,
$\{Q,\phi_a\}\approx 0$ for all $a$. It is called second-class, if at least one Poisson bracket is not weakly equal to zero:
$\{Q,\phi_a\}\not\approx 0$ for not less than a single $a$. The Dirac bracket can be defined from the latter, which allows for
imposing second-class constraints strongly to zero~\cite{Hanson:1976}. The total number of physical degrees of freedom then
corresponds to (see, e.g., p.~29 in \cite{Henneaux:1992}):
\begin{equation}
\label{eq:number-degrees-freedom}
N_{\mathrm{dof}}=\frac{1}{2}(N_{\mathrm{ph}}-2N_1-N_2)\,,
\end{equation}
where $N_{\mathrm{ph}}$ is the number of phase space variables and $N_1$ ($N_2$) is the number of first-class (second-class)
constraints. Note that the number of first-class constraints has a weight factor of 2 showing that these
contribute differently to the number of degrees of freedom than second-class constraints. In particular, GR involves 20 phase
space variables in total (10 metric components and 10 conjugate momenta). Equations~(\ref{eq:primary-constraints-GR}),
(\ref{eq:secondary-constraints-GR}) comprise a set of 8 first-class constraints, whereas there are no second-class constraints.
Then, $N_{\mathrm{ph}}=20$, $N_1=8$, and $N_2=0$, which implies $N_{\mathrm{dof}}=2$ corresponding to the correct number of
physical, propagating degrees of freedom, as expected.

In gravity, the Hamiltonian and momentum constraints, which are first-class, play an essential role in the context of diffeomorphisms. In general,
diffeomorphisms are generated by vector fields $\psi$. A representation of the diffeomorphism algebra in the tangent bundle of the spacetime manifold, where tensor fields of arbitrary rank are defined, is given by the Lie derivative $\mathcal{L}_{\psi}$. In what follows, we will compute Poisson brackets of the canonical variables $q_{ij},\pi^{ij}$ with the Hamiltonian and momentum constraints according to \eqref{eq:poisson-bracket}. The constraints will be integrated over with the lapse function and the shift vector chosen as smearing functions. The following important results can then be derived within GR~\cite{Gambini:1996,Menotti:2017}:
\begin{subequations}
\label{eq:spatial-diffeomorphisms-gr}
\begin{align}
\label{eq:spatial-diffeomorphisms-1}
\left\{q_{ij}(x),\int_{\Sigma_t} \mathrm{d}^3y\,C_i(y)N^i(y)\right\}&=q_{jk}(x)D_iN^k(x)+q_{ik}(x)D_jN^k(x)=\mathcal{L}_{\mathbf{N}}q_{ij}(x)\,, \\[2ex]
\label{eq:spatial-diffeomorphisms-2}
\left\{\pi_{ij}(x),\int_{\Sigma_t}\mathrm{d}^3y\,C_i(y)N^i(y)\right\}&=N^k(x)D_k\pi^{ij}(x)+\pi^{ij}(x)D_kN^k(x) \notag \\
&\phantom{{}={}}-\pi^{ik}(x)D_kN^j(x)-\pi^{jk}(x)D_kN^i(x)=\mathcal{L}_{\mathbf{N}}\pi^{ij}(x)\,.
\end{align}
\end{subequations}
For the second of these Poisson brackets it is crucial to take into account that $\pi^{ij}$ transforms as a tensor density. These findings mean that the momentum constraint is the generator of spatial diffeomorphisms in the spacelike hypersurfaces $\Sigma_t$, as these are connected to the shift vector~$\mathbf{N}$ \cite{Gambini:1996}. Furthermore, we obtain
\begin{equation}
\label{eq:spacetime-diffeomorphism-gr}
\left\{q_{ij}(x),\int_{\Sigma_t}\mathrm{d}^3y\,C_0(y)N(y)\right\}=2N(x)K_{ij}(x)=\dot{q}_{ij}(x)-\mathcal{L}_{\mathbf{N}}q_{ij}(x)=\mathcal{L}_mq_{ij}(x)\,,
\end{equation}
where Eqs.~(\ref{eq:extrinsic-curvature}), (\ref{eq:extrinsic-curvature-canonical-momentum-GR}) are understood to be used here. This relation means that the Hamiltonian constraint generates spacetime diffeomorphisms connected to the four-vector $m^{\mu}$ \cite{Gambini:1996}. Finally, we can confirm the validity of the first set of Hamilton's field equations:
\begin{equation}
\label{eq:hamilton-field-equation-1}
\dot{q}_{ij}(x)=\left\{q_{ij}(x),\int_{\Sigma_t}\mathrm{d}^3y\,\mathcal{H}^{(0)}(y)\right\}\,,
\end{equation}
with $\mathcal{H}^{(0)}$ given by \eqref{eq:hamilton-density-EH-constraints}. The latter Hamilton equations are interpreted as geometrical identities and should not be modified as long as Riemannian geometry is taken as the foundation of a modified-gravity theory. The second of Hamilton's field equations involves the canonical momentum $\pi^{ij}$ and conveys information on the dynamics.

\subsection{Minimal gravitational SME: $s^{\mu\nu}$ term}

By adding suitable modified GHY boundary terms to the action (see Sec.~\ref{sec:GHY-boundary-term-generalized} for their construction), we were able to move all additional time derivatives acting on the extrinsic curvature to the Lorentz-violating background fields. Now, a further support of the consistency conditions in \eqref{eq:consistency-conditions} is provided by the following argument. As long as these requirements are satisfied, the modifications $\mathcal{H}^{(1,3,4)}$ of the Hamilton densities provided by Eqs.~(\ref{eq:hamiltonian-part1}), (\ref{eq:hamiltonian-part3}), and (\ref{eq:hamiltonian-part4}) neither involve time derivatives of the lapse function nor of the shift vector. Hence, \eqref{eq:primary-constraints-GR} remain primary constraints in a gravity theory modified by the presence of $s^{ij}$, $s^{\mathbf{nn}}$, and $u$. Furthermore, the structure of the modified Hamilton densities in terms of $N$ and $N^i$ remains standard, whereupon \eqref{eq:poisson-brackets-constraints-EH} still holds with the corresponding $C_0^{(1,3,4)}$ of Eqs.~(\ref{eq:hamiltonian-constraint-C01}), (\ref{eq:hamiltonian-constraint-C03}), (\ref{eq:hamiltonian-constraint-C04}) and $C^{(1,3,4)}_k$ of Eqs.~(\ref{eq:momentum-constraint-Ci1}), (\ref{eq:momentum-constraint-Ci3}), (\ref{eq:momentum-constraint-Ci4}) inserted. Then the Hamiltonian and momentum constraints remain secondary constraints and the constraint structure proper is not modified by diffeomorphism violation, as proposed right at the beginning of the paper. Let us now compute the Poisson brackets previously considered for EH theory in the diffeomorphism-violating setting under consideration.
\subsubsection{Purely spacelike sector}
\label{sec:purely-spatial-sector-constraints}
For the purely spacelike sector, the Poisson bracket of \eqref{eq:hamiltonian-constraint-C01} with the induced metric implies:
\begin{align}
\left\{q_{ij}(x),\int_{\Sigma_t}\mathrm{d}^3y\,C_0^{(1)}N(y)\right\}&=N\frac{2\kappa}{\sqrt{q}}\left[2(\pi_{ij}+s_i^{\phantom{i}k}\pi_{jk}+s_j^{\phantom{j}k}\pi_{ik}-s_{ij}\pi)\right. \notag \\
&\phantom{{}={}}\hspace{1.2cm}\left.-(\pi-s^k_{\phantom{k}k}\pi+2s^{kl}\pi_{kl})q_{ij}\right] \notag \\
&=2NG_{ijab}\left(\frac{2\kappa}{\sqrt{q}}\pi^{ab}\right)=2NK_{ij}\,.
\end{align}
The variables on the right-hand side of the first equality sign in the latter relation are understood to depend on the coordinates $x$. From now on, such dependencies are omitted for brevity. As the momentum constraint remains unmodified at first order in the controlling coefficients, \eqref{eq:spatial-diffeomorphisms-1} can be taken over. Therefore, the first of Hamilton's field equations of \eqref{eq:hamilton-field-equation-1} remains valid for the purely spacelike sector --- at least at first order in diffeomorphism violation.

\subsubsection{Mixed sector}

We are already aware of the mixed sector involving mere gauge degrees of freedom. Nevertheless, we will take a brief look at the constraint structure that has to reduce to that of EH theory when the analysis of Sec.~\ref{sec:hamiltonian-mixed-sector} is correct. We will consider the Poisson bracket of the induced metric with the smeared Hamiltonian constraint $C_0^{(2)}$ given by \eqref{eq:hamiltonian-constraint-mixed-sector}:
\begin{equation}
\left\{q_{ij}(x),\int_{\Sigma_t}\mathrm{d}^3y\,C_0^{(2)}(y)N(y)\right\}=\frac{4N\kappa}{\sqrt{q}}\left(\pi_{ij}-\frac{\pi}{2}q_{ij}\right)+N(\tilde{\pi} q_{ij}-2\tilde{\pi}_{ij})=2NK_{ij}\,.
\end{equation}
For the momentum constraint we can directly reproduce \eqref{eq:spatial-diffeomorphisms-gr} with $C_0$ replaced by $C_0^{(2)}$. Clearly, these outcomes are expected when employing the redefined momentum density $P^{ij}$ of \eqref{eq:momentum-constraint-mixed-sector}.

\subsubsection{Purely timelike sector}
Let us now evaluate the Poisson bracket of the smeared Hamiltonian constraint $C_0^{(3)}$ of \eqref{eq:hamiltonian-constraint-C03} with the induced metric according to Eq.~(\ref{eq:spacetime-diffeomorphism-gr}):
\begin{equation}
\left\{q_{ij}(x),\int_{\Sigma_t}\mathrm{d}^3y\,C_0^{(3)}(y)N(y)\right\}=\frac{2N}{1-s^{\mathbf{nn}}}\left[\frac{2\kappa}{\sqrt{q}}\left(\pi_{ij}-\frac{\pi}{2}q_{ij}\right)+\frac{\Xi}{4}q_{ij}\right]=2NK_{ij}\,,
\end{equation}
which corresponds to the expected result when the exterior curvature of Eq.~(\ref{eq:extrinsic-curvature-timelike-case}) is taken into account. As the momentum constraint remains unmodified when expressed in terms of the canonical momentum density, \eqref{eq:spatial-diffeomorphisms-gr} remains valid when $C_0$ is substituted by $C_0^{(3)}$. Furthermore, the first of Hamilton's field equations given by~\eqref{eq:hamilton-field-equation-1} still applies.

\subsection{Minimal gravitational SME: $u$ term}
Repeating the procedure employed for the three sectors of $s^{\mu\nu}$ for \eqref{eq:hamiltonian-constraint-C04} implies
\begin{equation}
\left\{q_{ij}(x),\int_{\Sigma_t}\mathrm{d}^3y\,C_0^{(4)}(y)N(y)\right\}=\frac{2N}{1-u}\left[\frac{2\kappa}{\sqrt{q}}\left(\pi_{ij}-\frac{\pi}{2}q_{ij}\right)+\frac{\Upsilon}{2}q_{ij}\right]=2NK_{ij}\,,
\end{equation}
as expected, when \eqref{eq:extrinsic-curvature-scalar-sector} is employed. The momentum constraint remains again unmodified, as reported in \eqref{eq:momentum-constraint-Ci4}. Under these conditions, both \eqref{eq:spatial-diffeomorphisms-gr} and \eqref{eq:hamilton-field-equation-1} are not in conflict with the current setting.

\subsection{Final remarks}

Thus, we conclude that the first set of Hamilton's field equations~(\ref{eq:hamilton-field-equation-1}) remains unmodified in the presence of the diffeomorphism-violating contributions that we have been focusing on in this article. This finding is reasonable, as we do not modify the geometrical setting. The second set of Hamilton's field equations is linked to the modified Einstein equations~(\ref{eq:einstein-equations-modified}), (\ref{eq:einstein-equations-modified-u}) and is expected to be modified. We will delve into this problem in a future work.

\section{Conclusions and outlook}
\label{sec:conclusions}
In the current work we employed the ADM formalism \cite{Arnowitt:1962hi,Arnowitt:2008} to investigate a modified theory of gravity characterized by the observer Lorentz tensor $s^{\mu\nu}$ as well the observer scalar $u$ of the minimal gravitational SME \cite{Kostelecky:2003fs}. We worked in the setting of explicit diffeomorphism violation, i.e., the background fields $s^{\mu\nu}$ and $u$ did not arise dynamically, but they were put into the action by hand. The main objective was to understand within the ADM formalism what kind of restrictions Riemannian geometry poses on $s^{\mu\nu}$ and $u$, in other words, how the no-go result in the context of the gravitational SME~\cite{Kostelecky:2003fs} becomes manifest in this formalism.

To do so, we decomposed the Lorentz-violating background field $s^{\mu\nu}$ into three sectors. The first was formed from the subset of six independent purely spacelike coefficients $s^{ij}$. The second contained the vector-valued object $s^{i\mathbf{n}}$ endowed with a single spacelike index only and the third involved the single Lorentz scalar $s^{\mathbf{nn}}$ without spacelike indices. We obtained the Hamiltonians associated with each of the three sectors of $s^{\mu\nu}$ as well as for $u$ and were able to identify modified Hamiltonian and momentum constraints. To derive the Hamiltonians, it turned out to be crucial to include modified GHY boundary terms in the action that allowed us to move time derivatives from the exterior-curvature tensor to the diffeomorphism-violating background field in an unambiguous manner.

One interesting outcome is that the mixed sector is governed by mere gauge degrees of freedom, i.e., the coefficients $s^{i\mathbf{n}}$ are unphysical, as they can be absorbed into a redefinition of the shift vector. This observation may even have implications for phenomenology. If $s^{i\mathbf{n}}$ is comprised, indeed, by gauge degrees of freedom, it is meaningless to constrain these coefficients by experiment. Instead, they should be disregarded in any phenomenological study of explicit diffeomorphism violation in gravity. Note that this finding can most probably not be taken over to spontaneous diffeomorphism violation, as then $s^{i\mathbf{n}}$ would be dynamical and it does not make sense to say that they are absorbed into the nondynamical shift vector. Another remarkable property is that the scalar background $u$ in the context of explicit diffeomorphism violation cannot be removed by a redefinition of the gravitational field. Hence, the latter is physical, in fact, and could be searched for in experiments. Therefore, we conclude that explicit and spontaneous diffeomorphism violation can be distinguished from each other in experimental searches.

We also tried to connect the Hamiltonian and momentum constraints to suitable functional derivatives of the ADM actions as well as to projections of the modified Einstein equations along directions orthogonal and parallel to the spacelike hypersurfaces. For the spacelike, timelike, and scalar sectors we found a match under a set of consistency requirements given by \eqref{eq:consistency-conditions} in the text: $\mathcal{L}_ms^{ij}=0$, $\mathcal{L}_ms^{\mathbf{nn}}=0$, and $\mathcal{L}_mu=0$. The latter are considered as the central results of this work. These conditions are interpreted as consequences of the no-go result \cite{Kostelecky:2003fs} applied to the pure-gravity sector.

However, \eqref{eq:consistency-conditions} also tells us that the no-go result is likely to be avoided as long as background fields are considered that satisfy these conditions. For example, in Gaussian normal coordinates, the first relation requires that the background field $s^{ij}$ do not exhibit an explicit dependence on the time coordinate. An arbitrary dependence on the spatial coordinates does not seem to be in conflict with Riemannian geometry, though. In summary, the benefits of using the ADM formalism in a setting of diffeomorphism violation in gravity are apparent. One has additional control over diffeomorphism violation and understands better whether or not there are unphysical sets of coefficients.

The analysis performed in the current article may pave new pathways of exciting research in the context of explicit diffeomorphism violation and/or local Lorentz violation in gravity. First, we intend to better understand the connection between the consistency conditions and the no-go result. To do so, it will be necessary to include matter fields into the pure-gravity sector and to treat them within the ADM formalism. Second, performing an analogous study for the coefficients $t^{\mu\nu\varrho\sigma}$ \cite{Kostelecky:2003fs} of the minimal gravitational SME will be reasonable. Such an investigation could provide further insights into the problem known as the ``t puzzle'' \cite{Bonder:2015maa,Bonder:2021gjo}. Third, our expectation is that the ADM formalism will also be valuable in the context of the nonminimal gravitational SME \cite{Kostelecky:2020hbb}. An intriguing question is whether nonminimal diffeomorphism violation in gravity implies additional consistency requirements linked to the higher derivatives in the action. Last but not least, the obtained results are highly promising to find applications in phenomenological studies of explicit diffeomorphism and local Lorentz violation in the context of cosmology or scenarios of strong gravitational fields such as black holes.

\medskip
\acknowledgments
The authors thank V.A.~Kosteleck\'{y} for clarifications on the modified Einstein equations
of~\cite{Kostelecky:2003fs,Bailey:2006fd}. In addition, we thank A.Y. Petrov for useful comments
on this work.
M.S. is indebted to FAPEMA Universal 01149/17; FAPEMA Universal 00830/19; CNPq Universal
421566/2016-7; CNPq Produtividade 312201/2018-4. Furthermore, M.S. is grateful to CAPES/Finance
Code 001. C.M.R. acknowledges partial support from project Fondecyt Regular 1191553, Chile.

\appendix
\section{Mathematical appendix}
\label{sec:mathematical-appendix}
Here we intend provide a compilation of the essential geometrical formulas that our work rests on.
The books and papers \cite{Poisson:2002,Poisson:2004,Gourgoulhon:2007ue,Gourgoulhon:2012} serve as
primary references for these results.

\subsection{Exterior curvature}
The covariant derivative in the spacetime manifold $\mathcal M$
is denoted as $\nabla_{\mu}$, whereas the covariant derivative
on a spacelike hypersurface $\Sigma_t$ is called $D_{\mu}$. The four-acceleration
$a_{\mu}$ associated with a foliation of a spacelike hypersurface is a measure for
how $n_{\mu}$ changes covariantly long itself \cite{Gourgoulhon:2007ue,Gourgoulhon:2012}. It holds that
\begin{equation}
\label{eq:acceleration}
a_{\mu}\equiv n^{\nu}\nabla_{\nu}n_{\mu}=\frac{D_{\mu}N}{N}=D_{\mu}\ln N\,,
\end{equation}
i.e., it can be expressed via the covariant derivative $D_{\mu}$ linked to the induced metric of
\eqref{eq:induced-metric} and the lapse function
$N$. This acceleration is tangent to the hypersurface:
\begin{equation}
\label{eq:orthogonality-acceleration}
n\cdot a=0\,.
\end{equation}
Note that $a_{\mu}=0$ in Gaussian normal coordinates. Next, we define the extrinsic-curvature tensor as
\begin{equation}
\label{eq:extrinsic-curvature-projections}
K_{\mu\nu}\equiv q^{\varrho}_{\phantom{\varrho}\mu}
q^{\sigma}_{\phantom{\sigma}\nu}\nabla_{\varrho}n_{\sigma}\,.
\end{equation}
Due to the property
\begin{equation}
\label{eq:extrinsic-curvature-in-hypersurface}
n^{\mu}K_{\mu\nu}=K_{\mu\nu}n^{\nu}=0\,,
\end{equation}
the extrinsic curvature lives completely in $\Sigma$. The extrinsic curvature is symmetric and can be expressed
via the Lie derivative of $q_{\mu\nu}$ with respect to $m^{\mu}\equiv Nn^{\nu}$~\cite{Thiemann:2007zz}:
\begin{equation}
K_{\mu\nu}=q^{\varrho}_{\phantom{\varrho}\mu}q^{\sigma}_{\phantom{\sigma}\nu}
\nabla_{(\varrho}n_{\sigma)}=\frac{1}{2N}\mathcal{L}_mq_{\mu\nu}\,.
\end{equation}
This quantity is a measure for the curvature of a hypersurface $\Sigma_t$ due
to its embedding in $\mathcal{M}$. It is in stark contrast to the intrinsic curvature of a
manifold that is given by the Riemann curvature tensor and that does not require
an embedding into an ambient (higher-dimensional) manifold.
By considering the spacelike part of \eqref{eq:extrinsic-curvature-projections}, we have
\begin{equation}
\label{eq:extrinsic-curvature-spatial}
K_{ij}=D_in_j=\partial_in_j-\Gamma^{\lambda}_{\phantom{\lambda}ij}n_{\lambda}\,.
\end{equation}
As the first term vanishes in the latter, we obtain Eq.~(\ref{eq:extrinsic-curvature}),
which is a highly valuable result in the Hamiltonian description of GR.

Both the acceleration of \eqref{eq:acceleration} and the extrinsic curvature of
\eqref{eq:extrinsic-curvature} play a pivotal role in projecting the intrinsic curvature of
$\mathcal{M}$ into the hypersurface $\Sigma_t$. Tensors defined in the ambient manifold $\mathcal{M}$
will be denoted by a label `(4)'. For brevity, the analogous label `(3)' for quantities defined
in $\Sigma_t$ will be dropped. We take the commonly used viewpoint that tensors
defined on a spacelike hypersurface $\Sigma_t$ can be extended into the spacetime
$\mathcal{M}$ via suitable pull-back and push-forward operations \cite{Gourgoulhon:2007ue,Gourgoulhon:2012}. By
doing so, the extrinsic curvature of \eqref{eq:extrinsic-curvature} is extended into $\mathcal{M}$ via~\cite{Gourgoulhon:2007ue,Gourgoulhon:2012}
\begin{equation}
\label{eq:extrinsic-curvature-M}
K_{\varrho\sigma}\equiv \nabla_{\sigma}n_{\varrho}+a_{\varrho}n_{\sigma}\,,
\end{equation}
where, for brevity, we omit the index `(4)'. Whenever the extrinsic curvature occurs with spatial
indices, \eqref{eq:extrinsic-curvature-spatial} is understood to be employed.
Note that we take over the sign convention of \cite{Thiemann:2007zz}, but different sign conventions are
also common (see, e.g., \cite{Misner:1973,Gourgoulhon:2007ue,Gourgoulhon:2012,ONeal-Ault:2020ebv}). 
In addition, \eqref{eq:extrinsic-curvature-in-hypersurface} also holds for \eqref{eq:extrinsic-curvature-M},
which follows from
\begin{equation}
0=\nabla_{\nu}(n_{\mu}n^{\mu})=2n^{\mu}\nabla_{\nu}n_{\mu}=2n^{\mu}K_{\mu\nu}\,.
\end{equation}
This property emphasizes again that the extrinsic curvature
lives in the spatial hypersurface $\Sigma_t$ entirely.
The trace $K$ of the extrinsic curvature is defined via the contraction of \eqref{eq:extrinsic-curvature-M} with the spacetime metric:
\begin{equation}
\label{eq:extrinsic-curvature-trace}
K\equiv g^{\varrho\sigma}K_{\varrho\sigma}=\nabla_{\mu}n^{\mu}\,.
\end{equation}
Alternatively, \eqref{eq:extrinsic-curvature-spatial} can be contracted with $q_{ij}$.

\subsection{Decomposition formula for curvature tensors}
As a starting point, we quote the Gauss relation (sometimes also called
the Gauss-Codazzi equation) that gives the projection of the Riemann
curvature tensor into $\Sigma_t$:
\begin{equation}
\label{eq:gauss-relation}
q^{\mu}_{\phantom{\mu}\alpha}q^{\nu}_{\phantom{\nu}\beta}q^{\varrho}_{\phantom{\varrho}\gamma}
q^{\sigma}_{\phantom{\sigma}\delta}{}^{(4)}R_{\mu\nu\varrho\sigma}=R_{\alpha\beta\gamma\delta}
+K_{\alpha\gamma}K_{\beta\delta}-K_{\beta\gamma}K_{\alpha\delta}\,.
\end{equation}
Thus, a projection of the Riemann curvature tensor in $\mathcal M$ into the spacelike hypersurface induces
 the curvature tensor in this hypersurface plus correction terms that involve the extrinsic curvature.
 This property is not a surprise, as the hypersurface $\Sigma_t$ is embedded into $\mathcal  M$.

The Gauss relation of \eqref{eq:gauss-relation} can be contracted once to provide a valuable
 relation for the projected Ricci curvature into the hypersurface:
\begin{equation}
\label{eq:gauss-relation-contracted}
q^{\mu}_{\phantom{\mu}\alpha}q^{\nu}_{\phantom{\nu}\beta}{}^{(4)}R_{\mu\nu}+q_{\alpha\mu}
n^{\nu}q^{\varrho}_{\phantom{\varrho}\beta}n^{\sigma}{}^{(4)}R^{\mu}_{\phantom{\mu}\nu\varrho\sigma}
=R_{\alpha\beta}+KK_{\alpha\beta}-K_{\alpha\mu}K^{\mu}_{\phantom{\mu}\beta}\,,
\end{equation}
with the trace $K$ of the extrinsic curvature stated in \eqref{eq:extrinsic-curvature-trace}.
In this relation, the Riemann curvature tensor cannot be eliminated, but it must be kept. Another
contraction results in a scalar Gauss relation that has the form
\begin{equation}
\label{eq:gauss-relation-scalar}
{}^{(4)}R+2{}^{(4)}R_{\mu\nu}n^{\mu}n^{\nu}=R+K^2-K_{\mu\nu}K^{\mu\nu}\,.
\end{equation}
We now compute the Ricci tensor from \eqref{eq:gauss-relation}:
\begin{align}
\label{eq:decomposition-ricci-tensor-preliminary}
R_{\beta\delta}&=R^{\alpha}_{\phantom{\alpha}\beta\alpha\delta}=R_{\alpha\beta\gamma\delta}q^{\alpha\gamma}=q^{\alpha\gamma}q^{\mu}_{\phantom{\mu}\alpha}q^{\nu}_{\phantom{\nu}\beta}q^{\varrho}_{\phantom{\varrho}\gamma}q^{\sigma}_{\phantom{\sigma}\delta}{}^{(4)}R_{\mu\nu\varrho\sigma}+K_{\gamma\beta}K_{\alpha\delta}q^{\alpha\gamma}-K_{\gamma\alpha}K_{\beta\delta}q^{\alpha\gamma} \notag \displaybreak[0]\\
&=q^{\alpha\gamma}q^{\mu}_{\phantom{\mu}\alpha}q^{\nu}_{\phantom{\nu}\beta}q^{\varrho}_{\phantom{\varrho}\gamma}q^{\sigma}_{\phantom{\sigma}\delta}{}^{(4)}R_{\mu\nu\varrho\sigma}+K^{\alpha}_{\phantom{\alpha}\beta}K_{\alpha\delta}-KK_{\beta\delta}\,.
\end{align}
At this point we would like to eliminate the first contribution. Consider
\begin{align}
\label{eq:useful-intermediate-result}
q^{\nu}_{\phantom{\nu}\beta}q^{\sigma}_{\phantom{\sigma}\delta}{}^{(4)}R_{\nu\sigma}&=q^{\nu}_{\phantom{\nu}\beta}q^{\sigma}_{\phantom{\sigma}\delta}{}^{(4)}R_{\alpha\nu\gamma\sigma}g^{\alpha\gamma}=\delta^{\mu}_{\phantom{\mu}\alpha}q^{\nu}_{\phantom{\nu}\beta}\delta^{\varrho}_{\phantom{\varrho}\gamma}q^{\sigma}_{\phantom{\sigma}\delta}{}^{(4)}R_{\mu\nu\varrho\sigma}g^{\alpha\gamma} \notag \displaybreak[0]\\
&=(q^{\mu}_{\phantom{\mu}\alpha}-n^{\mu}n_{\alpha})q^{\nu}_{\phantom{\nu}\beta}(q^{\varrho}_{\phantom{\varrho}\gamma}-n^{\varrho}n_{\gamma})q^{\sigma}_{\phantom{\sigma}\delta}{}^{(4)}R_{\mu\nu\varrho\sigma}(q^{\alpha\gamma}-n^{\alpha}n^{\gamma})
 \notag \displaybreak[0]\\
&=(q^{\mu}_{\phantom{\mu}\alpha}q^{\nu}_{\phantom{\nu}\beta}q^{\varrho}_{\phantom{\varrho}\gamma}q^{\sigma}_{\phantom{\sigma}\delta}-q^{\mu}_{\phantom{\mu}\alpha}q^{\nu}_{\phantom{\nu}\beta}n^{\varrho}n_{\gamma}q^{\sigma}_{\phantom{\sigma}\delta} \notag \\
&\phantom{{}={}}-n^{\mu}n_{\alpha}q^{\nu}_{\phantom{\nu}\beta}q^{\varrho}_{\phantom{\varrho}\gamma}q^{\sigma}_{\phantom{\sigma}\delta}+n^{\mu}n_{\alpha}q^{\nu}_{\phantom{\nu}\beta}n^{\varrho}n_{\gamma}q^{\sigma}_{\phantom{\sigma}\delta}){}^{(4)}R_{\mu\nu\varrho\sigma}(q^{\alpha\gamma}-n^{\alpha}n^{\gamma})
 \notag \displaybreak[0]\\
&=q^{\alpha\gamma}q^{\mu}_{\phantom{\mu}\alpha}q^{\nu}_{\phantom{\nu}\beta}q^{\varrho}_{\phantom{\varrho}\gamma}q^{\sigma}_{\phantom{\sigma}\delta}{}^{(4)}R_{\mu\nu\varrho\sigma}-n^{\mu}q^{\nu}_{\phantom{\nu}\beta}n^{\varrho}q^{\sigma}_{\phantom{\sigma}\delta}{}^{(4)}R_{\mu\nu\varrho\sigma} \notag \displaybreak[0]\\
&=q^{\alpha\gamma}q^{\mu}_{\phantom{\mu}\alpha}q^{\nu}_{\phantom{\nu}\beta}q^{\varrho}_{\phantom{\varrho}\gamma}q^{\sigma}_{\phantom{\sigma}\delta}{}^{(4)}R_{\mu\nu\varrho\sigma}+q^{\nu}_{\phantom{\nu}\beta}q^{\sigma}_{\phantom{\sigma}\delta}n^{\mu}{}^{(4)}R_{\mu\nu\sigma\varrho}n^{\varrho} \notag \displaybreak[0]\\
&=q^{\alpha\gamma}q^{\mu}_{\phantom{\mu}\alpha}q^{\nu}_{\phantom{\nu}\beta}q^{\varrho}_{\phantom{\varrho}\gamma}q^{\sigma}_{\phantom{\sigma}\delta}{}^{(4)}R_{\mu\nu\varrho\sigma}+q^{\nu}_{\phantom{\nu}\beta}q^{\sigma}_{\phantom{\nu}\delta}n^{\mu}[\nabla_{\mu},\nabla_{\nu}]n_{\sigma}\,. \end{align}
Now we use that
\begin{align}
n^{\mu}[\nabla_{\mu},\nabla_{\nu}]n_{\sigma}&=(\nabla_{\nu}n^{\mu})(\nabla_{\mu}n_{\sigma})-(\nabla_{\mu}n^{\mu})(\nabla_{\nu}n_{\sigma})+\nabla_{\mu}(n^{\mu}\nabla_{\nu}n_{\sigma})-\nabla_{\nu}(n^{\mu}\nabla_{\mu}n_{\sigma}) \notag \displaybreak[0]\\
&=g^{\mu\varrho}(K_{\varrho\nu}-a_{\varrho}n_{\nu})(K_{\sigma\mu}-a_{\sigma}n_{\mu})-K(K_{\sigma\nu}-a_{\sigma}n_{\nu})-\nabla_{\nu}a_{\sigma} \notag \\
&\phantom{{}={}}+\nabla_{\mu}(n^{\mu}K_{\sigma\nu}-n^{\mu}a_{\sigma}n_{\nu}) \notag \displaybreak[0]\\
&=q^{\mu\varrho}K_{\varrho\nu}K_{\sigma\mu}-KK_{\sigma\nu}+\nabla_{\mu}(n^{\mu}K_{\sigma\nu})-\nabla_{\nu}a_{\sigma}-a_{\nu}a_{\sigma} \notag \\
&\phantom{{}={}}-n_{\nu}(K_{\sigma\mu}a^{\mu}+n^{\mu}\nabla_{\mu}a_{\sigma})\,,
\end{align}
which follows from Eqs.~(\ref{eq:orthogonality-acceleration}), (\ref{eq:extrinsic-curvature-in-hypersurface}), and (\ref{eq:extrinsic-curvature-M}). Furthermore,
\begin{align}
q^{\nu}_{\phantom{\nu}\beta}q^{\sigma}_{\phantom{\sigma}\delta}n^{\mu}[\nabla_{\mu},\nabla_{\nu}]n_{\sigma}&=K^{\mu}_{\phantom{\mu}\beta}K_{\mu\delta}-KK_{\beta\delta}+q^{\nu}_{\phantom{\nu}\beta}q^{\sigma}_{\phantom{\sigma}\delta}[\nabla_{\mu}(n^{\mu}K_{\sigma\nu})-\nabla_{\nu}a_{\sigma}-a_{\nu}a_{\sigma}] \notag \displaybreak[0]\\
&=K^{\mu}_{\phantom{\mu}\beta}K_{\mu\delta}+q^{\nu}_{\phantom{\nu}\beta}q^{\sigma}_{\phantom{\sigma}\delta}[n^{\mu}\nabla_{\mu}K_{\sigma\nu}-\nabla_{\nu}a_{\sigma}-a_{\nu}a_{\sigma}] \notag \displaybreak[0]\\
&=\frac{1}{N}\mathcal{L}_mK_{\beta\delta}-K^{\mu}_{\phantom{\mu}\beta}K_{\mu\delta}-D_{\beta}a_{\delta}-a_{\beta}a_{\delta} \notag \displaybreak[0]\\
&=\frac{1}{N}\mathcal{L}_mK_{\beta\delta}-K^{\mu}_{\phantom{\mu}\beta}K_{\mu\delta}-\frac{1}{N}D_{\beta}D_{\delta}N\,,
\end{align}
where we employed
\begin{align}
\label{eq:relation-derivative-acceleration-lapse}
D_{\beta}a_{\sigma}+a_{\beta}a_{\sigma}&=D_{\beta}D_{\sigma}\ln N+D_{\beta}\ln ND_{\sigma}\ln N=D_{\beta}\left(\frac{1}{N}D_{\sigma}N\right)+\frac{1}{N^2}D_{\beta}ND_{\sigma}N \notag \\
&=-\frac{1}{N^2}D_{\beta}ND_{\sigma}N+\frac{1}{N}D_{\beta}D_{\sigma}N+\frac{1}{N^2}D_{\beta}ND_{\sigma}N=\frac{1}{N}D_{\beta}D_{\sigma}N\,,
\end{align}
as well as Eq.~(3.42) of \cite{Gourgoulhon:2007ue,Gourgoulhon:2012}.
Then,
\begin{align}
R_{\beta\delta}&=K^{\mu}_{\phantom{\mu}\beta}K_{\mu\delta}-KK_{\beta\delta}+q^{\nu}_{\phantom{\nu}\beta}q^{\sigma}_{\phantom{\sigma}\delta}{}^{(4)}R_{\nu\sigma}-q^{\nu}_{\phantom{\nu}\beta}q^{\sigma}_{\phantom{\sigma}\delta}n^{\mu}[\nabla_{\mu},\nabla_{\nu}]n_{\sigma} \notag \\
&=K^{\mu}_{\phantom{\mu}\beta}K_{\mu\delta}-KK_{\beta\delta}+q^{\nu}_{\phantom{\nu}\beta}q^{\sigma}_{\phantom{\sigma}\delta}{}^{(4)}R_{\nu\sigma}-\left(\frac{1}{N}\mathcal{L}_mK_{\beta\delta}-K^{\mu}_{\phantom{\mu}\beta}K_{\mu\delta}-\frac{1}{N}D_{\beta}D_{\delta}N\right) \notag \\
&=-\frac{1}{N}\mathcal{L}_mK_{\beta\delta}+2K^{\mu}_{\phantom{\mu}\beta}K_{\mu\delta}-KK_{\beta\delta}+\frac{1}{N}D_{\beta}D_{\delta}N+q^{\nu}_{\phantom{\nu}\beta}q^{\sigma}_{\phantom{\sigma}\delta}{}^{(4)}R_{\nu\sigma}\,,
\end{align}
which is why
\begin{equation}
\label{eq:projection-ricci-tensor-into-hypersurface}
q^{\nu}_{\phantom{\nu}\beta}q^{\sigma}_{\phantom{\sigma}\delta}{}^{(4)}R_{\nu\sigma}=\frac{1}{N}\mathcal{L}_mK_{\beta\delta}-\frac{1}{N}D_{\beta}D_{\delta}N+R_{\beta\delta}+KK_{\beta\delta}-2K^{\mu}_{\phantom{\mu}\beta}K_{\mu\delta}\,.
\end{equation}
The contracted Gauss relation of \eqref{eq:gauss-relation-contracted} implies
\begin{align}
q^{\mu}_{\phantom{\mu}\beta}n^{\nu}q^{\varrho}_{\phantom{\varrho}\delta}n^{\sigma}{}^{(4)}R_{\mu\nu\varrho\sigma}&=R_{\beta\delta}+KK_{\beta\delta}-K^{\mu}_{\phantom{\mu}\beta}K_{\mu\delta}-q^{\mu}_{\phantom{\mu}\beta}q^{\nu}_{\phantom{\nu}\delta}{}^{(4)}R_{\mu\nu} \notag \\
&=-\frac{1}{N}\mathcal{L}_mK_{\beta\delta}+\frac{1}{N}D_{\beta}D_{\delta}N+K^{\mu}_{\phantom{\mu}\beta}K_{\mu\delta}\,.
\end{align}
A suitable contraction of the left-hand side of the latter equation implies
\begin{align}
q^{\mu\beta}n^{\nu}q^{\varrho}_{\phantom{\varrho}\beta}n^{\sigma}{}^{(4)}R_{\mu\nu\varrho\sigma}&=q^{\mu\varrho}n^{\nu}n^{\sigma}{}^{(4)}R_{\mu\nu\varrho\sigma}=(g^{\mu\varrho}+n^{\mu}n^{\varrho})n^{\nu}n^{\sigma}{}^{(4)}R_{\mu\nu\varrho\sigma} \notag \\
&=n^{\nu}n^{\sigma}{}^{(4)}R_{\nu\sigma}\,,
\end{align}
which together with
\begin{equation}
q^{\beta\delta}\mathcal{L}_mK_{\beta\delta}=\mathcal{L}_mK+2NK^{\mu\nu}K_{\mu\nu}\,,
\end{equation}
leads to
\begin{equation}
\label{eq:ricci-totally-orthogonal-projection}
n^{\nu}n^{\sigma}{}^{(4)}R_{\nu\sigma}=-\frac{1}{N}\mathcal{L}_mK+\frac{1}{N}D_{\beta}D^{\beta}N-K^{\mu\nu}K_{\mu\nu}\,.
\end{equation}
The Ricci scalar can be decomposed by applying Eqs.~(\ref{eq:gauss-relation-scalar}), (\ref{eq:ricci-totally-orthogonal-projection}):
\begin{align}
\label{eq:decomposition-ricci-scalar}
{}^{(4)}R&=R+K^2-K_{ij}K^{ij}-2{}^{(4)}R_{\mu\nu}n^{\mu}n^{\nu} \notag \\
&=R+K^2-K_{ij}K^{ij}-2\left(-\frac{1}{N}\mathcal{L}_mK+\frac{1}{N}D_iD^iN-K^{ij}K_{ij}\right) \notag \\
&=\frac{2}{N}\mathcal{L}_mK-\frac{2}{N}D_iD^iN+R+K^2+K_{ij}K^{ij}\,.
\end{align}
Now, by using
\begin{subequations}
\begin{align}
\frac{1}{N}\mathcal{L}_mK&=n^{\mu}\nabla_{\mu}K=\nabla_{\mu}(n^{\mu}K)-K^2\,, \\[2ex]
D_iD^iN&=D_i(Na^i)=N(a_ia^i+D_ia^i)=N\nabla_{\mu}a^{\mu}\,,
\end{align}
\end{subequations}
it is clear that Eq.~(\ref{eq:decomposition-ricci-scalar}) can be brought into the form occurring in the ADM-decomposed EH action of Eq.~(\ref{eq:einstein-hilbert-action-ADM-decomposed}):
\begin{align}
\label{eq:decomposition-ricci-scalar-covariant-derivative}
{}^{(4)}R&=2\nabla_{\mu}(n^{\mu}K)-\frac{2}{N}D_iD^iN+R-K^2+K_{ij}K^{ij} \notag \\
&=R-K^2+K_{ab}K^{ab}+2\nabla_{\mu}(n^{\mu}K-a^{\mu})\,.
\end{align}
Finally, it is possible to project the Riemann curvature tensor partially into the hypersurface $\Sigma_t$~\cite{Gourgoulhon:2007ue,Gourgoulhon:2012}. The result involves covariant derivatives of the extrinsic curvature defined within the hypersurface:
\begin{subequations}
\begin{equation}
q^{\gamma}_{\phantom{\gamma}\varrho}n^{\sigma}q^{\mu}_{\phantom{\mu}\alpha}q^{\nu}_{\phantom{\nu}\beta}{}^{(4)}R^{\varrho}_{\phantom{\varrho}\sigma\mu\nu}=D_{\alpha}K^{\gamma}_{\phantom{\gamma}\beta}-D_{\beta}K^{\gamma}_{\phantom{\gamma}\alpha}\,. \end{equation}
The latter bears the name Codazzi-Mainardi relation. Contracting it once implies
\begin{equation}
\label{eq:codazzi-mainardi-contracted}
q^{\mu}_{\phantom{\mu}\beta}n^{\nu}{}^{(4)}R_{\mu\nu}=D_{\mu}K^{\mu}_{\phantom{\mu}\beta}-D_{\beta}K\,.
\end{equation}
\end{subequations}
Note that all relations derived before are identities, as they stand. In principle, tensors (or parts of tensors) live in the spacelike hypersurface, if contractions of the corresponding Lorentz indices with $n^{\mu}$ give zero. These indices can, in principle, be interpreted as spatial ones. In particular, if $n^{\mu}\Psi_{\mu\nu\dots}=0$ of a spacetime tensor $\Psi_{\mu\nu\dots}$, we do not lose any information by considering $\Psi_{i\nu\dots}$. For example, this holds for the exterior curvature $K_{\mu\nu}$, the acceleration $a_{\mu}$, and the covariant derivative $D_{\mu}$ on the hypersurface.
\section{Hamiltonian formulation}
\label{sec:hamiltonian-formulation-computations}

The current section will provide computational details on how to perform Legendre transformations to obtain Hamiltonians starting from the Lagrange densities of the (modified) gravity theory under consideration.

\subsection{General Relativity}
\label{sec:hamiltonian-einstein-hilbert}
We are going to start with the Hamiltonian of GR. In this case, the canonical momentum follows from the ADM-decomposed Lagrange density given by the integrand of \eqref{eq:einstein-hilbert-action-ADM-decomposed} with the surface term is discarded:
\begin{align}
\label{eq:canonical-momentum-general-relativity}
\pi^{ij}&=\frac{\partial\mathcal{L}^{(0)}}{\partial\dot{q}_{ij}}=\frac{N\sqrt{q}}{2\kappa}\frac{\partial}{\partial\dot{q}_{ij}}(R-K^2+K_{ab}K^{ab}) \notag \\
&=\frac{N\sqrt{q}}{2\kappa}\left[2K^{ab}\frac{\partial K_{ab}}{\partial\dot{q}_{ij}}-2K\frac{\partial K}{\partial\dot{q}_{ij}}\right]=\frac{\sqrt{q}}{2\kappa}(K^{ij}-q^{ij}K)\,,
\end{align}
where we used that
\begin{subequations}
\label{eq:derivatives-extrinsic-curvature}
\begin{align}
K^{ab}\frac{\partial K_{ab}}{\partial\dot{q}_{ij}}&=K^{ab}\frac{\delta^i_{\phantom{i}a}\delta^j_{\phantom{j}b}}{2N}=\frac{1}{2N}K^{ij}\,, \displaybreak[0]\\[2ex]
\frac{\partial K}{\partial\dot{q}_{ij}}&=\frac{\partial K^a_{\phantom{a}a}}{\partial\dot{q}_{ij}}=\frac{q^{ai}\delta_a^{\phantom{a}j}}{2N}=\frac{1}{2N}q^{ij}\,.
\end{align}
The trace of the canonical momentum is given by
\begin{equation}
\label{eq:trace-canonical-momentum-general-relativity}
\pi=\pi^a_{\phantom{a}a}=\frac{\sqrt{q}}{2\kappa}(K^a_{\phantom{a}a}-\delta^a_{\phantom{a}a}K)=\frac{\sqrt{q}}{2\kappa}(K-3K)=-\frac{\sqrt{q}}{\kappa}K\,.
\end{equation}
\end{subequations}
We employ \eqref{eq:canonical-momenta-q-dot} to obtain the Hamilton density of GR:
\begin{align}
\mathcal{H}^{(0)}&=\pi^{ab}\dot{q}_{ab}-\mathcal{L}^{(0)}= \notag \\
&=\frac{\sqrt{q}}{2\kappa}(K^{ab}-q^{ab}K)\left[2NK_{ab}+D_aN_b+D_bN_a\right]-\frac{N\sqrt{q}}{2\kappa}(R-K^2+K_{ab}K^{ab}) \notag \displaybreak[0]\\
&=-\frac{\sqrt{q}}{2\kappa}\left[N(R+K^2-K_{ab}K^{ab})+2(KD_aN^a-K^{ab}D_aN_b)\right] \notag \\
&\overset{\text{p.i.}}{=}-\frac{\sqrt{q}}{2\kappa}\left[N(R+K^2-K_{ab}K^{ab})+2(D_bK^b_{\phantom{b}a}-D_aK)N^a\right]\,.
\end{align}
In the last step we performed a partial integration of the three-dimensional covariant derivative $D_i$ with respect to the measure $\mathrm{d}^3x\,\sqrt{q}$ that is used when the Hamilton density is integrated to obtain the Hamiltonian. Surface terms are safely discarded for these integrations. The resulting Hamiltonian can then be written as \eqref{eq:hamiltonian-einstein-hilbert}.

Finally, we would like to express this Hamiltonian in terms of the canonical momentum. We solve \eqref{eq:trace-canonical-momentum-general-relativity} for $K$ and insert this result into \eqref{eq:canonical-momentum-general-relativity}:
\begin{equation}
\label{eq:extrinsic-curvature-canonical-momentum-GR}
K^{ij}=\frac{2\kappa}{\sqrt{q}}\pi^{ij}+q^{ij}K=\frac{2\kappa}{\sqrt{q}}\left(\pi^{ij}-\frac{\pi}{2}q^{ij}\right)\,.
\end{equation}
Then, the contraction of the extrinsic curvature with itself provides
\begin{equation}
K_{ij}K^{ij}=\frac{4\kappa^2}{q}\left(\pi^{ij}\pi_{ij}-\pi\pi^{ij}q_{ij}+\frac{\pi^2}{4}q^{ij}q_{ij}\right)=\frac{4\kappa^2}{q}\left(\pi^{ij}\pi_{ij}-\frac{1}{4}\pi^2\right)\,.
\end{equation}
By inserting these findings into Eqs.~(\ref{eq:parameter-c0-GR}), (\ref{eq:parameter-ci-GR}), we arrive at
\begin{subequations}
\begin{align}
C_0&=-\frac{\sqrt{q}}{2\kappa}\left[R+\frac{\kappa^2}{q}\pi^2-\frac{4\kappa^2}{q}\left(\pi^{ij}\pi_{ij}-\frac{1}{4}\pi^2\right)\right]=\frac{2\kappa}{\sqrt{q}}\left(\pi^{ij}\pi_{ij}-\frac{\pi^2}{2}\right)-\frac{\sqrt{q}}{2\kappa}R\,, \\[2ex]
C_a&=-\frac{\sqrt{q}}{\kappa}\frac{2\kappa}{\sqrt{q}}\left(D_b\pi^b_{\phantom{b}a}-\frac{1}{2}D_a\pi+\frac{1}{2}D_a\pi\right)=-2D_b\pi^b_{\phantom{b}a}\,.
\end{align}
\end{subequations}
The latter are employed to obtain Eqs.~(\ref{eq:hamiltonian-constraint}), (\ref{eq:momentum-constraint}).
\subsection{Standard-Model Extension}
\label{sec:hamiltonian-SME}

We will now turn to obtaining the Hamiltonians given by Eqs.~(\ref{eq:hamiltonian-part1}), (\ref{eq:hamiltonian-part2}), (\ref{eq:hamiltonian-part3}), and (\ref{eq:hamiltonian-part4}).

\subsubsection{Purely spacelike sector}
\label{sec:hamiltonian-SME-purely-spacelike-details}
We employ the Lie derivative stated in \eqref{eq:lie-derivative-s}. As the latter result only involves quantities and derivatives defined in the spatial hypersurface, it is clear that the Lie derivative $\mathcal{L}_ms^{ij}$ does not depend on $\dot{q}_{kl}$. Based on Eqs.~(\ref{eq:f1-reformulated}), (\ref{eq:f1-mod-total-derivative}), the canonical momentum of $\mathcal{L}^{(1)}$ is given by
\begin{subequations}
\begin{equation}
\pi^{(1)rs}=\frac{N\sqrt{q}}{2\kappa}\frac{\partial f^{(1)}}{\partial\dot{q}_{rs}}\,,
\end{equation}
with
\begin{align}
\frac{\partial f^{(1)}}{\partial\dot{q}_{rs}}&=-\frac{\partial}{\partial\dot{q}_{rs}}\left\{\frac{1}{N}[K_{ij}(\dot{s}^{ij}-\mathcal{L}_Ns^{ij})+s^{ij}D_iD_jN]+s^{ij}(2K_i^{\phantom{i}l}K_{lj}-R_{ij})\right\} \notag \\
&=-\frac{1}{N}\frac{\partial K_{ij}}{\partial\dot{q}_{rs}}(\dot{s}^{ij}-\mathcal{L}_Ns^{ij})-2s^{ij}\frac{K_i^{\phantom{i}l}K_{lj}}{\partial\dot{q}_{rs}}\,.
\end{align}
\end{subequations}
At this point we will benefit from the results of Eqs.~(\ref{eq:canonical-momenta-q-dot}), (\ref{eq:derivatives-extrinsic-curvature}) that are still valid in the presence of explicit diffeomorphism violation as long as Riemannian geometry is imposed. In a more general Finsler setting, these relations would probably be subject to modifications. Thus,
\begin{align}
\label{eq:canonicel-momentum-sector1}
\frac{\partial f^{(1)}}{\partial\dot{q}_{rs}}&=-\frac{\delta^r_{\phantom{r}i}\delta^s_{\phantom{s}j}}{2N^2}\mathcal{L}_ms^{ij}-2s^{ij}\left(\frac{\delta^r_{\phantom{r}i}q^{ls}}{2N}K_{lj}+K_i^{\phantom{i}l}\frac{\delta^r_{\phantom{r}l}\delta^s_{\phantom{s}j}}{2N}\right) \notag \\
&=-\frac{1}{N}\left(\frac{1}{2N}\mathcal{L}_ms^{rs}+s^{rj}K^s_{\phantom{s}j}+s^{is}K_i^{\phantom{i}r}\right)\,.
\end{align}
To obtain the Hamilton density $\mathcal{H}^{(1)}$ we perform a Legendre transformation. Note that the presence of the integral measure $\mathrm{d}^3x\,\sqrt{q}$ allows us to carry out partial integrations of the spatial covariant derivative $D_i$. By doing so, we obtain
\begin{align}
\mathcal{H}^{(1)}&=\pi^{(1)ij}\dot{q}_{ij}-\mathcal{L}^{(1)} \notag \displaybreak[0]\\
&=-\frac{\sqrt{q}}{2\kappa}\left[\frac{1}{2N}\mathcal{L}_ms^{ij}+s^{il}K^j_{\phantom{j}l}+s^{jl}K^i_{\phantom{i}l}\right](2NK_{ij}+D_iN_j+D_jN_i) \notag \displaybreak[0]\\
&\phantom{{}={}}+\frac{\sqrt{q}}{2\kappa}\Big[K_{ij}\mathcal{L}_ms^{ij}+s^{ij}(D_iD_jN-NR_{ij}+2NK^l_{\phantom{l}i}K_{lj})\Big] \notag \displaybreak[0]\\
&=\frac{\sqrt{q}}{2\kappa}\Big\{-\left[\frac{1}{2N}\mathcal{L}_ms^{ij}+s^{il}K^j_{\phantom{j}l}+s^{jl}K^i_{\phantom{i}l}\right](D_iN_j+D_jN_i) \notag \\
&\phantom{{}={}}+s^{ij}(D_iD_jN-NR_{ij}-2NK^l_{\phantom{l}i}K_{lj})\Big\} \notag \displaybreak[0]\\
&=\frac{\sqrt{q}}{2\kappa}\Big\{-\left[\frac{1}{N}\mathcal{L}_ms^{ij}+2(s^{il}K^j_{\phantom{j}l}+s^{jl}K^i_{\phantom{i}l})\right]D_iN_j \notag \\
&\phantom{{}={}}+s^{ij}(D_iD_jN-NR_{ij}-2NK^l_{\phantom{l}i}K_{lj})\Big\} \notag \displaybreak[0]\\
&\overset{\text{p.i.}}{=}\frac{\sqrt{q}}{2\kappa}\Big\{D_i\left[\frac{1}{N}\mathcal{L}_ms^{ij}+2(s^{il}K^j_{\phantom{j}l}+s^{jl}K^i_{\phantom{i}l})\right]N_j \notag \\
&\phantom{{}={}}+[D_iD_js^{ij}-s^{ij}(R_{ij}+2K^l_{\phantom{l}i}K_{lj})]N\Big\}\,.
\end{align}
The latter result implies \eqref{eq:hamiltonian-part1}.

\subsubsection{Mixed sector}
\label{sec:derivation-hamiltonian-mixed-sector}
We derive the canonical momentum density from $\mathcal{L}^{(2)}$ stated in \eqref{eq:eq:lagrange-density-LV-2-with-boundary-term}:
\begin{align}
\label{eq:canonical-momentum-mixed-sector}
\pi^{(2)rs}=\frac{\partial\mathcal{L}^{(2)}}{\partial\dot{q}_{rs}}&=\frac{\sqrt{q}}{\kappa}N\left[\frac{q^{ir}\delta_j^{\phantom{j}s}}{2N}D_is^{j\mathbf{n}}-\frac{q^{rs}}{2N}D_is^{i\mathbf{n}}+a_i\left(\frac{q^{ir}\delta_j^{\phantom{j}s}}{2N}s^{j\mathbf{n}}-\frac{q^{rs}}{2N}s^{i\mathbf{n}}\right)\right] \notag \displaybreak[0]\\
&=\frac{\sqrt{q}}{2\kappa}\left[(a^r+D^r)s^{s\mathbf{n}}-q^{rs}(a_i+D_i)s^{i\mathbf{n}}\right] \notag \displaybreak[0]\\
&\overset{\text{sym}}{=}\frac{\sqrt{q}}{4\kappa}\left[(a^r+D^r)s^{s\mathbf{n}}+(a^s+D^s)s^{r\mathbf{n}}-2q^{rs}(a_i+D_i)s^{i\mathbf{n}}\right]\,,
\end{align}
where the latter has been symmetrized in the last step. A Legendre transformation provides the corresponding Hamilton density:
\begin{align}
\mathcal{H}^{(2)}&=\pi^{(2)rs}\dot{q}_{rs}-\mathcal{L}^{(2)} \notag \\
&=\frac{\sqrt{q}}{4\kappa}\left[(a^r+D^r)s^{s\mathbf{n}}+(a^s+D^s)s^{r\mathbf{n}}-2q^{rs}(a_i+D_i)s^{i\mathbf{n}}\right](2NK_{rs}+D_rN_s+D_sN_r) \notag \displaybreak[0]\\
&\phantom{{}={}}-\frac{\sqrt{q}}{\kappa}N\left[K^i_{\phantom{i}j}(a_i+D_i)s^{j\mathbf{n}}-K(a_i+D_i)s^{i\mathbf{n}}\right] \notag \displaybreak[0]\\
&=\frac{\sqrt{q}}{2\kappa}\left[(a^r+D^r)s^{s\mathbf{n}}+(a^s+D^s)s^{r\mathbf{n}}-2q^{rs}(a_i+D_i)s^{i\mathbf{n}}\right]D_rN_s \notag \displaybreak[0]\\
&\overset{\text{p.i.}}{=} \frac{\sqrt{q}}{2\kappa}\left[2(D^sa_i+D^sD_i)s^{i\mathbf{n}}-(D_ra^r+D_rD^r)s^{s\mathbf{n}}-(D_ra^s+D_rD^s)s^{r\mathbf{n}}\right]N_s\,.
\end{align}
This finding leads to the result quoted in \eqref{eq:hamiltonian-part2}.

\subsubsection{Purely timelike sector}
\label{sec:derivation-hamiltonian-purely-timelike}
We compute the canonical momentum of $\mathcal{L}^{(3)}$ by using Eqs.~(\ref{eq:f3-reformulated}), (\ref{eq:f3-mod-total-derivative}):
\begin{subequations}
\label{eq:canonical-momentum-purely-timelike}
\begin{equation}
\pi^{(3)rs}=\frac{N\sqrt{q}}{2\kappa}\frac{\partial f^{(3)}}{\partial\dot{q}_{rs}}\,,
\end{equation}
where
\begin{align}
\frac{\partial f^{(3)}}{\partial\dot{q}_{rs}}&=\frac{\partial K}{\partial\dot{q}_{rs}}n^{\mu}\nabla_{\mu}s^{\mathbf{nn}}+s^{\mathbf{nn}}\frac{\partial}{\partial\dot{q}_{rs}}(K^2-K^{ij}K_{ij}) \notag \\
&=\frac{1}{N}\left[\frac{q^{rs}}{2N}\mathcal{L}_ms^{\mathbf{nn}}+s^{\mathbf{nn}}(q^{rs}K-K^{rs})\right]\,.
\end{align}
\end{subequations}
A suitable Legendre transformation results in
\begin{align}
\mathcal{H}^{(3)}&=\pi^{(3)ij}\dot{q}_{ij}-\mathcal{L}^{(3)} \notag \displaybreak[0]\\
&=\frac{\sqrt{q}}{2\kappa}\left[\frac{1}{2N}q^{ij}\mathcal{L}_ms^{\mathbf{nn}}+s^{\mathbf{nn}}(q^{ij}K-K^{ij})\right](2NK_{ij}+D_iN_j+D_jN_i) \notag \\
&\phantom{{}={}}-\frac{\sqrt{q}}{2\kappa}\Big[K\mathcal{L}_ms^{\mathbf{nn}}+s^{\mathbf{nn}}(D_iD^iN-NK^{ij}K_{ij}+NK^2)\Big] \notag \displaybreak[0]\\
&=\frac{\sqrt{q}}{2\kappa}\left[\left(\frac{1}{N}\mathcal{L}_ms^{\mathbf{nn}}+2s^{\mathbf{nn}}K\right)D_iN^i+2s^{\mathbf{nn}}(NK^2-NK^{ij}K_{ij}-K^{ij}D_iN_j)\right. \notag \\
&\phantom{{}={}}\hspace{0.8cm}\left.{}-s^{\mathbf{nn}}(D_iD^iN-NK^{ij}K_{ij}+NK^2)\right] \notag \displaybreak[0]\\
&\overset{\text{p.i.}}{=}\frac{\sqrt{q}}{2\kappa}\left\{-D_i\left(\frac{1}{N}\mathcal{L}_ms^{\mathbf{nn}}+2s^{\mathbf{nn}}K\right)N^i+2D_i(s^{\mathbf{nn}}K^{ij})N_j\right. \notag \\
&\phantom{{}={}}\hspace{0.8cm}\left.{}-N\left[D_iD^is^{\mathbf{nn}}+s^{\mathbf{nn}}(K^{ij}K_{ij}-K^2)\right]\right\}\,.
\end{align}
The Hamiltonian of \eqref{eq:hamiltonian-part3} is a direct implication of the latter result.

\subsubsection{Scalar sector}
\label{sec:derivation-hamiltonian-scalar}

Finally, the canonical momentum density associated with the Lagrange density of \eqref{eq:lagrange-density-u} reads
\begin{equation}
\label{eq:canonical-momentum-u}
\pi^{(4)ij}=\frac{\partial\mathcal{L}^{(4)}}{\partial\dot{q}_{ij}}=\frac{\sqrt{q}}{2\kappa}\left[\frac{q^{ij}}{N}\mathcal{L}_mu+(q^{ij}K-K^{ij})u\right]\,,
\end{equation}
whereupon we can compute the Hamilton density:
\begin{align}
\mathcal{H}^{(4)}&=\pi^{(4)kl}\dot{q}_{kl}-\mathcal{L}^{(u)} \notag \\
&=\frac{\sqrt{q}}{2\kappa}\left[\frac{1}{N}q^{kl}\mathcal{L}_mu+(q^{kl}K-K^{kl})u\right](2NK_{kl}+D_kN_l+D_lN_k) \notag \\
&\phantom{{}={}}-\frac{\sqrt{q}}{2\kappa}\left[2(K\mathcal{L}_mu+uD_iD^iN)-N(R-K^2+K_{ij}K^{ij})u\right] \notag \\
&=\frac{\sqrt{q}}{2\kappa}\left[(R+K^2-K_{ij}K^{ij})uN-2uD_iD^iN\right. \notag \\
&\phantom{{}={}}\left.{}+2\left(\frac{1}{N}\mathcal{L}_mu+uK\right)D_lN^l-2K^{kl}uD_kN_l\right]\,.
\end{align}
Carrying out suitable partial integrations implies \eqref{eq:hamiltonian-part4}.

\section{Modified ADM decomposition}
\label{sec:modified-ADM}

To understand the mixed sector based on \eqref{eq:lagrange-density-LV-2} better, we perform an ADM decomposition with an effective shift vector
\begin{equation}
\label{eq:effective-shift-vector}
\tilde{N}^i\equiv N^i-Ns^{i\mathbf{n}}\,,
\end{equation}
i.e., the diffeomorphism-violating degrees of freedom of this sector are put into the shift vector. The corresponding effective extrinsic-curvature tensor is now defined as
\begin{equation}
\tilde{K}_{ij}\equiv \frac{1}{2N}(\dot{q}_{ij}-D_i\tilde{N}_j-D_j\tilde{N}_i)\,.
\end{equation}
Besides, we define a Lagrange density that has a form analogous to that of the ADM-decomposed EH Lagrange density~(\ref{eq:einstein-hilbert-action-ADM-decomposed}):
\begin{equation}
\label{eq:lagrange-density-L-tilde}
\tilde{\mathcal{L}}^{(0)}\equiv \frac{N\sqrt{q}}{2\kappa}(R-\tilde{K}^2+\tilde{K}_{ij}\tilde{K}^{ij})\,,
\end{equation}
where the conventional shift vector is replaced by the effective one in \eqref{eq:effective-shift-vector}. Boundary terms are disregarded. We then evaluate
\begin{align}
\tilde{K}_{ij}\tilde{K}^{ij}&=\frac{1}{4N^2}\Big\{\dot{q}_{ij}\dot{q}^{ij}-2\dot{q}_{ij}[D^i(N^j-Ns^{j\mathbf{n}})+D^j(N^i-Ns^{i\mathbf{n}})] \notag \\
&\phantom{{}={}}+[D_i(N_j-Ns_j^{\phantom{j}\mathbf{n}})+D_j(N_i-Ns_i^{\phantom{i}\mathbf{n}})][D^i(N^j-Ns^{j\mathbf{n}})+D^j(N^i-Ns^{i\mathbf{n}})]\Big\} \notag \\
&\simeq\frac{1}{4N^2}\Big\{\dot{q}_{ij}\dot{q}^{ij}-2\dot{q}_{ij}[D^i(N^j-Ns^{j\mathbf{n}})+D^j(N^i-Ns^{i\mathbf{n}})] \notag \\
&\phantom{{}={}}-2(D_iN_j+D_jN_i)[D^i(Ns^{j\mathbf{n}})+D^j(Ns^{i\mathbf{n}})]\,,
\end{align}
where we have dropped terms beyond linear order in the controlling coefficients. Thereupon,
\begin{align}
\tilde{K}_{ij}\tilde{K}^{ij}&=K_{ij}K^{ij}+\frac{1}{2N^2}(\dot{q}_{ij}-D_iN_j-D_jN_i)[D^i(Ns^{j\mathbf{n}})+D^j(Ns^{i\mathbf{n}})] \notag \\
&=K_{ij}K^{ij}+\frac{1}{N}K_{ij}[D^i(Ns^{j\mathbf{n}})+D^j(Ns^{i\mathbf{n}})]=K_{ij}K^{ij}+\frac{2}{N}K_{ij}D^i(Ns^{j\mathbf{n}})\,.
\end{align}
In an analogous manner we obtain
\begin{equation}
\tilde{K}^2=K^2+\frac{2}{N}KD_i(Ns^{i\mathbf{n}})\,.
\end{equation}
Hence, it is possible to write
\begin{align}
N(\tilde{K}^2-\tilde{K}_{ij}\tilde{K}^{ij})&\simeq N(K^2-K_{ij}K^{ij})+2[KD_i(Ns^{i\mathbf{n}})-K_{ij}D^i(Ns^{j\mathbf{n}})] \notag \\
&\overset{\text{p.i.}}{=}N(K^2-K_{ij}K^{ij})-2Ns^{i\mathbf{n}}(D_iK-D_jK^j_{\phantom{j}i})]\,,
\end{align}
after suitable partial integrations where the surface terms are discarded again. Therefore, at first order in diffeomorphism violation, the following correspondence holds:
\begin{equation}
\tilde{\mathcal{L}}^{(0)}=\mathcal{L}^{(0)}-\mathcal{L}^{(2)}\,,
\end{equation}
with $\mathcal{L}^{(0)}$ of \eqref{eq:einstein-hilbert-action-ADM-decomposed} and the Lorentz-violating piece $\mathcal{L}^{(2)}$ of \eqref{eq:lagrange-density-LV-2}. Hence, we have shown that at first order in diffeomorphism violation, the coefficients of the mixed sector can be absorbed into a redefined shift vector. This demonstration is another argument for $s^{i\mathbf{n}}$ being gauge degrees of freedom (see the discussion in Sec.~\ref{sec:hamiltonian-mixed-sector}).

\section{Boundary terms in the action}
\label{sec:boundary-terms}

In this section we present detailed computations on how to obtain the (modified) GHY boundary terms that play a crucial role when moving time derivatives from the exterior curvature to the background fields in the Lagrange densities of Eqs.~(\ref{eq:lagrange-density-LV-1}), (\ref{eq:lagrange-density-LV-3}), and (\ref{eq:lagrange-density-LV-4}). The corresponding results are presented and interpreted in Sec.~\ref{sec:GHY-boundary-term-generalized}.

\subsection{General Relativity: Gibbons–Hawking–York boundary term}
\label{sec:gibbons-hawking-york}
We would like to compute the variation of the second term on the right-hand side of Eq.~(\ref{eq:einstein-hilbert-action-extracting-time-derivatives-1}).
In local-frame coordinates we have that $g_{\mu\nu}=\eta_{\mu\nu}$, $\partial_{\varrho}g_{\mu\nu}=0$, and $\Gamma^{\mu}_{\phantom{\mu}\nu\varrho}=0$, but $\partial_{\varrho}\partial_{\sigma}g_{\mu\nu}\neq 0$ and $\partial_{\sigma}\Gamma^{\mu}_{\phantom{\mu}\nu\varrho}\neq 0$. Therefore, in these coordinates we can express the variation of the second term as
\begin{equation}
\delta\int_{\mathcal{M}}\mathrm{d}^4x\,\frac{\partial(\sqrt{-\eta}w^{\lambda})}{\partial x^{\lambda}}=\int_{\mathcal{M}}\mathrm{d}^4x\,\frac{\partial}{\partial x^{\lambda}}\left[\sqrt{-\eta}(\eta^{\alpha\beta}\delta\Gamma^{\lambda}_{\phantom{\lambda}\alpha\beta}-\eta^{\lambda\alpha}\delta\Gamma^{\nu}_{\phantom{\nu}\alpha\nu})\right]\,.
\end{equation}
Note that the Minkowski metric is a nondynamical object. The contributions that transform $\Gamma^{\mu}_{\phantom{\mu}\nu\varrho}$ nonlinearly under general coordinate transformations cancel when the variation of the Christoffel symbols are considered, whereupon $\delta\Gamma^{\mu}_{\phantom{\mu}\nu\varrho}$ transforms as a tensor. Then the above integrand is a Lorentz scalar in a local frame, which means that it is a Lorentz scalar in an arbitrary frame. Hence, we can generalize this expression to arbitrary coordinates and obtain:
\begin{subequations}
\begin{align}
\label{eq:variation-boundary-term}
\delta\int_{\mathcal{M}}\mathrm{d}^4x\,\frac{\partial(\sqrt{-g}w^{\lambda})}{\partial x^{\lambda}}&=\int_{\mathcal{M}}\mathrm{d}^4x\,\sqrt{-g}\,\nabla_{\lambda}V^{\lambda}\,, \\[2ex]
V^{\lambda}&=g^{\alpha\beta}\delta\Gamma^{\lambda}_{\phantom{\lambda}\alpha\beta}-g^{\lambda\alpha}\delta\Gamma^{\nu}_{\phantom{\nu}\alpha\nu}\,.
\end{align}
\end{subequations}
Inserting the variation of the Christoffel symbols expressed in terms of covariant derivatives leads to
\begin{equation}
V^{\lambda}=g^{\alpha\beta}g^{\lambda\varrho}(\nabla_{\alpha}\delta g_{\varrho\beta}-\nabla_{\varrho}\delta g_{\alpha\beta})\,.
\end{equation}
To apply Gauss' theorem to the right-hand side of Eq.~(\ref{eq:variation-boundary-term}), the integrand must be contracted with the normal vector $n^{\mu}$ associated with the boundary. Therefore,
\begin{equation}
\int_{\mathcal{M}}\mathrm{d}^4x\,\sqrt{-g}\,\nabla_{\lambda}V^{\lambda}=\oint_{\partial\mathcal{M}} \mathrm{d}^3y\,\sqrt{q}\,\varepsilon n_{\lambda}V^{\lambda}\,,
\end{equation}
where $\varepsilon=-1$ for timelike $n_{\mu}$ (spacelike boundary) and $\varepsilon=1$ for spacelike $n_{\mu}$ (timelike boundary). Coordinates denoted as $y$ are used on the boundary. We now decompose the four-metric on the boundary into the induced metric and a combination of normal vectors according to Eq.~(\ref{eq:induced-metric}). Then,
\begin{align}
\label{eq:result-boundary-term}
n_{\lambda}V^{\lambda}&=g^{\alpha\beta}n^{\varrho}(\nabla_{\alpha}\delta g_{\varrho\beta}-\nabla_{\varrho}\delta g_{\alpha \beta})=(q^{\alpha\beta}n^{\varrho}\mp n^{\alpha}n^{\beta}n^{\varrho})(\nabla_{\alpha}\delta g_{\varrho\beta}-\nabla_{\varrho}\delta g_{\alpha\beta}) \notag \\
&=q^{\alpha \beta}n^{\varrho} (\nabla_{\alpha}\delta g_{\varrho \beta}-\nabla_{\varrho}\delta g_{\alpha\beta})=-q^{\alpha\beta}n^{\varrho}\nabla_{\varrho}\delta g_{\alpha\beta}\,.
\end{align}
The contribution involving three normal vectors is eliminated, as it is contracted with an antisymmetric term. In the last step we took into account that the induced metric is fixed on the boundary as is $g_{\mu\nu}$: $\delta q_{\mu\nu}|_{\partial\mathcal{M}}=0$. Therefore, directional derivatives of the variation within the boundary can safely be set zero, which eliminates the first term. However, taking assumptions on the derivative of the variation along directions perpendicular to the boundary is beyond Hamilton's principle applied to field theory. Thus, the remaining term provides a nonvanishing contribution on the boundary that reads
\begin{equation}
\label{eq:result-boundary-term-final}
-\varepsilon \oint_{\partial\mathcal{M}}\mathrm{d}^3y\,\sqrt{q}\,q^{\alpha\beta}n^{\varrho}\nabla_{\varrho}\delta g_{\alpha\beta}\,.
\end{equation}
In fact, this term is canceled by the Gibbons–Hawking–York (GHY) boundary term added to the action. We employ the extrinsic curvature defined in Eq.~(\ref{eq:extrinsic-curvature}) to obtain:
\begin{align}
\label{eq:variation-extrinsic-curvature-trace}
\delta K&=-q^{\alpha\beta}\delta\Gamma^{\lambda} _{\phantom{\lambda}\alpha\beta} n_{\lambda}=-q^{\alpha \beta}\frac{1}{2}g^{\lambda\varrho}(\nabla_{\alpha}\delta g_{\varrho\beta}+\nabla_{\beta}\delta g_{\varrho\alpha}-\nabla_{\varrho}\delta g_{\alpha\beta})n_{\lambda} \notag \\
&=-q^{\alpha\beta}\frac{1}{2}n^{\varrho}(\nabla_{\alpha}\delta g_{\varrho \beta}+\nabla_{\beta}\delta g_{\varrho\alpha}-\nabla_{\varrho}\delta g_{\alpha\beta})=\frac{1}{2}q^{\alpha\beta}n^{\varrho}\nabla_{\varrho}\delta g_{\alpha \beta}\,,
\end{align}
where we again used that derivatives of variations along the boundary vanish. So we identify $n_{\lambda}V^{\lambda}=-\delta K/2$ based on Eq.~(\ref{eq:result-boundary-term}). Therefore, to cancel Eq.~(\ref{eq:result-boundary-term-final}), we must add the GHY boundary term of Eq.~(\ref{eq:GHY-boundary-term}) to the EH action.

\subsection{Gravitational SME: Modified boundary terms}
\label{sec:gibbons-hawking-york-modified}
Now we would like to evaluate the variation of the second term on the right-hand side of Eq.~(\ref{eq:modified-action-extracting-time-derivatives-1}). As before, we employ local coordinates:
\begin{equation}
\delta\int_{\mathcal{M}}\mathrm{d}^4x\,\frac{\partial(\sqrt{-\eta}w^{(s)\lambda})}{\partial x^{\lambda}}=\int_{\mathcal{M}}\mathrm{d}^4x\,\frac{\partial}{\partial x^{\lambda}}\left[\sqrt{-\eta}(s^{\alpha\beta}\delta\Gamma^{\lambda}_{\phantom{\lambda}\alpha\beta}-s^{\lambda\alpha}\delta\Gamma^{\nu}_{\phantom{\nu}\alpha\nu})\right]\,,
\end{equation}
which is a scalar with respect to general coordinate transformations. Thus, in general coordinates it can be written as
\begin{subequations}
\begin{align}
\delta\int_{\mathcal{M}}\mathrm{d}^4x\,\frac{\partial(\sqrt{-g}w^{(s)\lambda})}{\partial x^{\lambda}}&=\int_{\mathcal{M}}\mathrm{d}^4x\,\sqrt{-g}\,\nabla_{\lambda}Q^{\lambda}\,, \\[2ex]
Q^{\lambda}&=s^{\alpha\beta}\delta\Gamma^{\lambda}_{\phantom{\lambda}\alpha\beta}-s^{\lambda\alpha}\delta\Gamma^{\nu}_{\phantom{\nu}\alpha\nu}\,.
\end{align}
\end{subequations}
Inserting the variations of the Cristoffel symbols leads to
\begin{equation}
Q^{\lambda}=s^{\alpha\beta}g^{\lambda\varrho}\nabla_{\alpha}\delta g_{\varrho\beta}-\frac{1}{2}(s^{\alpha\beta}g^{\lambda\varrho}\nabla_{\varrho}\delta g_{\alpha\beta}+s^{\alpha\lambda} g^{\nu\gamma}\nabla_{\alpha}\delta g_{\gamma\nu})\,,
\end{equation}
and a subsequent contraction with $n_{\mu}$ implies:
\begin{equation}
n_{\lambda}Q^{\lambda}=s^{\alpha\beta}n^{\varrho}\nabla_{\alpha}\delta g_{\varrho\beta}-\frac{1}{2}(s^{\alpha\beta}n^{\varrho}\nabla_{\varrho}\delta g_{\alpha \beta}+n_{\lambda}s^{\alpha \lambda}g^{\nu\gamma}\nabla_{\alpha}\delta g_{\gamma\nu})\,.
\end{equation}
We can now benefit again from the decomposition of $s^{\mu\nu}$ stated in Eq.~(\ref{eq:decomposition-s}). For each of the three terms this decomposition gives rise to
\begin{subequations}
\begin{align}
s^{\alpha\beta}n^{\varrho}\nabla_{\alpha}\delta g_{\varrho\beta}&=[q^{\alpha}_{\phantom{\alpha}\mu}q^{\beta}_{\phantom{\beta}\nu}s^{\mu\nu}-(q^{\alpha}_{\phantom{\alpha}\nu}n^{\beta}+q^{\beta}_{\phantom{\beta}\nu} n^{\alpha})s^{\nu\mathbf{n}}+n^{\alpha}n^{\beta}s^{\mathbf{nn}}]n^{\varrho}\nabla_{\alpha}\delta g_{\varrho\beta} \notag \\
&=(n^{\alpha}n^{\beta}s^{\mathbf{nn}}-q^{\beta}_{\phantom{\beta}\nu}n^{\alpha}s^{\nu\mathbf{n}})n^{\varrho}\nabla_{\alpha}\delta g_{\varrho\beta}\,, \displaybreak[0]\\[2ex]
s^{\alpha\beta}n^{\varrho}\nabla_{\varrho}\delta g_{\alpha \beta}&=[q^{\alpha}_{\phantom{\alpha}\mu}q^{\beta}_{\phantom{\beta}\nu}s^{\mu\nu}-2q^{\alpha}_{\phantom{\alpha}\nu}n^{\beta}s^{\nu\mathbf{n}}+n^{\alpha}n^{\beta}s^{\mathbf{nn}}]n^{\varrho}\nabla_{\varrho}\delta g_{\alpha\beta}\,, \displaybreak[0]\\[2ex]
n_{\lambda}s^{\alpha \lambda}g^{\nu\gamma}\nabla_{\alpha}\delta g_{\gamma\nu}&=n_{\lambda}[q^{\alpha}_{\phantom{\alpha}\mu}q^{\lambda}_{\phantom{\lambda}\nu}s^{\mu\nu}-(q^{\alpha}_{\phantom{\alpha}\nu}n^{\lambda}+q^{\lambda}_{\phantom{\lambda}\nu} n^{\alpha})s^{\nu\mathbf{n}}+n^{\alpha}n^{\lambda}s^{\mathbf{nn}}]g^{\nu\gamma}\nabla_{\alpha}\delta g_{\gamma\nu} \notag \\
&=-n^{\alpha}(q^{\nu\gamma}-n^{\nu}n^{\gamma})s^{\mathbf{nn}}\nabla_{\alpha}\delta g_{\gamma\nu}=(n^{\nu}n^{\gamma}-q^{\nu\gamma})n^{\alpha}s^{\mathbf{nn}}\nabla_{\alpha}\delta g_{\gamma\nu}\,.
\end{align}
\end{subequations}
Computing the linear combination of these terms that forms $n_{\lambda}Q^{\lambda}$, many contributions cancel each other. In particular, cancelations occur for all terms involving the mixed coefficients $s^{\nu\mathbf{n}}$ and for the purely timelike ones $s^{\mathbf{nn}}$ multiplied by $n^{\alpha}n^{\beta}\nabla_{\varrho}\delta g_{\alpha\beta}$. What remains is
\begin{equation}
n_{\lambda}Q^{\lambda}=\frac{1}{2}(q^{\nu\gamma}s^{\mathbf{nn}}n^{\alpha}\nabla_{\alpha}\delta g_{\gamma\nu}-q^{\alpha}_{\phantom{\alpha}\mu}q^{\beta}_{\phantom{\beta}\nu}s^{\mu\nu}n^{\varrho}\nabla_{\varrho}\delta g_{\alpha\beta})\,.
\end{equation}
So there is a nonvanishing contribution on the boundary given by
\begin{equation}
\int_{\mathcal{M}}\mathrm{d}^4x\,\sqrt{-g}\,\nabla_{\lambda}Q^{\lambda}=\oint_{\partial\mathcal{M}} \mathrm{d}^3y\,\sqrt{q}\,\varepsilon n_{\lambda}Q^{\lambda}\,.
\end{equation}
As in the case of GR, we try to reproduce this contribution via variations of suitable coordinate scalars on the boundary that are formed from the controlling coefficients and the extrinsic-curvature tensor. There are not too many possibilities, but we can consider $s^{ij}K_{ij}$ and the trace $K$. The variation of the first contraction gives
\begin{align}
\delta(s^{ij}K_{ij})&=s^{ij}\delta K_{ij}=s^{ij}q^{\alpha}_{\phantom{\alpha}i}q^{\beta}_{\phantom{\beta}j}\delta(\nabla_{(\alpha}n_{\beta)})=s^{ij}q^{\alpha}_{\phantom{\alpha}i}q^{\beta}_{\phantom{\beta}j}(-\delta\Gamma^{\lambda}_{\phantom{\lambda}\alpha\beta}n_{\lambda}) \notag \\
&=-\frac{1}{2}s^{ij}q^{\alpha}_{\phantom{\alpha}i}q^{\beta}_{\phantom{\beta}j}(q^{\lambda\delta}-n^{\lambda}n^{\delta})(\nabla_{\alpha}\delta g_{\beta\delta}+\nabla_{\beta}\delta g_{\alpha\delta}-\nabla_{\delta}\delta g_{\alpha\beta})n_{\lambda} \notag \\
&=\frac{1}{2}q^{\alpha}_{\phantom{\alpha}i}q^{\beta}_{\phantom{\beta}j}s^{ij}n^{\delta}\nabla_{\delta}\delta g_{\alpha\beta}\,.
\end{align}
Therefore, by employing the previous result of Eq.~(\ref{eq:variation-extrinsic-curvature-trace}), we deduce that
\begin{equation}
n_{\lambda}Q^{\lambda}=s^{\mathbf{nn}}\delta K-\delta(s^{\mu\nu}K_{\mu\nu})\,,
\end{equation}
which implies the boundary terms stated in Eq.~(\ref{eq:gibbons-hawking-york-modified}). Note that the boundary contribution associated with the $u$ term simply follows from scaling the GHY boundary contribution by the factor $(1-u)$.

\section{Derivation of modified Einstein equations}
\label{eq:modified-field-equations-derivation}

For clarification, we will provide a brief derivation of the modified field equations stated in Eqs.~(\ref{eq:einstein-equations-modified}), (\ref{eq:einstein-equations-modified-u}). All quantities are defined in the spacetime manifold $\mathcal{M}$ and, for brevity, the superscript `(4)' is omitted throughout this section. In what follows, we will benefit from the variation of the metric determinant:
\begin{equation}
\delta\sqrt{-g}=-\frac{1}{2}\sqrt{-g}g_{\alpha\beta}\delta g^{\alpha\beta}\,.
\end{equation}
Furthermore, we need the variation of the Ricci tensor given by the Palatini identity:
\begin{subequations}
\begin{align}
\label{varR}
\delta R_{\alpha\beta}&=\nabla_{\lambda}\delta\Gamma^{\lambda}_{\phantom{\lambda}\alpha\beta}-\nabla_{\beta}\delta\Gamma^{\lambda}_{\phantom{\lambda}\alpha\lambda}\,, \displaybreak[0]\\[2ex]
\label{eq:variation-christoffel}
\delta\Gamma^{\lambda}_{\phantom{\lambda}\mu\nu}&=\frac{1}{2}g^{\lambda\rho}\left(\nabla_{\mu}\delta g_{\rho\nu}+\nabla_{\nu}\delta g_{\rho\mu}-\nabla_{\rho}\delta g_{\mu\nu}\right)\,.
\end{align}
\end{subequations}
Last but not least, the contracted Palatini identity
\begin{equation}
g^{\alpha\beta}\delta R_{\alpha\beta}=(\nabla^{\alpha}\nabla^{\beta}-g^{\alpha\beta}\nabla^2)\delta g_{\alpha\beta}\,,
\end{equation}
as well as the variations involving the inverse metric,
\begin{subequations}
\begin{align}
\delta g^{\alpha\beta}&=-g^{\alpha\mu}\delta g_{\mu\nu}g^{\nu\beta}\,, \\[2ex]
\label{eq:variation-inverse-metric}
g_{\alpha\beta}\delta g^{\alpha\beta}&=-g^{\alpha\beta}\delta g_{\alpha\beta}\,,
\end{align}
\end{subequations}
will also turn out to be valuable. Now, a variation of the EH action implies
\begin{align}
\delta S^{(0)}&=\frac{1}{2\kappa}\int_{\mathcal{M}}\mathrm{d}^4x\,\left[\left(\frac{1}{2}\sqrt{-g} g^{\alpha\beta}\delta g_{\alpha\beta}\right)R-\sqrt{-g}R^{\alpha\beta}\delta g_{\alpha\beta}\right. \notag \\
&\phantom{{}={}}\left.{}\hspace{2.1cm}+\sqrt{-g}(\nabla^{\alpha}\nabla^{\beta}-g^{\alpha\beta}\nabla^2)\delta g_{\alpha\beta}\right]\,.
\end{align}
The last term is a total derivative. However, its treatment is subtle, as it involves second-order derivatives of the metric variation. In accordance with Hamilton's principle, this contribution can only be discarded when taking into account a boundary term as described in Sec.~\ref{sec:GHY-boundary-term-generalized}. Then,
\begin{equation}
\delta S^{(0)}=\int_{\mathcal{M}}\mathrm{d}^4x\,\frac{\sqrt{-g}}{2\kappa}(-G^{\alpha\beta})\delta g_{\alpha\beta}\,,\quad G^{\alpha\beta}=R^{\alpha\beta}-\frac{1}{2}g^{\alpha\beta}R\,,
\end{equation}
with the Einstein tensor $G^{\mu\nu}$. The latter finding implies the Einstein equations without matter
\begin{equation}
G^{\alpha\beta}=0\,.
\end{equation}
Varying the Lorentz-violating contribution of the action given by Eqs.~(\ref{eq:minimal-gravity-contribution-u}), (\ref{eq:action-u-term}) leads to:
\begin{equation}
\delta S^{(u)}=\int_{\mathcal{M}}\mathrm{d}^4x\,\frac{\sqrt{-g}}{2\kappa}\left[uG^{\alpha\beta}\delta g_{\alpha\beta}-u(\nabla^{\alpha}\nabla^{\beta}-g^{\alpha\beta}\nabla^2)\delta g_{\alpha\beta}\right]\,,
\end{equation}
where fluctuations $\delta u$ of the nondynamical background do not occur. A double partial integration with suitable boundary terms taken into account implies
\begin{equation}
\delta S^{(u)}=\int_{\mathcal{M}}\mathrm{d}^4x\,\frac{\sqrt{-g}}{2\kappa}\left(uG^{\alpha\beta}-\nabla^{\alpha}\nabla^{\beta}u+g^{\alpha\beta}\nabla^2u\right)\delta g_{\alpha\beta}\,.
\end{equation}
Hence, we arrive at \eqref{eq:TuLIV}. Finally, the variation of the action given by Eqs.~(\ref{eq:minimal-gravity-contribution-s}), (\ref{eq:action-s-term}) is considered:
\begin{equation}
\delta S^{(s)}=\frac{1}{2\kappa}\int_{\mathcal{M}}\mathrm{d}^4x\,\left[(\delta\sqrt{-g})s^{\mu\nu}R_{\mu\nu}+\sqrt{-g}s^{\mu\nu}(\delta R_{\mu\nu})+\sqrt{-g}(\delta s^{\mu\nu})R_{\mu\nu} \right]\,.
\end{equation}
The last term vanishes in the setting of explicit diffeomorphism violation, because the background field does not exhibit fluctuations in this case: $\delta s^{\mu \nu}=0$. We then employ \eqref{eq:variation-inverse-metric} to obtain:
\begin{subequations}
\begin{align}
\delta S^{(s)}&=\sum_{i=1}^3 \delta S^{(s)}_i\,, \displaybreak[0]\\[2ex]
\delta S^{(s)}_1&=\int_{\mathcal{M}}\mathrm{d}^4x\,\frac{\sqrt{-g}}{2\kappa}\left(\frac{1}{2}g^{\alpha\beta}s^{\mu\nu}R_{\mu\nu}\right)\delta g_{\alpha\beta}\,, \displaybreak[0]\\[2ex]
\label{eq:action-modification-2}
\delta S^{(s)}_2&=\int_{\mathcal{M}}\mathrm{d}^4x\,\frac{\sqrt{-g}}{2\kappa}s^{\mu\nu}(\nabla_{\lambda}\delta\Gamma^{\lambda}_{\phantom{\lambda}\mu\nu})\,, \displaybreak[0]\\[2ex]
\delta S^{(s)}_3&=-\int_{\mathcal{M}}\mathrm{d}^4x\,\frac{\sqrt{-g}}{2\kappa}s^{\mu\nu}(\nabla_{\nu}\delta \Gamma^{\lambda}_{\phantom{\lambda}\mu \lambda})\,.
\end{align}
\end{subequations}
To treat \eqref{eq:action-modification-2}, we perform an integration by parts and use \eqref{eq:variation-christoffel}:
\begin{equation}
\delta S^{(s)}_2=\int_{\mathcal{M}}\mathrm{d}^4x\,\frac{\sqrt{-g}}{2\kappa}(-\nabla_{\lambda}s^{\mu\nu})\frac{1}{2}g^{\lambda\rho}\left(\nabla_{\mu}\delta g_{\rho\nu}+\nabla_{\nu}\delta g_{\rho \mu}-\nabla_{\rho}\delta g_{\mu\nu}\right)\,,
\end{equation}
where we have integrated by parts. Another integration by parts and renaming some indices gives
\begin{equation}
\delta S^{(s)}_2=\int_{\mathcal{M}}\mathrm{d}^4x\,\frac{\sqrt{-g}}{2\kappa}\frac{1}{2}\left(\nabla_{\mu}\nabla^{\alpha}s^{\mu\beta}+\nabla_{\nu}\nabla^{\alpha}s^{\beta\nu}-\nabla^2 s^{\alpha\beta}\right)\delta g_{\alpha\beta}\,.
\end{equation}
We now apply the same procedure to $\delta S^{(s)}_3$:
\begin{align}
\delta S^{(s)}_3&=\int_{\mathcal{M}}\mathrm{d}^4x\,\frac{\sqrt{-g}}{2\kappa}\frac{1}{2}\left(-\nabla_{\mu}\nabla_{\nu}s^{\mu\nu}g^{\alpha\beta}-\nabla_{\lambda}\nabla_{\nu}s^{\beta\nu} g^{\lambda \alpha}+\nabla_{\rho}\nabla_{\nu} s^{\alpha\nu}g^{\beta\rho}\right)\delta g_{\alpha\beta} \notag \\
&=\int_{\mathcal{M}}\mathrm{d}^4x\,\frac{\sqrt{-g}}{2\kappa}\frac{1}{2}(-\nabla_{\mu}\nabla_{\nu}s^{\mu\nu}g^{\alpha \beta})\delta g_{\alpha\beta}\,.
\end{align}
Summing all the contributions and performing a symmetrization leads to \eqref{eq:TsLIV}.

\section{Projections of modified Einstein equations}
\label{sec:projection-einstein-equations-modified}

In the current section we are going to present detailed computations showing how to arrive at the results presented in Sec.~\ref{sec:projections-field-equations-s-term}. For brevity, we will introduce the following observer two-tensors:
\begin{subequations}
\label{eq:two-tensors-modified-einstein-equations}
\begin{align}
(T_1^{Rs})^{\alpha\beta}&=g^{\alpha\beta}s^{\mu\nu}{}^{(4)}R_{\mu\nu}\,,
\quad (T_2^{Rs})^{\alpha\beta}=s^{\alpha\mu}{}^{(4)}R_{\mu}^{\phantom{\mu}\beta}\,,
\quad (T_3^{Rs})^{\alpha\beta}=s^{\beta\mu}{}^{(4)}R_{\mu}^{\phantom{\mu}\alpha}\,,
\displaybreak[0]\\[2ex]
(T_4^{Rs})^{\alpha\beta}&=\nabla_{\mu}\nabla^{\alpha}s^{\mu\beta}\,,\quad (T_5^{Rs})^{\alpha\beta}
=\nabla_{\mu}\nabla^{\beta}s^{\mu\alpha}\,, \displaybreak[0]\\[2ex] (T_6^{Rs})^{\alpha\beta}&
=\nabla_{\mu}\nabla^{\mu}s^{\alpha\beta}\,, \quad (T_7^{Rs})^{\alpha\beta}=g^{\alpha\beta}
\nabla_{\mu}\nabla_{\nu}s^{\mu\nu}\,.
\end{align}
\end{subequations}
Now, the tensor $(T^{Rs})^{\alpha\beta}$ given in \eqref{eq:TsLIV} is expressed in terms of these quantities as follows:
\begin{align}
\label{eq:einstein-equation-modification-upper-indices}
(T^{Rs})^{\alpha\beta}&=\frac{1}{2}(T_1^{Rs})^{\alpha\beta}+\frac{1}{2}\left[(T_4^{Rs})^{\alpha\beta}+(T_5^{Rs})^{\alpha\beta}-(T_6^{Rs})^{\alpha\beta}-(T_7^{Rs})^{\alpha\beta}\right]\,.
\end{align}
For completeness, note that the analogous tensor occurring in the modified Einstein equations in \cite{Kostelecky:2003fs} reads
\begin{align}
\label{eq:einstein-equation-modification-lower-indices}
(\tilde{T}^{Rs})^{\alpha\beta}&=\frac{1}{2}(T_1^{Rs})^{\alpha\beta}-(T_2^{Rs})^{\alpha\beta}-(T_3^{Rs})^{\alpha\beta} \notag \\
&\phantom{{}={}}+\frac{1}{2}\left[(T_4^{Rs})^{\alpha\beta}+(T_5^{Rs})^{\alpha\beta}-(T_6^{Rs})^{\alpha\beta}-(T_7^{Rs})^{\alpha\beta}\right]\,.
\end{align}
Obviously, there are two additional terms involving the tensors $(T_{2,3}^{Rs})^{\alpha\beta}$ showing that the theory based on the lower-index background field $s_{\mu\nu}$ in the setting of explicit diffeomorphism violation is different from the theory described by (\ref{eq:minimal-gravity-contribution-s}) that we consider in this paper. Hence, we will be working with \eqref{eq:einstein-equation-modification-upper-indices} instead of \eqref{eq:einstein-equation-modification-lower-indices}.

\subsection{Purely spacelike sector}
\label{eq:projections-purely-spacelike-details}

For this sector, it is paramount to employ the key equation
\begin{subequations}
\label{eq:key-equation-purely-spacelike}
\begin{align}
q^{\mu}_{\phantom{\mu}\alpha}q^{\nu}_{\phantom{\nu}\beta}\nabla_{\mu}\nabla_{\nu}s^{\alpha\beta}&=D_iD_js^{ij}-K^l_{\phantom{l}i}K_{jl}s^{ij}+KK_{ij}s^{ij}-\frac{1}{N}K_{ij}\mathcal{L}_ms^{ij}\,, \\[2ex]
\mathcal{L}_ms^{ij}&=m^{\alpha}\partial_{\alpha}s^{ij}+s^{il}\partial_lN^j+s^{jl}\partial_lN^i\,,
\end{align}
\end{subequations}
which is to be derived as follows. We start by expressing the covariant derivatives in $\Sigma_t$ in terms of projected covariant derivatives of the four-dimensional spacetime manifold:
\begin{equation}
D_{\alpha}(D_{\beta}s^{\alpha\beta})=q^{\lambda}_{\phantom{\lambda}\alpha}q^{\mu}_{\phantom{\mu}\beta}q^{\alpha}_{\phantom{\alpha}\nu}q^{\beta}_{\phantom{\beta}\sigma}\nabla_{\lambda}(D_{\mu}s^{\nu\sigma})=q^{\lambda}_{\phantom{\lambda}\nu}q^{\mu}_{\phantom{\mu}\sigma}\nabla_{\lambda}(q^{\alpha}_{\phantom{\alpha}\mu}q^{\nu}_{\phantom{\nu}\beta}q^{\sigma}_{\phantom{\sigma}\gamma}\nabla_{\alpha}s^{\beta\gamma})\,.
\end{equation}
Now we apply the outer covariant derivative providing four terms:
\begin{align}
\nabla_{\lambda}(q^{\alpha}_{\phantom{\alpha}\mu}q^{\nu}_{\phantom{\nu}\beta}q^{\sigma}_{\phantom{\sigma}\gamma}\nabla_{\alpha}s^{\beta\gamma})&=(\nabla_{\lambda}q^{\alpha}_{\phantom{\alpha}\mu})q^{\nu}_{\phantom{\nu}\beta}q^{\sigma}_{\phantom{\sigma}\gamma}\nabla_{\alpha}s^{\beta\gamma}+q^{\alpha}_{\phantom{\alpha}\mu}(\nabla_{\lambda}q^{\nu}_{\phantom{\nu}\beta})q^{\sigma}_{\phantom{\sigma}\gamma}\nabla_{\alpha}s^{\beta\gamma} \notag \\
&\phantom{{}={}}+q^{\alpha}_{\phantom{\alpha}\mu}q^{\nu}_{\phantom{\nu}\beta}(\nabla_{\lambda}q^{\sigma}_{\phantom{\sigma}\gamma})\nabla_{\alpha}s^{\beta\gamma}+q^{\alpha}_{\phantom{\alpha}\mu}q^{\nu}_{\phantom{\nu}\beta}q^{\sigma}_{\phantom{\sigma}\gamma}\nabla_{\lambda}\nabla_{\alpha}s^{\beta\gamma}\,.
\end{align}
We evaluate each of these four contributions:
\begin{subequations}
\begin{align}
r_1&=q^{\lambda}_{\phantom{\lambda}\nu}q^{\mu}_{\phantom{\mu}\sigma}[(\nabla_{\lambda}q^{\alpha}_{\phantom{\alpha}\mu})q^{\nu}_{\phantom{\nu}\beta}q^{\sigma}_{\phantom{\sigma}\gamma}\nabla_{\alpha}s^{\beta\gamma}]=q^{\lambda}_{\phantom{\lambda}\beta}q^{\mu}_{\phantom{\mu}\gamma}\nabla_{\lambda}(n^{\alpha}n_{\mu})\nabla_{\alpha}s^{\beta\gamma} \notag \\
&=q^{\lambda}_{\phantom{\lambda}\beta}q^{\mu}_{\phantom{\mu}\gamma}n^{\alpha}(\nabla_{\lambda}n_{\mu})\nabla_{\alpha}s^{\beta\gamma}=q^{\lambda}_{\phantom{\lambda}\beta}q^{\mu}_{\phantom{\mu}\gamma}n^{\alpha}K_{\lambda\mu}\nabla_{\alpha}s^{\beta\gamma}=K_{\beta\gamma}n^{\alpha}\nabla_{\alpha}s^{\beta\gamma}\,, \displaybreak[0]\\[2ex]
r_2&=q^{\lambda}_{\phantom{\lambda}\nu}q^{\mu}_{\phantom{\mu}\sigma}[q^{\alpha}_{\phantom{\alpha}\mu}(\nabla_{\lambda}q^{\nu}_{\phantom{\nu}\beta})q^{\sigma}_{\phantom{\sigma}\gamma}\nabla_{\alpha}s^{\beta\gamma}]=q^{\lambda}_{\phantom{\lambda}\nu}q^{\alpha}_{\phantom{\alpha}\gamma}\nabla_{\lambda}(n^{\nu}n_{\beta})\nabla_{\alpha}s^{\beta\gamma} \notag \\
&=q^{\lambda}_{\phantom{\lambda}\nu}q^{\alpha}_{\phantom{\alpha}\gamma}(\nabla_{\lambda}n^{\nu})n_{\beta}\nabla_{\alpha}s^{\beta\gamma}=q^{\lambda}_{\phantom{\lambda}\nu}q^{\alpha}_{\phantom{\alpha}\gamma}K_{\lambda}^{\phantom{\lambda}\nu}n_{\beta}\nabla_{\alpha}s^{\beta\gamma}=K_{\nu}^{\phantom{\nu}\nu}q^{\alpha}_{\phantom{\alpha}\gamma}n_{\beta}\nabla_{\alpha}s^{\beta\gamma} \notag \\
&=Kq^{\alpha}_{\phantom{\alpha}\gamma}n_{\beta}\nabla_{\alpha}s^{\beta\gamma}\,, \displaybreak[0]\\[2ex]
r_3&=q^{\lambda}_{\phantom{\lambda}\nu}q^{\mu}_{\phantom{\mu}\sigma}[q^{\alpha}_{\phantom{\alpha}\mu}q^{\nu}_{\phantom{\nu}\beta}(\nabla_{\lambda}q^{\sigma}_{\phantom{\sigma}\gamma})\nabla_{\alpha}s^{\beta\gamma}]=q^{\lambda}_{\phantom{\lambda}\beta}q^{\alpha}_{\phantom{\alpha}\sigma}\nabla_{\lambda}(n^{\sigma}n_{\gamma})\nabla_{\alpha}s^{\beta\gamma} \notag \\
&=q^{\lambda}_{\phantom{\lambda}\beta}q^{\alpha}_{\phantom{\alpha}\sigma}(\nabla_{\lambda}n^{\sigma})n_{\gamma}\nabla_{\alpha}s^{\beta\gamma}=q^{\lambda}_{\phantom{\lambda}\beta}q^{\alpha}_{\phantom{\alpha}\sigma}K_{\lambda}^{\phantom{\lambda}\sigma}n_{\gamma}\nabla_{\alpha}s^{\beta\gamma}=K_{\beta}^{\phantom{\beta}\alpha}n_{\gamma}\nabla_{\alpha}s^{\beta\gamma}\,, \displaybreak[0]\\[2ex]
r_4&=q^{\lambda}_{\phantom{\lambda}\nu}q^{\mu}_{\phantom{\mu}\sigma}[q^{\alpha}_{\phantom{\alpha}\mu}q^{\nu}_{\phantom{\nu}\beta}q^{\sigma}_{\phantom{\sigma}\gamma}\nabla_{\lambda}\nabla_{\alpha}s^{\beta\gamma}]=q^{\lambda}_{\phantom{\lambda}\beta}q^{\alpha}_{\phantom{\alpha}\gamma}\nabla_{\lambda}\nabla_{\alpha}s^{\beta\gamma}\,.
\end{align}
\end{subequations}
Now, considering a purely spacelike $s^{\mu\nu}$, we employ
\begin{subequations}
\begin{align}
0&=\nabla_{\alpha}(Kq^{\alpha}_{\phantom{\alpha}\gamma}n_{\beta}s^{\beta\gamma})=Kq^{\alpha}_{\phantom{\alpha}\gamma}(\nabla_{\alpha}n_{\beta})s^{\beta\gamma}+Kq^{\alpha}_{\phantom{\alpha}\gamma}n_{\beta}\nabla_{\alpha}s^{\beta\gamma}\,, \\[2ex]
0&=\nabla_{\alpha}(K_{\beta}^{\phantom{\beta}\alpha}n_{\gamma}s^{\beta\gamma})=K_{\beta}^{\phantom{\beta}\alpha}(\nabla_{\alpha}n_{\gamma})s^{\beta\gamma}+K_{\beta}^{\phantom{\beta}\alpha}n_{\gamma}\nabla_{\alpha}s^{\beta\gamma}\,,
\end{align}
\end{subequations}
to reformulate the second and third term:
\begin{subequations}
\begin{align}
r_2&=-Kq^{\alpha}_{\phantom{\alpha}\gamma}(\nabla_{\alpha}n_{\beta})s^{\beta\gamma}=-Kq^{\alpha}_{\phantom{\alpha}\gamma}K_{\alpha\beta}s^{\beta\gamma}=-KK_{\gamma\beta}s^{\beta\gamma}\,, \\[2ex]
r_3&=-K_{\beta}^{\phantom{\beta}\alpha}(\nabla_{\alpha}n_{\gamma})s^{\beta\gamma}=-K_{\beta}^{\phantom{\beta}\alpha}K_{\alpha\gamma}s^{\beta\gamma}\,.
\end{align}
\end{subequations}
Finally, by organizing all contributions, we have
\begin{align}
q^{\lambda}_{\phantom{\lambda}\beta}q^{\alpha}_{\phantom{\alpha}\gamma}\nabla_{\lambda}\nabla_{\alpha}s^{\beta\gamma}&=D_{\alpha}D_{\beta}s^{\alpha\beta}+K^{\alpha}_{\phantom{\alpha}\beta}K_{\alpha\gamma}s^{\beta\gamma}+KK_{\beta\gamma}s^{\beta\gamma}-K_{\beta\gamma}n^{\alpha}\nabla_{\alpha}s^{\beta\gamma}\,.
\end{align}
Introducing the Lie derivative on $s^{ij}$ and considering only spacelike indices on the right-hand side of the latter relation implies Eq.~(\ref{eq:key-equation-purely-spacelike}).

\subsubsection{Orthogonal projection}

First of all, based on Eqs.~(\ref{eq:relation-deruelle-1}), (\ref{eq:projection-ricci-tensor-into-hypersurface}), we obtain the following contraction of the background tensor with the projected Ricci tensor into $\Sigma_t$:
\begin{align}
s^{\lambda\sigma}q^{\mu}_{\phantom{\mu}\lambda}q^{\nu}_{\phantom{\nu}\sigma}{}^{(4)}R_{\mu\nu}&=-\frac{1}{N}K_{ij}\mathcal{L}_m s^{ij}+\nabla_{\mu}(n^{\mu}K_{ij}s^{ij})-s^{ij}\frac{1}{N}D_iD_jN \notag \\
&\phantom{{}={}}+s^{ij}R_{ij}-2s^{ij}K_{il}K^l_{\phantom{l}j}\,.
\end{align}
We also employ
\begin{equation}
\frac{1}{N}K_{\mu\nu}\mathcal{L}_ms^{\mu\nu}=K_{\mu\nu}n^{\alpha}\nabla_{\alpha}s^{\mu\nu}-2K_{\mu\nu}K^{\nu}_{\phantom{\nu}\lambda}s^{\mu\lambda}\,.
\end{equation}
Then,
\begin{subequations}
\begin{align}
n_{\alpha}n_{\beta}(T_1^{Rs})^{\alpha\beta}&=n_{\alpha}n_{\beta}g^{\alpha\beta}q^{\mu}_{\phantom{\mu}\lambda}q^{\nu}_{\phantom{\nu}\sigma}s^{\lambda\sigma}{}^{(4)}R_{\mu\nu} \notag \\
&=-\left[\nabla_{\mu}(n^{\mu}K_{\alpha\beta}s^{\alpha\beta})-K_{\mu\nu}n^{\alpha}\nabla_{\alpha}s^{\mu\nu}-s^{ij}\frac{1}{N}D_iD_jN+s^{ij}R_{ij}\right]\,, \displaybreak[0]\\[2ex]
n_{\alpha}n_{\beta}(T_4^{Rs})^{\alpha\beta}&=n_{\alpha}n_{\beta}\nabla_{\nu}\nabla^{\alpha}(q^{\nu}_{\phantom{\nu}\lambda}q^{\beta}_{\phantom{\beta}\sigma}s^{\lambda\sigma}) \notag \\
&=-a_{\lambda}a_{\sigma}s^{\lambda\sigma}+K_{\nu\alpha}s^{\nu\sigma}K^{\alpha}_{\phantom{\alpha}\sigma}-s^{\nu\sigma}\nabla_{\nu}a_{\sigma}-a_{\sigma}\nabla_{\lambda}s^{\lambda\sigma}-n^{\alpha}K_{\lambda \sigma}\nabla_{\alpha}s^{\lambda\sigma} \notag \\
&=n_{\alpha}n_{\beta}(T_5^{Rs})^{\alpha\beta}\,, \displaybreak[0]\\[2ex]
n_{\alpha}n_{\beta}(T_6^{Rs})^{\alpha\beta}&=n_{\alpha}n_{\beta}\nabla^2(q^{\alpha}_{\phantom{\alpha}\mu}q^{\beta}_{\phantom{\beta}\nu}s^{\mu\nu})=2(K^{\mu}_{\phantom{\mu}\lambda}K_{\mu\sigma}s^{\lambda\sigma}-a_{\lambda}a_{\sigma}s^{\lambda\sigma})\,, \displaybreak[0]\\[2ex]
n_{\alpha}n_{\beta}(T_7^{Rs})^{\alpha\beta}&=n_{\alpha}n_{\beta}g^{\alpha\beta}\nabla_{\mu}\nabla_{\nu}(q^{\mu}_{\phantom{\mu}\kappa}q^{\nu}_{\phantom{\nu}\lambda}s^{\kappa\lambda}) \notag \\
&=-\left[\nabla_{\mu}(n^{\mu}K_{\alpha\beta}s^{\alpha\beta})+(\nabla_{\mu}a_{\beta})s^{\mu\beta}+a_{\beta}\nabla_{\mu}s^{\mu\beta}-KK_{\alpha\beta}s^{\alpha\beta}\right. \notag \\
&\phantom{{}={}}\hspace{0.5cm}+a_{\alpha}\nabla_{\beta}s^{\alpha\beta}-a_{\alpha}a_{\beta}s^{\alpha\beta}-K^{\nu}_{\phantom{\nu}\alpha}K_{\beta\nu}s^{\alpha\beta}+n^{\nu}K_{\alpha\beta}\nabla_{\nu}s^{\alpha\beta} \notag \\
&\phantom{{}={}}\left.{}\hspace{0.5cm}+q^{\mu}_{\phantom{\mu}\alpha} q^{\nu}_{\phantom{\nu}\beta}\nabla_{\mu}\nabla_{\nu}s^{\alpha\beta}\right]\,.
\end{align}
\end{subequations}
Summing all these contributions implies
\begin{align}
2n_{\alpha}n_{\beta}(T^{Rs})^{\alpha\beta}&=n_{\alpha}n_{\beta}\left[(T_1^{Rs})^{\alpha\beta}+2(T_4^{Rs})^{\alpha\beta}-(T_6^{Rs})^{\alpha\beta}-(T_7^{Rs})^{\alpha\beta}\right] \notag \\
&=-a_{\alpha}a_{\beta}s^{\alpha \beta}-K^{\mu}_{\phantom{\mu}\lambda}K_{\mu\sigma}s^{\lambda\sigma}-KK_{\alpha\beta}s^{\alpha\beta} -s^{\lambda\sigma}\nabla_{\lambda}a_{\sigma} \notag \\
&\phantom{{}={}}+s^{ij}\frac{1}{N}D_{i}D_{j}N-s^{ij}R_{ij}+q^{\mu}_{\phantom{\mu}\alpha} q^{\nu}_{\phantom{\nu}\beta}\nabla_{\mu}\nabla_{\nu}s^{\alpha\beta}\,.
\end{align}
We now use that
\begin{equation}
s^{\lambda\sigma}D_{\lambda}a_{\sigma}=s^{\lambda\sigma}q^{\alpha}_{\phantom{\alpha}\lambda}q^{\beta}_{\phantom{\beta}\sigma}\nabla_{\alpha}a_{\beta}=s^{\alpha\beta} \nabla_{\alpha}a_{\beta}\,,
\end{equation}
in combination with Eq.~(\ref{eq:relation-derivative-acceleration-lapse}) to obtain
\begin{equation}
s^{\alpha\beta}\nabla_{\alpha}a_{\beta}=s^{\lambda\sigma}\frac{1}{N}D_{\lambda}D_{\sigma}N-s^{\lambda\sigma}a_{\lambda}a_{\sigma}\,.
\end{equation}
Finally, by employing Eq.~(\ref{eq:key-equation-purely-spacelike}), we deduce
\begin{equation}
2n_{\alpha}n_{\beta}(T^{Rs})^{\alpha\beta}=\frac{1}{2}\left[-2K^j_{\phantom{j}i}K_{jk}s^{ik}-s^{ij}R_{ij}+D_iD_js^{ij}-\frac{1}{N}K_{ij}\mathcal{L}_ms^{ij}\right]\,.
\end{equation}
The latter corresponds to Eq.~(\ref{eq:orthogonal-projection-purely-spacelike}).

\subsubsection{Mixed projection}

Here it is reasonable to employ the following form of the Lie derivative of the background:
\begin{equation}
\label{Lie_der1}
\frac{1}{N}\mathcal{L}_ms^{\lambda\sigma}=n^{\alpha}\nabla_{\alpha}s^{\lambda\sigma}-(a_{\mu}n^{\lambda }+K^{\lambda}_{\phantom{\lambda}\mu})s^{\mu\sigma}-(a_{\mu}n^{\sigma}+ K^{\sigma}_{\phantom{\sigma}\mu})s^{\lambda\mu}\,.
\end{equation}
For any tensor $T_{\lambda\sigma}$ with spacelike lower indices we have
\begin{equation}
T_{\lambda\sigma}n^{\alpha}\nabla_{\alpha}s^{\lambda \sigma}=T_{\lambda\sigma}(K^{\lambda}_{\phantom{\lambda}\mu}s^{\mu\sigma}+K^{\sigma}_{\phantom{\sigma}\mu}s^{\lambda\mu})+T_{\lambda\sigma}\frac{1}{N}\mathcal{L}_ms^{\lambda\sigma}\,.
\end{equation}
In the forthcoming analysis we only consider those terms that are nonzero when contracted with $q^k_{\phantom{k}\alpha}n_{\beta}$. We also take $s^{\lambda\sigma}$ as purely spacelike, which tells us that $n_{\lambda}s^{\lambda\sigma}=0$. Then,
\begin{align}
\label{eq:mixed-projection-T4}
q^k_{\phantom{k}\alpha}n_{\beta}(T_4^{Rs})^{\alpha\beta}&=q^k_{\phantom{k}\alpha}n_{\beta}\nabla_{\nu}\nabla^{\alpha}(q^{\nu}_{\phantom{\nu}\lambda}q^{\beta}_{\phantom{\beta}\sigma}s^{\lambda\sigma}) \notag \\
&=q^k_{\phantom{k}\alpha}n_{\beta}\nabla_{\nu} \left[(\nabla^{\alpha}q^{\nu}_{\phantom{\nu}\lambda})q^{\beta}_{\phantom{\beta}\sigma}s^{\lambda\sigma}+q^{\nu}_{\phantom{\nu}\lambda}(\nabla^{\alpha}q^{\beta}_{\phantom{\beta}\sigma}) s^{\lambda\sigma}+q^{\nu}_{\phantom{\nu}\lambda}q^{\beta}_{\phantom{\beta}\sigma}(\nabla^{\alpha}s^{\lambda \sigma})\right] \notag \\
&=q^k_{\phantom{k}\alpha}n_{\beta}\left[(\nabla^{\alpha}q^{\nu}_{\phantom{\nu}\lambda})(\nabla_{\nu}q^{\beta}_{\phantom{\beta}\sigma})s^{\lambda\sigma}+(\nabla_{\nu} q^{\nu}_{\phantom{\nu}\lambda})(\nabla^{\alpha}q^{\beta}_{\phantom{\beta}\sigma})s^{\lambda\sigma}\right. \notag \\
&\phantom{{}={}}\hspace{1.2cm}+q^{\nu}_{\phantom{\nu}\lambda}(\nabla_{\nu}\nabla^{\alpha}q^{\beta}_{\phantom{\beta}\sigma})s^{\lambda \sigma}+q^{\nu}_{\phantom{\nu}\lambda}(\nabla^{\alpha}q^{\beta}_{\phantom{\beta}\sigma})(\nabla_{\nu}s^{\lambda\sigma}) \notag \\
&\phantom{{}={}}\hspace{1.1cm}\left.{}+q^{\nu}_{\phantom{\nu}\lambda}(\nabla_{\nu}   q^{\beta}_{\phantom{\beta}\sigma})(\nabla^{\alpha}s^{\lambda\sigma})\right] \notag \\
&=q^k_{\phantom{k}\alpha}n_{\beta}\left[K^{\alpha}_{\phantom{\alpha}\lambda}n^{\beta}a_{\sigma}s^{\lambda\sigma}+a_{\lambda}n^{\beta}K^{\alpha}_{\phantom{\alpha}\sigma}s^{\lambda\sigma}+q^{\nu}_{\phantom{\nu}\lambda}n^{\beta}(\nabla_{\nu}K^{\alpha}_{\phantom{\alpha}\sigma}-K^{\alpha}_{\phantom{\alpha}\nu}a_{\sigma})s^{\lambda\sigma}\right. \notag \\
&\phantom{{}={}}\hspace{1.1cm}\left.{}+q^{\nu}_{\phantom{\nu}\lambda} n^{\beta}K^{\alpha}_{\phantom{\alpha}\sigma}\nabla_{\nu}s^{\lambda\sigma}+q^{\nu}_{\phantom{\nu}\lambda}n^{\beta}K_{\nu\sigma}\nabla^{\alpha}s^{\lambda\sigma}\right] \notag \\
&=-K^k_{\phantom{k}\lambda}a_{\sigma}s^{\lambda \sigma}-D_{\lambda}(K^k_{\phantom{k}\sigma}s^{\lambda \sigma})-K_{\lambda\sigma}D^{k}s^{\lambda\sigma}\,.
\end{align}
In an analogous manner, we obtain
\begin{align}
\label{eq:mixed-projection-T5}
q^k_{\phantom{k}\alpha}n_{\beta}(T_5^{Rs})^{\alpha\beta}&=q^k_{\phantom{k}\beta}n_{\alpha}\nabla_{\nu}\nabla^{\alpha}(q^{\nu}_{\phantom{\nu}\lambda}q^{\beta}_{\phantom{\beta}\sigma} s^{\lambda\sigma}) \notag \\
&=Ka_{\lambda}s^{\lambda k}-a_{\alpha}K^{\alpha}_{\lambda}s^{\lambda k}+n^{\nu}\nabla_{\nu}a_{\lambda}s^{\lambda k}+n^{\nu}q^k_{\phantom{k}\beta}a_{\lambda}\nabla_{\nu} s^{\lambda\beta}+K^k_{\phantom{k}\nu}a_{\sigma}s^{\nu\sigma} \notag \\
&\phantom{{}={}}+n_{\alpha}q^k_{\phantom{k}\sigma}(Kn_{\lambda}+a_{\lambda})\nabla^{\alpha}s^{\lambda\sigma}+n_{\alpha}K^k_{\phantom{k}\lambda}n_{\sigma}\nabla^{\alpha}s^{\lambda\sigma} \notag \\
&\phantom{{}={}}+n_{\alpha}q^{\nu}_{\phantom{\nu}\lambda}q^{k} _{\phantom{k}\sigma}\nabla _{\nu}\nabla^{\alpha}s^{\lambda\sigma}\,,
\end{align}
as well as
\begin{align}
\label{eq:mixed-projection-T6}
q^k_{\phantom{k}\alpha}n_{\beta}(T_6^{Rs})^{\alpha\beta}&=q^k_{\phantom{k}\alpha}n_{\beta}\nabla^2(q^{\alpha}_{\phantom{\alpha}\lambda}q^{\beta}_{\phantom{\beta}\sigma}s^{\lambda \sigma}) \notag \\
&=-q^k_{\phantom{k}\lambda}[\nabla_{\mu}K^{\mu}_{\phantom{\mu}\sigma}-\nabla_{\mu}(n^{\mu}a_{\sigma})]s^{\lambda\sigma}-2q^k_{\phantom{k}\lambda}(K^{\mu}_{\phantom{\mu}\sigma}-n^{\mu} a_{\sigma})\nabla_{\mu}s^{\lambda\sigma}\,,
\end{align}
whereby
\begin{equation}
\label{eq:mixed-projection-T17}
q^k_{\phantom{k}\alpha}n_{\beta}(T_1^{Rs})^{\alpha\beta}=q^k_{\phantom{k}\alpha}n_{\beta}(T_7^{Rs})^{\alpha\beta}=0\,.
\end{equation}
Performing the sum of Eqs.~(\ref{eq:mixed-projection-T4}) -- (\ref{eq:mixed-projection-T17}), leads to Eq.~(\ref{eq:mixed-projection-purely-spacelike-intermediate}). To simplify this result, it is important to find
\begin{eqnarray}
n_{\alpha}q^{\nu}_{\phantom{\nu}\lambda}q^{k}_{\phantom{k}\sigma}\nabla_{\nu}\nabla^{\alpha}s^{\lambda\sigma}&=\nabla_{\nu}(n_{\alpha}q^{\nu} _{\phantom{\nu}\lambda}q^{k}_{\phantom{k}\sigma}\nabla^{\alpha}s^{\lambda\sigma})-\nabla_{\nu}(n_{\alpha}q^{\nu} _{\phantom{\nu}\lambda}q^{k} _{\phantom{k}\sigma})(\nabla^{\alpha}s^{\lambda\sigma})\,.
\end{eqnarray}
We need to consider the two pieces:
\begin{subequations}
\begin{align}
n_{\alpha}q^{\nu}_{\phantom{\nu}\lambda}q^{k}_{\phantom{k}\sigma}\nabla^{\alpha}s^{\lambda\sigma}&=q^{\nu}_{\phantom{\nu}\lambda}q^{k}_{\phantom{k}\sigma}\left(\frac{1}{N}\mathcal{L}_m s^{\lambda\sigma}+K^{\lambda}_{\phantom{\lambda}\mu}s^{\mu\sigma}+K^{\sigma}_{\phantom{\sigma}\mu}s^{\lambda\mu}\right)\,, \\[2ex]
\nabla_{\nu}(n_{\alpha}q^{\nu}_{\phantom{\nu}\lambda}q^{k}_{\phantom{k}\sigma})(\nabla^{\alpha}s^{\lambda\sigma})&=\left[K_{\lambda\alpha}q^k_{\phantom{k}\sigma}+Kn_{\alpha}n_{\lambda}   q^k_{\phantom{k}\sigma}+n_{\alpha}a_{\lambda}q^k_{\phantom{k}\sigma}\right. \notag \\
&\phantom{{}={}}\left.{}+n_{\alpha}K^k_{\phantom{k}\lambda}n_{\sigma}+n_{\alpha}n^{k}K_{\lambda\sigma}\right]\nabla^{\alpha}s^{\lambda\sigma}\,.
\end{align}
\end{subequations}
Finally, some algebra gives rise to
\begin{align}
n_{\alpha}q^{\nu}_{\phantom{\nu}\lambda}q^{k}_{\phantom{k}\sigma}\nabla_{\nu}\nabla^{\alpha}s^{\lambda\sigma}&=\nabla_{\nu}(K^{\nu}_{\phantom{\nu}\mu}s^{\mu k}+K^{k}_{\phantom{k}\mu}s^{\nu \mu})-K_{\lambda\alpha}q^k_{\phantom{k}\sigma}\nabla^{\alpha}s^{\lambda\sigma} \notag \\
&\phantom{{}={}}+Ka_{\mu}s^{\mu k}-a_{\lambda}K^{\lambda}_{\phantom{\lambda}\mu}s^{\mu k}-n^k(K_{\lambda\sigma}K^{\lambda}_{\phantom{\lambda}\mu}s^{\mu\sigma}+K_{\lambda\sigma}K^{\sigma}_{\phantom{\sigma}\mu}s^{\lambda\mu}) \notag \\
&\phantom{{}={}}+D_i\left(\frac{1}{N}\mathcal{L}_ms^{ik}\right)\,.
\end{align}

\subsection{Purely timelike sector}
\label{eq:projections-purely-timelike-details}

Here we must compute the projections of $(T^{Rs})^{\alpha\beta}+s^{\mathbf{nn}}R^{\alpha\beta}$ with $(T^{Rs})^{\alpha\beta}$ given by \eqref{eq:einstein-equation-modification-upper-indices}.

\subsubsection{Orthogonal projection}

We start by deriving the complete projection along the direction orthogonal to $\Sigma_t$ (see also \eqref{eq:purely-timelike-sector-T-extra-term}). The individual contributions amount to
\begin{subequations}
\begin{align}
n_{\alpha}n_{\beta}(T_1^{Rs})^{\alpha\beta}&=s^{\mathbf{nn}}n_{\alpha}n_{\beta}R^{\alpha\beta}=s^{\mathbf{nn}}(K_{ij}K^{ij}+n^{\mu}\nabla_{\mu}K-\nabla_{\mu}a^{\mu})\,, \displaybreak[0]\\[2ex]
n_{\alpha}n_{\beta}(T_4^{Rs})^{\alpha\beta}&=n_{\alpha}n_{\beta}\left[(\nabla_{\nu}\nabla^{\alpha}s^{\mathbf{nn}})n^{\nu}n^{\beta}+(\nabla^{\alpha}s^{\mathbf{nn}})(\nabla_{\nu}n^{\nu})n^{\beta}+(\nabla_{\nu}s^{\mathbf{nn}})(\nabla^{\alpha}n^{\nu})n^{\beta}\right. \notag \\
&\phantom{{}={}}\hspace{1cm}\left.{}+s^{\mathbf{nn}}(\nabla_{\nu}\nabla^{\alpha}n^{\nu})n^{\beta}+s^{\mathbf{nn}}n^{\nu}\nabla_{\nu}\nabla^{\alpha}n^{\beta}\right] \notag \\
&=n_{\alpha}n_{\beta}\left[(\nabla_{\nu}\nabla^{\alpha}s^{\mathbf{nn}})n^{\nu}n^{\beta}+(\nabla^{\alpha}s^{\mathbf{nn}})Kn^{\beta}+\nabla_{\nu}s^{\mathbf{nn}}(K^{\alpha\nu}-n^{\alpha}a^{\nu})n^{\beta}\right. \notag \\
&\phantom{{}={}}\hspace{1cm}\left.{}+s^{\mathbf{nn}}(\nabla_{\nu}K^{\alpha\nu}-n^{\alpha}\nabla_{\nu}a^{\nu})n^{\beta}+s^{\mathbf{nn}}n^{\nu}(\nabla_{\nu}K^{\alpha\beta}-n^{\alpha}\nabla_{\nu}a^{\beta})\right] \notag \\
&=-(n_{\alpha}n^{\nu}\nabla_{\nu}\nabla^{\alpha}s^{\mathbf{nn}}+n_{\alpha}\nabla^{\alpha}s^{\mathbf{nn}}K+a^{\nu}\nabla_{\nu}s^{\mathbf{nn}} \notag \\
&\phantom{{}={}}\hspace{0.5cm} -s^{\mathbf{nn}}K_{\alpha\nu}K^{\alpha\nu}+s^{\mathbf{nn}}\nabla_{\nu}a^{\nu}+s^{\mathbf{nn}}a_{\beta}a^{\beta})\,, \displaybreak[0]\\[2ex]
n_{\alpha}n_{\beta}(T_6^{Rs})^{\alpha\beta}&=n_{\alpha}n_{\beta}\left[(\nabla_{\mu}\nabla^{\mu}n^{\alpha})n^{\beta}s^{\mathbf{nn}}+n^{\alpha}(\nabla_{\mu}\nabla^{\mu}n^{\beta})s^{\mathbf{nn}}+n^{\alpha}n^{\beta}\nabla_{\mu}\nabla^{\mu}s^{\mathbf{nn}}\right] \notag \\
&=n_{\alpha}n_{\beta}\left[(\nabla_{\mu}K^{\mu\alpha})n^{\beta}s^{\mathbf{nn}}-n^{\mu}(\nabla_{\mu}a^{\alpha})n^{\beta}s^{\mathbf{nn}}+n^{\alpha}(\nabla_{\mu}K^{\mu\beta})s^{\mathbf{nn}}\right. \notag \\
&\phantom{{}={}}\hspace{1cm}\left.{}-n^{\alpha}(\nabla_{\mu}a^{\beta})n^{\mu}s^{\mathbf{nn}}+n^{\alpha}n^{\beta}\nabla_{\mu}\nabla^{\mu}s^{\mathbf{nn}}\right] \notag \\
&=-n_{\alpha}(\nabla_{\mu}K^{\mu\alpha})s^{\mathbf{nn}}+n^{\mu}n_{\alpha}(\nabla_{\mu}a^{\alpha})s^{\mathbf{nn}}-n_{\beta}(\nabla_{\mu}K^{\mu\beta})s^{\mathbf{nn}} \notag \\
&\phantom{{}={}}+n_{\beta}n^{\mu}(\nabla_{\mu}a^{\beta})s^{\mathbf{nn}}+\nabla_{\mu}\nabla^{\mu}s^{\mathbf{nn}} \notag \\
&=2s^{\mathbf{nn}}(K_{\mu\alpha}K^{\mu\alpha}-a_{\alpha}a^{\alpha})+\nabla_{\mu}\nabla^{\mu}s^{\mathbf{nn}}\,, \displaybreak[0]\\[2ex]
n_{\alpha}n_{\beta}(T_7^{Rs})^{\alpha\beta}&=-\nabla_{\mu}\left[(a^{\mu}+n^{\mu}K)s^{\mathbf{nn}}+n^{\mu}n^{\nu}\nabla_{\nu}s^{\mathbf{nn}}\right] \notag \\
&=-\left[(\nabla_{\mu}a^{\mu})s^{\mathbf{nn}}+a^{\mu}\nabla_{\mu}s^{\mathbf{nn}}+(\nabla_{\mu}n^{\mu})Ks^{\mathbf{nn}}+n^{\mu}(\nabla_{\mu}K)s^{\mathbf{nn}}+n^{\mu}K\nabla_{\mu}s^{\mathbf{nn}}\right. \notag \\
&\phantom{{}={}}\hspace{0.3cm}\left.{}+(\nabla_{\mu}n^{\mu})n^{\nu}\nabla_{\nu}s^{\mathbf{nn}}+n^{\mu}(\nabla_{\mu}n^{\nu})\nabla_{\nu}s^{\mathbf{nn}}+n^{\mu}n^{\nu}\nabla_{\mu}\nabla_{\nu}s^{\mathbf{nn}}\right] \notag \\
&=-\left[s^{\mathbf{nn}}\nabla_{\nu}a^{\nu}+2a^{\mu}\nabla_{\mu}s^{\mathbf{nn}}+s^{\mathbf{nn}}\nabla_{\mu}(n^{\mu}K)\right. \notag \\
&\phantom{{}={}}\left.{}\hspace{0.3cm}+2Kn^{\nu} \nabla_{\nu}s^{\mathbf{nn}}+n^{\mu}n^{\nu}\nabla_{\mu}\nabla_{\nu}s^{\mathbf{nn}}\right]\,.
\end{align}
\end{subequations}
Here we used that
\begin{subequations}
\begin{align}
0&=\nabla_{\nu}(n_{\alpha}n_{\beta}s^{\mathbf{nn}}n^{\nu}K^{\alpha\beta})=n_{\alpha}n_{\beta}s^{\mathbf{nn}}n^{\nu}\nabla_{\nu}K^{\alpha\beta}\,, \\[2ex]
0&=\nabla_{\nu}(n_{\alpha}s^{\mathbf{nn}}K^{\alpha\nu})=(\nabla_{\nu}n_{\alpha})s^{\mathbf{nn}}K^{\alpha\nu}+n_{\alpha}s^{\mathbf{nn}}\nabla_{\nu}K^{\alpha\nu} \notag \\
&=s^{\mathbf{nn}}K_{\alpha\nu}K^{\alpha\nu}+n_{\alpha}s^{\mathbf{nn}}\nabla_{\nu}K^{\alpha\nu}\,, \\[2ex]
0&=\nabla_{\nu}(s^{\mathbf{nn}}n^{\nu}n_{\beta}a^{\beta})=s^{\mathbf{nn}}n^{\nu}(\nabla_{\nu}n_{\beta})a^{\beta}+s^{\mathbf{nn}}n^{\nu}n_{\beta}\nabla_{\nu}a^{\beta} \notag \\
&=s^{\mathbf{nn}}a_{\beta}a^{\beta}+s^{\mathbf{nn}}n^{\nu}n_{\beta}\nabla_{\nu}a^{\beta}\,.
\end{align}
\end{subequations}
Summing the individual terms implies
\begin{equation}
2n_{\alpha}n_{\beta}\left[(T^{Rs})^{\alpha\beta}+s^{\mathbf{nn}}R^{\alpha\beta}\right]=s^{\mathbf{nn}}(K^2-K_{ij} K^{ij})-q^{\nu}_{\phantom{\nu}\alpha}\nabla_{\nu}\nabla^{\alpha}s^{\mathbf{nn}}\,.
\end{equation}
By employing
\begin{equation}
\label{eq:key-equation-purely-timelike-1}
q^{\lambda}_{\phantom{\lambda}\sigma}\nabla_{\lambda}\nabla^{\sigma}s^{\mathbf{nn}}=q^{\lambda}_{\phantom{\lambda}\alpha}q_{\sigma}^{\phantom{\sigma}\alpha}   \nabla_{\lambda}\nabla^{\sigma}s^{\mathbf{nn}}=D_{\alpha}D^{\alpha}s^{\mathbf{nn}}-\frac{1}{N}K\mathcal{L}_ms^{\mathbf{nn}}\,,
\end{equation}
we obtain
\begin{equation}
2n_{\alpha}n_{\beta}\left[(T^{Rs})^{\alpha\beta}+s^{\mathbf{nn}}R^{\alpha\beta}\right]=s^{\mathbf{nn}}(K^2-K_{ij} K^{ij})-D_iD^is^{\mathbf{nn}}+\frac{1}{N}K\mathcal{L}_ms^{\mathbf{nn}}\,,
\end{equation}
which corresponds to \eqref{eq:orthogonal-projection-purely-timelike}. We still have prove \eqref{eq:key-equation-purely-timelike-1}. It is reasonable to proceed as for \eqref{eq:key-equation-purely-spacelike}, although it turns out that the current computations are much less involved:
\begin{align}
D_{\alpha}D^{\alpha}s^{\mathbf{nn}}&=q^{\beta}_{\phantom{\beta}\alpha}\nabla_{\beta}(D^{\alpha}s^{\mathbf{nn}})=q^{\beta}_{\phantom{\beta}\alpha}\nabla_{\beta}(q^{\alpha}_{\phantom{\alpha}\gamma}\nabla^{\gamma}s^{\mathbf{nn}}) \notag \\
&=q^{\beta}_{\phantom{\beta}\alpha}(\nabla_{\beta}q^{\alpha}_{\phantom{\alpha}\gamma})\nabla^{\gamma}s^{\mathbf{nn}}+q^{\beta}_{\phantom{\beta}\alpha}q^{\alpha}_{\phantom{\alpha}\gamma}\nabla_{\beta}\nabla^{\gamma}s^{\mathbf{nn}}\,.
\end{align}
The second term just corresponds to the left-hand side of \eqref{eq:key-equation-purely-timelike-1}. The first contribution can be reformulated as
\begin{align}
q^{\beta}_{\phantom{\beta}\alpha}(\nabla_{\beta}q^{\alpha}_{\phantom{\alpha}\gamma})\nabla^{\gamma}s^{\mathbf{nn}}&=q^{\beta}_{\phantom{\beta}\alpha}(\nabla_{\beta}n^{\alpha}n_{\gamma})\nabla^{\gamma}s^{\mathbf{nn}}=q^{\beta}_{\phantom{\beta}\alpha}(\nabla_{\beta}n^{\alpha})n_{\gamma}\nabla^{\gamma}s^{\mathbf{nn}} \notag \\
&=q^{\beta}_{\phantom{\beta}\alpha}(K^{\alpha}_{\phantom{\alpha}\beta}-a^{\alpha}n_{\beta})n_{\gamma}\nabla^{\gamma}s^{\mathbf{nn}}=Kn^{\gamma}\nabla_{\gamma}s^{\mathbf{nn}}=\frac{1}{N}K\mathcal{L}_ms^{\mathbf{nn}}\,,
\end{align}
which directly implies \eqref{eq:key-equation-purely-timelike-1}.

\subsubsection{Mixed projection}

The mixed projections are given as follows:
\begin{subequations}
\begin{align}
n_{\alpha}q^k_{\phantom{k}\beta}(T_1^{Rs})^{\alpha\beta}&=n_{\alpha}q^k_{\phantom{k}\beta}(T_7^{Rs})^{\alpha\beta}=0\,, \displaybreak[0]\\[2ex]
n_{\alpha}q^k_{\phantom{k}\beta}(T_4^{Rs})^{\alpha\beta}&=n_{\alpha}q^k_{\phantom{k}\beta}\nabla_{\nu}[(\nabla^{\alpha}n^{\nu})n^{\beta}s^{\mathbf{nn}}+n^{\nu}(\nabla^{\alpha}n^{\beta})s^{\mathbf{nn}}+n^{\nu}n^{\beta}(\nabla^{\alpha}s^{\mathbf{nn}})] \notag \\
&=n_{\alpha}q^k_{\phantom{k}\beta}\left[(\nabla^{\alpha}n^{\nu})(\nabla_{\nu}n^{\beta})s^{\mathbf{nn}}+(\nabla_{\nu}n^{\nu})(\nabla^{\alpha}n^{\beta})s^{\mathbf{nn}}+n^{\nu}(\nabla_{\nu}\nabla^{\alpha}n^{\beta})s^{\mathbf{nn}}\right. \notag \\
&\phantom{{}={}}\left.{}\hspace{1.1cm}+n^{\nu}(\nabla^{\alpha}n^{\beta})\nabla_{\nu}s^{\mathbf{nn}}+n^{\nu}(\nabla_{\nu}n^{\beta})\nabla^{\alpha}s^{\mathbf{nn}}\right] \notag \\
&=n_{\alpha}q^k_{\phantom{k}\beta}\left[-n^{\alpha}a^{\nu}K_{\nu}^{\phantom{\nu}\beta}s^{\mathbf{nn}}-Kn^{\alpha}a^{\beta}s^{\mathbf{nn}}+n^{\nu}(\nabla_{\nu}K^{\alpha\beta})s^{\mathbf{nn}} \right. \notag \\
&\phantom{{}={}}\left.{}\hspace{1.1cm}-n^{\alpha}n^{\nu}(\nabla_{\nu}a^{\beta})s^{\mathbf{nn}}-n^{\alpha}n^{\nu}a^{\beta}\nabla_{\nu}s^{\mathbf{nn}}+a^{\beta}\nabla^{\alpha}s^{\mathbf{nn}}\right] \notag \\
&=a^k(s^{\mathbf{nn}}K+2n^{\nu}\nabla_{\nu}s^{\mathbf{nn}})+s^{\mathbf{nn}}n^{\nu}\nabla_{\nu}a^k\,, \displaybreak[0]\\[2ex]
n_{\alpha}q^k_{\phantom{k}\beta}(T_5^{Rs})^{\alpha\beta}&=n_{\alpha}q^k_{\phantom{k}\beta}\nabla_{\nu}\left[(\nabla^{\beta}n^{\nu})n^{\alpha}s^{\mathbf{nn}}+n^{\nu}(\nabla^{\beta}n^{\alpha})s^{\mathbf{nn}}+n^{\nu}n^{\alpha}(\nabla^{\beta}s^{\mathbf{nn}})\right] \notag \\
&=n_{\alpha}q^k_{\phantom{k}\beta}\left[(\nabla_{\nu}\nabla^{\beta}n^{\nu})n^{\alpha}n^{\mathbf{nn}}+(\nabla^{\beta}n^{\nu})n^{\alpha}\nabla_{\nu}s^{\mathbf{nn}}+n^{\nu}(\nabla_{\nu}\nabla^{\beta}n^{\alpha})s^{\mathbf{nn}}\right. \notag \\
&\phantom{{}={}}\hspace{1.1cm}\left.{}+(\nabla_{\nu}n^{\nu})n^{\alpha}\nabla^{\beta}s^{\mathbf{nn}}+n^{\nu}n^{\alpha}\nabla_{\nu}\nabla^{\beta}s^{\mathbf{nn}}\right] \notag \\
&=n_{\alpha}q^k_{\phantom{k}\beta}\left[(\nabla_{\nu}K^{\beta\nu}-K_{\nu}^{\phantom{\nu}\beta}a^{\nu})n^{\alpha}s^{\mathbf{nn}}+K^{\beta\nu}n^{\alpha}\nabla_{\nu}s^{\mathbf{nn}}+n^{\nu}\nabla_{\nu}K^{\beta\alpha}s^{\mathbf{nn}}\right. \notag \\
&\phantom{{}={}}\left.{}\hspace{1.1cm}+Kn^{\alpha}\nabla^{\beta}s^{\mathbf{nn}}+n^{\nu}n^{\alpha}\nabla_{\nu}\nabla^{\beta}s^{\mathbf{nn}}\right] \notag \\
&=-s^{\mathbf{nn}}q^k_{\phantom{k}\beta}\nabla_{\nu}K^{\beta\nu}+s^{\mathbf{nn}}a^{\nu}K_{\nu}^{\phantom{\nu}k}-K^{k\nu}\nabla_{\nu}s^{\mathbf{nn}} \notag \\
&\phantom{{}={}}+s^{\mathbf{nn}}n_{\alpha}q^k_{\phantom{k}\beta}n^{\nu}\nabla_{\nu}K^{\alpha\beta}-Kq^k_{\phantom{k}\beta}\nabla^{\beta}s^{\mathbf{nn}}-n^{\nu}q^k_{\phantom{k}\beta}\nabla_{\nu}\nabla^{\beta}s^{\mathbf{nn}} \notag \\
&=-s^{\mathbf{nn}}\nabla_{\nu}K^{k\nu}+s^{\mathbf{nn}}a_iK^{ik}-K^{ki}D_is^{\mathbf{nn}} \notag \\
&\phantom{{}={}}+s^{\mathbf{nn}}n_{\alpha}n^{\nu}\nabla_{\nu}K^{\alpha k}-KD^ks^{\mathbf{nn}}-n^{\nu}q^k_{\phantom{k}\beta}\nabla_{\nu}\nabla^{\beta}s^{\mathbf{nn}}\,, \displaybreak[0]\\[2ex]
n_{\alpha}q^k_{\phantom{k}\beta}(T_6^{Rs})^{\alpha\beta}&=n_{\alpha}q^k_{\phantom{k}\beta}\nabla_{\mu}\left[(\nabla^{\mu}n^{\alpha})n^{\beta}s^{\mathbf{nn}}+n^{\alpha}(\nabla^{\mu}n^{\beta})s^{\mathbf{nn}}+n^{\alpha}n^{\beta}\nabla^{\mu}s^{\mathbf{nn}}\right] \notag \\
&=n_{\alpha}q^k_{\phantom{k}\beta}\left[n^{\alpha}(\nabla_{\mu}\nabla^{\mu}n^{\beta})s^{\mathbf{nn}}+2n^{\alpha}(\nabla^{\mu}n^{\beta})\nabla_{\mu}s^{\mathbf{nn}}\right] \notag \\
&=n_{\alpha}q^k_{\phantom{k}\beta}\left[n^{\alpha}(\nabla_{\mu}K^{\mu\beta}-Ka^{\beta}-n^{\mu}\nabla_{\mu}a^{\beta})s^{\mathbf{nn}}+2n^{\alpha}(K^{\mu\beta}-n^{\mu}a^{\beta})\nabla_{\mu}s^{\mathbf{nn}}\right] \notag \\
&=s^{\mathbf{nn}}(-q^k_{\phantom{k}\beta}\nabla_{\mu}K^{\mu\beta}+Ka^k+q^k_{\phantom{k}\beta}n^{\mu}\nabla_{\mu}a^{\beta}) \notag \\
&\phantom{{}={}}+2(a^kn^{\mu}-K^{\mu k})\nabla_{\mu}s^{\mathbf{nn}}\,.
\end{align}
\end{subequations}
By summing all the terms, we obtain
\begin{align}
2n_{\alpha}q^k_{\phantom{k}\beta}(T^{Rs})^{\alpha\beta}&=s^{\mathbf{nn}}a_iK^{ik}+K^{ki}D_is^{\mathbf{nn}}+s^{\mathbf{nn}}n_{\alpha}n^{\nu}\nabla_{\nu}K^{\alpha k} \notag \\
&\phantom{{}={}}-KD^ks^{\mathbf{nn}}-n^{\nu}q^k_{\phantom{k}\beta}\nabla_{\nu}\nabla^{\beta}s^{\mathbf{nn}}\,.
\end{align}
To evaluate the contraction of the extra contribution given by \eqref{eq:purely-timelike-sector-T-extra-term}, we employ the contracted Codazzi-Mainardi relation stated in \eqref{eq:codazzi-mainardi-contracted}:
\begin{equation}
n_{\alpha}q^k_{\phantom{k}\beta}(s^{\mathbf{nn}}R^{\alpha\beta})=s^{\mathbf{nn}}(D_iK^{ik}-D^kK)\,.
\end{equation}
Furthermore, we use
\begin{equation}
0=s^{\mathbf{nn}}n^{\nu}\nabla_{\nu}(n_{\alpha}K^{\alpha k})=s^{\mathbf{nn}}a_iK^{\alpha i}+s^{\mathbf{nn}}n_{\alpha}n^{\nu}\nabla_{\nu}K^{\alpha k}\,,
\end{equation}
as well as
\begin{equation}
\label{eq:key-equation-purely-timelike-2}
n^{\nu}q^k_{\phantom{k}\beta}\nabla_{\nu}\nabla^{\beta}s^{\mathbf{nn}}=D^k\left(\frac{1}{N}\mathcal{L}_ms^{\mathbf{nn}}\right)-K^{ki}D_is^{\mathbf{nn}}\,,
\end{equation}
to cast the mixed projection into its final form:
\begin{align}
2n_{\alpha}q^k_{\phantom{k}\beta}\left[(T^{Rs})^{\alpha\beta}+s^{\mathbf{nn}}R^{\alpha\beta}\right]&=-D^k\left(\frac{1}{N}\mathcal{L}_ms^{\mathbf{nn}}\right)+2s^{\mathbf{nn}}(D_iK^{ik}-D^kK) \notag \\
&\phantom{{}={}}+2K^{ki}D_is^{\mathbf{nn}}-KD^ks^{\mathbf{nn}} \notag \\
&=-D^k\left(\frac{1}{N}\mathcal{L}_ms^{\mathbf{nn}}+2s^{\mathbf{nn}}K\right) \notag \\
&\phantom{{}={}}+2D_j(s^{\mathbf{nn}}K^i_{\phantom{i}k})+KD^ks^{\mathbf{nn}}\,.
\end{align}
The latter result corresponds to \eqref{eq:mixed-projection-purely-timelike}. The validity of \eqref{eq:key-equation-purely-timelike-2} remains to be shown. We start by considering the covariant derivative of the Lie derivative:
\begin{align}
D^k\left(\frac{1}{N}\mathcal{L}_ms^{\mathbf{nn}}\right)&=D^k(n^{\mu}\nabla_{\mu}s^{\mathbf{nn}})=q^k_{\phantom{k}\nu}\nabla^{\nu}(n^{\mu}\nabla_{\mu}s^{\mathbf{nn}}) \notag \\
&=q^k_{\phantom{k}\nu}(\nabla^{\nu}n^{\mu})\nabla_{\mu}s^{\mathbf{nn}}+q^k_{\phantom{k}\nu}n^{\mu}\nabla^{\nu}\nabla_{\mu}s^{\mathbf{nn}} \notag \\
&=q^k_{\phantom{k}\nu}(K^{\mu\nu}-a^{\mu}n^{\nu})\nabla_{\mu}s^{\mathbf{nn}}+q^k_{\phantom{k}\nu}n^{\mu}\nabla^{\nu}\nabla_{\mu}s^{\mathbf{nn}} \notag \\
&=K^{ki}D_is^{\mathbf{nn}}+q^k_{\phantom{k}\nu}n^{\mu}\nabla_{\mu}\nabla^{\nu}s^{\mathbf{nn}}\,,
\end{align}
as the indices of the double covariant derivative in the last term can be switched:
\begin{align}
\nabla_{\nu}\nabla_{\mu}s^{\mathbf{nn}}&=\nabla_{\nu}\partial_{\mu}s^{\mathbf{nn}}=\partial_{\nu}\partial_{\mu}s^{\mathbf{nn}}-\Gamma^{\lambda}_{\phantom{\lambda}\nu\mu}\partial_{\lambda}s^{\mathbf{nn}} \notag \\
&=\partial_{\mu}\partial_{\nu}s^{\mathbf{nn}}-\Gamma^{\lambda}_{\phantom{\lambda}\mu\nu}\partial_{\lambda}s^{\mathbf{nn}}=\nabla_{\mu}\nabla_{\nu}s^{\mathbf{nn}}\,.
\end{align}
Thus, \eqref{eq:key-equation-purely-timelike-2} is confirmed.

\subsection{Scalar sector}
\label{eq:projections-scalar-details}

The scalar sector rests on the modification $(T^{Ru})^{\alpha\beta}$ of the Einstein equations given by \eqref{eq:TuLIV}. Its purely orthogonal projection follows from
\begin{align}
n_{\alpha}n_{\beta}(T^{Ru})^{\alpha\beta}&=-n_{\alpha}n_{\beta}\nabla^{\alpha}\nabla^{\beta}u-g_{\alpha\beta}\nabla^{\alpha}\nabla^{\beta}u+u\left(R_{\mathbf{nn}}+\frac{1}{2}{}^{(4)}R\right) \notag \\
&=-q_{\alpha\beta}\nabla^{\alpha}\nabla^{\beta}u+u\left(R_{\mathbf{nn}}+\frac{1}{2}{}^{(4)}R\right)\,.
\end{align}
Here we can use Eqs.~(\ref{eq:ricci-totally-orthogonal-projection}), (\ref{eq:decomposition-ricci-scalar}), and (\ref{eq:key-equation-purely-timelike-1}) to obtain:
\begin{align}
2n_{\alpha}n_{\beta}(T^{Ru})^{\alpha\beta}&=-2D_iD^iu+\frac{2}{N}K\mathcal{L}_mu+2u\left[-\frac{1}{N}\mathcal{L}_mK+\frac{1}{N}D_iD^iN-K^{ij}K_{ij}\right. \notag \\
&\phantom{{}={}}\left.{}+\frac{1}{N}\mathcal{L}_mK-\frac{1}{N}D_iD^iN+\frac{1}{2}(R+K^2+K_{ij}K^{ij})\right] \notag \\
&=(R+K^2-K_{ij}K^{ij})u-2D_iD^iu+\frac{2}{N}K\mathcal{L}_mu\,,
\end{align}
which results in \eqref{eq:orthogonal-projection-scalar}. Furthermore, we compute the mixed projection of the modification in benefitting from \eqref{eq:key-equation-purely-timelike-2} as well as the contracted Codazzi-Mainardi relation of \eqref{eq:codazzi-mainardi-contracted}. Then,
\begin{align}
2n_{\alpha}q^k_{\phantom{k}\beta}(T^{Ru})^{\alpha\beta}&=-2n_{\alpha}q^k_{\phantom{k}\beta}(\nabla^{\alpha}\nabla^{\beta}u+\nabla^{\beta}\nabla^{\alpha}u)+2un_{\alpha}q^k_{\phantom{k}\beta}R^{\alpha\beta} \notag \\
&=2K^{ki}D_iu-2D^k\left(\frac{1}{N}\mathcal{L}_mu\right)+2u(D_iK^{ik}-D^kK) \notag \\
&=2\left[-D^k\left(\frac{1}{N}\mathcal{L}_mu\right)+D_i(uK^{ik})-uD^kK\right]\,.
\end{align}
The latter corresponds to \eqref{eq:mixed-projection-scalar}.

\section{Functional derivatives of ADM action}
\label{sec:functional-derivatives-ADM-action-computations}

In the current section we intend to present some details on the computation of functional derivatives of ADM-decomposed actions with respect to the lapse function and the shift vector. The corresponding results serve as a base for Sec.~\ref{sec:functional-derivatives-ADM-action}.

\subsection{General Relativity}

To compute functional derivatives of the ADM action for the lapse function, we benefit from the result that $K$ and $K_{ij}$ scale with $1/N$. Therefore,
\begin{equation}
\label{eq:basic-functional-derivatives}
\frac{\delta(N K_{ij}K^{ij})}{\delta N}=-K_{ij}K^{ij}\,,\quad \frac{\delta(N K^2)}{\delta N}=-K^2\,.
\end{equation}
So the result of \eqref{eq:functional-derivative-EH-lapse} immediately follows. The derivatives of the shift vector are a bit more involved. To evaluate them, implicit partial integrations with respect to the measure $\mathrm{d}^3x\sqrt{q}$ are performed where surface terms are discarded. A partial integration having been carried out is indicated by a covariant derivative acting to the left. Then we obtain the following general result:
\begin{align}
\label{eq:derivative-extrinsic-curvature-shift}
Nf\frac{\delta}{\delta N^k}K^{ij}&=\frac{f}{2}\frac{\delta}{\delta N^k}(\dot{q}^{ij}-D^iN^j-D^j N^i)=\frac{f}{2}(\delta_k^{\phantom{k}j}\overleftarrow{D}^i+\delta_k^{\phantom{k}i}\overleftarrow{D}^j) \notag \\
&=\frac{1}{2}(\delta_k^{\phantom{k}j}D^if+\delta_k^{\phantom{k}i}D^jf)\,,
\end{align}
with a spacetime-dependent function $f$, which can also be a tensor (at which its indices are omitted for simplicity). Then,
\begin{equation}
\frac{\delta(NK_{ij}K^{ij})}{\delta N^k}=2NK_{ij}\frac{\delta}{\delta N^k}K^{ij}=\delta_k^{\phantom{k}j}D^iK_{ij}+\delta_k^{\phantom{k}i}D^jK_{ij}=2D^iK_{ik}\,.
\end{equation}
Furthermore,
\begin{equation}
\frac{\delta (NK^2)}{\delta N^k}=2NK\frac{\delta}{\delta N^k}K^i_{\phantom{i}i}=q_{ik}D^iK+\delta_k^{\phantom{k}i}D_iK=2D_kK\,.
\end{equation}
A combination of the latter results implies \eqref{eq:functional-derivative-EH-shift}.

\subsection{Standard-Model Extension: $s^{\mu\nu}$ term}

We perform the analogous computations for the ADM-decomposed actions based on the three Lagrange densities of Eqs.~(\ref{eq:lagrange-density-LV-1}) -- (\ref{eq:lagrange-density-LV-3}).

\subsubsection{Purely spacelike sector}

We employ the following results:
\begin{subequations}
\begin{align}
-\frac{\delta}{\delta N}(K_{ij}\mathcal{L}_m s^{ij})&=\frac{1}{N}(K_{ij}\mathcal L_m s^{ij})\,, \displaybreak[0]\\[2ex]
\frac{\delta}{\delta N}(-s^{ij}D_{i}D_{j}N)&=-D_{j}D_{i}s^{ij}\,, \displaybreak[0]\\[2ex]
\frac{\delta}{\delta N}(Ns^{ij}R_{ij})&=s^{ij}R_{ij}\,, \displaybreak[0]\\[2ex]
\frac{\delta}{\delta N}(-2Ns^{ij}K_{il}K^{l}_{\phantom{l}j})&=2s^{ij}K_{il}K^{l}_{\phantom{l}j}\,,
\end{align}
\end{subequations}
leading to \eqref{eq:functional-derivatives-purely-spacelike}. For the derivatives with respect to the shift vector we use
\begin{subequations}
\begin{align}
\frac{\delta}{\delta N^k}(K_{ij}\mathcal{L}_m s^{ij})&=\frac{\delta(K_{ij})}{\delta N^k}\mathcal{L}_m s^{ij}+K_{ij}\frac{\delta}{\delta N^k}(\mathcal{L}_m s^{ij})\,, \\[2ex]
\frac{\delta}{\delta N^k}(2Ns^{ij}K_{il}K^{l}_{\phantom{l}j})&=(2Ns^{ij})\frac{\delta}{\delta N^k}(K_{il}K^{l}_{\phantom{l}j})\,.
\end{align}
\end{subequations}
From \eqref{eq:derivative-extrinsic-curvature-shift} we deduce that
\begin{align}
\frac{\delta(K_{ij})}{\delta N^k}\mathcal{L}_m s^{ij}&=N\left(\frac{1}{N}\mathcal{L}_ms^{ij}\right)\frac{\delta}{\delta N^k}K_{ij}=\frac{1}{2}(q_{jk}D_i+q_{ik}D_j)\left(\frac{1}{N}\mathcal{L}_ms^{ij}\right) \notag \\
&=q_{jk}{D}_i\left(\frac{1}{N}\mathcal{L}_m s^{ij}\right)\,.
\end{align}
We now evaluate the functional derivative of the Lie derivative:
\begin{align}
K_{ij}\frac{\delta}{\delta N^k}(\mathcal{L}_m s^{ij})&=-K_{ij}\frac{\delta}{\delta N^k}(\mathcal{L}_N s^{ij}) \notag \\
&=-K_{ij}\frac{\delta}{\delta N^k}\left[N^m D_ms^{ij}-(D_pN^i)s^{pj}-(D_p N^j)s^{ip}\right] \notag \\
&=-K_{ij}\left(D_ks^{ij}+\delta_k^{\phantom{k}i}s^{pj}\overleftarrow{D}_p+\delta_k^{\phantom{k}j}s^{ip}\overleftarrow{D}_p\right) \notag \\
&=-\left(K_{ij}D_ks^{ij}+K_{kj}s^{pj}\overleftarrow{D}_p+K_{ik}s^{ip}\overleftarrow{D}_p\right) \notag \\
&=-\left[K_{ij}D_ks^{ij}+2{D}_p (K_{kj}s^{pj})\right]\,.
\end{align}
Finally, applying \eqref{eq:derivative-extrinsic-curvature-shift} again, results in
\begin{align}
(2Ns^{ij})\frac{\delta}{\delta N^k}(K_{il}K^{l}_{\phantom{l}j})&=2Ns^{ij}\left(K^l_{\phantom{l}j}\frac{\delta}{\delta N^k}K_{il}+K_{il}\frac{\delta}{\delta N^k}K^l_{\phantom{l}j}\right) \notag \\
&=(q_{kl}D_i+q_{ik}D_l)(s^{ij}K^l_{\phantom{l}j})+(q_{kj}D^l+\delta_k^{\phantom{k}l}D_j)(s^{ij}K_{il}) \notag \\
&=2\left[D_i(s^{ij}K_{jk})+q_{ik}D_l(s^{ij}K^l_{\phantom{l}j})\right]\,.
\end{align}
Using these findings provides \eqref{eq:functional-derivative-purely-spacelike-shift}.

\subsection{Mixed sector}

Based on Eq.~(\ref{eq:basic-functional-derivatives}), we perform implicit partial integrations to obtain
\begin{align}
Ns^{i\mathbf{n}}\frac{\delta}{\delta N}(D_iK-D_jK^j_{\phantom{j}i})&=-D_i(Ns^{i\mathbf{n}})\frac{\delta}{\delta N}K+D_j(Ns^{i\mathbf{n}})\frac{\delta}{\delta N}K^j_{\phantom{j}i}  \notag \\
&=\frac{1}{N}D_i(Ns^{i\mathbf{n}})K-\frac{1}{N}D_j(Ns^{i\mathbf{n}})K^j_{\phantom{j}i} \notag \\
&=\frac{1}{N}D_i\left[N(s^{i\mathbf{n}}K-s^{j\mathbf{n}}K^i_{\phantom{i}j})\right]\,.
\end{align}
Adding the other contribution with the lapse function eliminated leads to
\begin{align}
\frac{\delta S^{(2)}}{\delta N}&=2\left[s^{i\mathbf{n}}D_iK-s^{j\mathbf{n}}D_iK^i_{\phantom{i}j}+\frac{1}{N}D_i(Ns^{i\mathbf{n}})K-\frac{1}{N}D_i(Ns^{j\mathbf{n}})K^i_{\phantom{i}j}\right] \notag \\
&=\frac{2}{N}D_i\left[N(s^{i\mathbf{n}}K-s^{j\mathbf{n}}K^i_{\phantom{i}j})\right]\,,
\end{align}
which implies Eq.~(\ref{eq:functional-derivative-mixed-lapse}). Furthermore, we employ Eq.~(\ref{eq:derivative-extrinsic-curvature-shift}) again to derive
\begin{subequations}
\begin{align}
2Ns^{i\mathbf{n}}\frac{\delta}{\delta N^k}D_iK&=-2N\left[\frac{1}{N}D_i(Ns^{i\mathbf{n}})\right]\frac{\delta}{\delta N^k}K^j_{\phantom{j}j} \notag \\
&=-\left\{q_{kj}D^j\left[\frac{1}{N}D_i(Ns^{i\mathbf{n}})\right]+\delta_k^{\phantom{k}j}D_j\left[\frac{1}{N}D_i(Ns^{i\mathbf{n}})\right]\right\} \notag \\
&=-2D_k\left[\frac{1}{N}D_i(Ns^{i\mathbf{n}})\right]=-2D_k(a_is^{i\mathbf{n}}+D_is^{i\mathbf{n}})\,,
\end{align}
as well as
\begin{align}
2Ns^{i\mathbf{n}}\frac{\delta}{\delta N^k}(D_jK_i^{\phantom{i}j})&=-2N\left[\frac{1}{N}D_j(Ns^{i\mathbf{n}})\right]\frac{\delta}{\delta N^k}K_i^{\phantom{i}j} \notag \\
&=-\left\{\delta_k^{\phantom{k}j}D_i\left[\frac{1}{N}D_j(Ns^{i\mathbf{n}})\right]+q_{ki}D^j\left[\frac{1}{N}D_j(Ns^{i\mathbf{n}})\right]\right\} \notag \\
&=-\left\{D_i\left[\frac{1}{N}D_k(Ns^{i\mathbf{n}})\right]+q_{ki}D^j\left[\frac{1}{N}D_j(Ns^{i\mathbf{n}})\right]\right\} \notag \\
&=-\left[D_i(a_ks^{i\mathbf{n}}+D_ks^{i\mathbf{n}})+q_{ki}D^j(a_js^{i\mathbf{n}}+D_js^{i\mathbf{n}})\right]\,.
\end{align}
\end{subequations}
These findings result in Eq.~(\ref{eq:functional-derivative-mixed-shift}).

\subsection{Purely timelike sector}

Here we use that
\begin{equation}
\frac{\delta}{\delta N}(K\mathcal{L}_ms^{\mathbf{nn}})=-\frac{K}{N}\mathcal{L}_ms^{\mathbf{nn}}\,,
\end{equation}
which is a consequence of Eq.~(\ref{eq:basic-functional-derivatives}). In addition,
\begin{equation}
\frac{\delta}{\delta N}(s^{\mathbf{nn}}D_iD^iN)=D_iD^is^{\mathbf{nn}}\,.
\end{equation}
These findings immediately provide Eq.~(\ref{eq:functional-derivaties-purely-timelike}). To obtain the derivatives for the shift vector, we need
\begin{equation}
\frac{\delta}{\delta N^k}(\mathcal{L}_ms^{\mathbf{nn}})=-\nabla_ks^{\mathbf{nn}}\,,
\end{equation}
as well as
\begin{align}
\mathcal{L}_ms^{\mathbf{nn}}\frac{\delta}{\delta N^k}K&=N\left(\frac{1}{N}\mathcal{L}_ms^{\mathbf{nn}}\right)\frac{\delta}{\delta N^k}K^i_{\phantom{i}i} \notag \\
&=\frac{1}{2}\left[q_{ki}D^i\left(\frac{1}{N}\mathcal{L}_ms^{\mathbf{nn}}\right)+\delta_k^{\phantom{k}i}D_i\left(\frac{1}{N}\mathcal{L}_ms^{\mathbf{nn}}\right)\right] \notag \\
&=D_k\left(\frac{1}{N}\mathcal{L}_ms^{\mathbf{nn}}\right)\,,
\end{align}
and
\begin{equation}
Ns^{\mathbf{nn}}\frac{\delta}{\delta N^k}(K^2-K_{ij}K^{ij})=2D_k(Ks^{\mathbf{nn}})-2D^i(s^{\mathbf{nn}}K_{ik})\,,
\end{equation}
which follow from Eq.~(\ref{eq:derivative-extrinsic-curvature-shift}). Thus, we arrive at Eq.~(\ref{eq:functional-derivative-purely-timelike-shift}).

\section{Counting scheme}
\label{sec:counting-scheme}

To check the consistency of expressions in the context of the ADM formalism, it turned out to be valuable to associate a set of ``units'' to the various quantities that play a role in this paper. These units count how often the induced metric $q_{ij}$ (or the four-dimensional spacetime metric $g_{\mu\nu}$) occurs in a certain expression. Hence, we start with
\begin{subequations}
\label{eq:units-metric}
\begin{align}
\label{eq:units-metric-3dim}
[q_{ij}]&=1\,,\quad [q^i_{\phantom{i}j}]=[q_i^{\phantom{i}j}]=0\,,\quad [q^{ij}]=-1\,, \displaybreak[0]\\[1ex]
[\sqrt{q}\,]&=\frac{3}{2}\,, \displaybreak[0]\\[1ex]
[g_{\mu\nu}]&=1\,,\quad [g^{\mu\nu}]=-1\,,\quad [\sqrt{-g}\,]=2\,.
\end{align}
\end{subequations}
Then, from Eqs.~(\ref{eq:spacetime-metric}), (\ref{eq:nmu-upper}) we deduce immediately that
\begin{subequations}
\label{eq:units-shift-lapse}
\begin{align}
[N_i]&=1\,,\quad [N^i]=0\,,\quad [N]=\frac{1}{2}\,, \displaybreak[0]\\[1ex]
\label{eq:units-flow}
[n_{\mu}]&=\frac{1}{2}\,,\quad [n^{\mu}]=-\frac{1}{2}\,.
\end{align}
\end{subequations}
We define both partial derivatives and covariant derivatives with lower indices as ``dimensionless,'' which is why
\begin{subequations}
\begin{align}
[\dot{q}_{ij}]&=1\,,\quad [\dot{N}]=\frac{1}{2}\,, \displaybreak[0]\\[1ex]
[D_iN_j]&=1\,,\quad [\dot{N}^i]=0\,,\quad [\dot{N}_i]=1\,, \displaybreak[0]\\[1ex]
[a_{\mu}]&=0\,,\quad [a^{\mu}]=-1\,.
\end{align}
\end{subequations}
The latter follow from \eqref{eq:acceleration}. From the definition of the Christoffel symbol, the Riemann tensor and its contractions as well as from Eqs.~(\ref{eq:units-metric-3dim}), we obtain quickly
\begin{subequations}
\begin{align}
[\Gamma^i_{\phantom{i}jk}]=0\,,\quad [R^i_{\phantom{i}jkl}]=0\,,\quad [R_{ijkl}]=1\,, \displaybreak[0]\\[1ex]
[R_{ij}]=0\,,\quad [R]=-1\,,\quad [R^{ij}]=-2\,.
\end{align}
\end{subequations}
These results make sense. First, the Christoffel symbol involves a metric and its inverse, which is why their dimensions cancel. The Riemann tensor $R^i_{\phantom{i}jkl}$ contains both products of Christoffel symbols and derivatives of the latter, whereupon no additional metric occurs. The Ricci tensor $R_{ij}$ directly follows from contracting the Riemann tensor without introducing additional metric tensors, as the indices are already at appropriate positions. Finally, to obtain the Ricci scalar from the Ricci tensor $R_{ij}$, we must introduce an inverse metric.

The form of the EH Lagrange density~(\ref{eq:einstein-hilbert-lagrange-density}) and Hamilton density~(\ref{eq:einstein-hilbert-hamilton-density}) implies immediately
\begin{equation}
\label{eq:units-lagrange-hamilton-densities}
[\mathcal{L}]=[\mathcal{H}]=1\,.
\end{equation}
Based on \eqref{eq:extrinsic-curvature-trace}, we choose
\begin{equation}
[K_{ij}]=\frac{1}{2}\,,\quad [K]=-\frac{1}{2}\,,\quad [K^{ij}]=-\frac{3}{2}\,,
\end{equation}
which is consistent with \eqref{eq:units-lagrange-hamilton-densities}. From the definitions of the canonical momenta we obtain
\begin{subequations}
\begin{align}
[\pi^{ij}]=0\,,\quad [\pi]=1\,,\quad [\pi_{ij}]=2\,, \displaybreak[0]\\[2ex]
[\pi_i]=1\,,\quad [\pi^i]=0\,,\quad [\pi_N]=\frac{1}{2}\,.
\end{align}
\end{subequations}
Finally, it is possible to assign the same units to the diffeomorphism-violating background fields. Equation~(\ref{eq:minimal-gravity-sme-reformulated}) provides
\begin{subequations}
\begin{align}
[s_{\mu\nu}]&=1\,,\quad [s_{\mu}^{\phantom{\mu}\nu}]=[s^{\mu}_{\phantom{\mu}\nu}]=0\,, \displaybreak[0]\\[2ex]
[s^{\mu\nu}]&=-1\,,\quad [u]=0\,,\quad [t^{\mu\nu\varrho\sigma}]=-2\,.
\end{align}
\end{subequations}
The reason for these results is that $s_{\mu\nu}$ plays an analogous role as the metric, i.e., it is contracted with $R^{\mu\nu}$. Also,
\begin{equation}
s^{\mathbf{nn}}=0\,,\quad s^{\mathbf{n}i}=-\frac{1}{2}\,,
\end{equation}
which follows from their definitions under \eqref{eq:contraction-s-ricci} as well as \eqref{eq:units-flow}. A generic rule is that each lower Lorentz index leads to a dimension of 1/2, whereas each upper one implies a dimension of $-1/2$. For consistency, each term in a sum of contributions must have the same dimension.


\bibliographystyle{JHEP}

%
%
%
%
%
%
%

\end{document}